\documentclass[aps,prb,floatfix,nopacs,superscriptaddress,twocolumn,reprint]{revtex4-2}
\usepackage[utf8]{inputenc}
\usepackage{amsmath}
\usepackage{amsfonts}
\usepackage{graphicx}
\usepackage{bm}
\usepackage[dvipsnames]{xcolor}
\usepackage{physics}

\newcommand{\bb}[1]{\mathbf{#1}}
\newcommand{\m}[1]{\mathcal{#1}}
\usepackage{url}
\usepackage{mathtools}

\usepackage[american,]{babel}
\usepackage[T1]{fontenc}
\usepackage{hyperref}
\usepackage{xcolor}
\hypersetup{colorlinks,bookmarksopen,bookmarksnumbered,
citecolor=[rgb]{0.255,0.412,0.882},
linkcolor=[rgb]{0.255,0.412,0.882},
pdfstartview=false,
urlcolor=[rgb]{0.255,0.412,0.882}}

\begin{document}

\title{Diagrammatic Monte Carlo for Dissipative Quantum Impurity Models}

\author{Matthieu Vanhoecke}
\email{matthieu.vanhoecke@college-de-france.fr}
\affiliation{JEIP, UAR 3573 CNRS, Coll\`ege de France,   PSL  Research  University, 11,  place  Marcelin  Berthelot,75231 Paris Cedex 05, France}
\author{Marco Schir\`o}
\affiliation{JEIP, UAR 3573 CNRS, Coll\`ege de France,   PSL  Research  University, 11,  place  Marcelin  Berthelot,75231 Paris Cedex 05, France}

\begin{abstract}
We develop a diagrammatic Monte Carlo method for the real-time dynamics of dissipative quantum impurity models. These are small open quantum systems with interaction and local Markovian dissipation, coupled to a large quantum bath. Our algorithm sample the hybridization expansion formulated on a single real-time contour, rather than on the double Keldysh one, as it naturally arises in the thermofield/vectorized representation of the Lindblad dynamics. We show that local Markovian dissipation generally helps the convergence of the diagrammatic Monte Carlo sampling by reducing the sign problem, thus allowing to reach longer time scales as compared to the conventional unitary case. We apply our method to an Anderson impurity model in presence of local dephasing and discuss its effect on the charge and spin dynamics of the impurity.
\end{abstract}

\maketitle

\section{Introduction}
\label{sec:intro}

Quantum impurity models represent the simplest non-trivial class of quantum many-body problems, where interaction and correlation effects involve only a finite number of degrees of freedom, the impurity. This is in turn coupled to an extended set of harmonic modes representing the bath or environment.  Examples of these models emerge ubiquitously in condensed matter, atomic physics and quantum optics, from the Caldeira-Leggett model of a dissipative two-level system~\cite{leggett1987dynamics} to the Kondo effect of magnetic impurities in metals or quantum dots in nanostructures~\cite{Pustilnik_2004} to the decay of a driven atom in a cavity~\cite{Haroche:993568}. 

While sharing the general setting of an open quantum system, 
much of the emergent low-energy, long-time physics in these models is controlled by the spectral properties of their respective environments. These can be rather different, ranging from a gapless bath with power-law correlations for the conduction electrons of a metal at zero temperature, to fast, featureless Markovian environments used to describe for example charge transport at high-temperature or photonic degrees of freedom in atomic physics and quantum optics platforms. As such, traditional studies have treated these two as rather separate classes of dissipative quantum systems~\cite{breuerPetruccione2010,weissquantumdissipative}.

The recent development of quantum simulators and noisy intermediate scale quantum devices has brought forth a variety of platforms where different types of dissipative environment can coexist and be controlled with high degree of tunability~\cite{google_dephasing_abanin}. Experiments with ultracold atoms, for example, have realised quantum transport through a dissipative quantum point contact~\cite{lebrat2019quantized,corman2019quantized,huang2023superfluid}, where the constriction between two quantum conductors is exposed to additional particle losses. 
Celebrate quantum impurity models such as the Anderson or the Kondo model have been realised with ultracold alkaline-earth atoms~\cite{riegger2018localized,zhang2020controlling} which are naturally exposed to correlated dissipative processes, such as dephasing due to spontaneous emission~\cite{gerbier2010heating,bouganne2020} or two-body losses due to inelastic scattering~\cite{garcia-ripoll2009,TomitaEtAlScienceAdv17,honda2022observation}. In solid state platforms one can couple quantum dots to a quantum point contact~\cite{avinun2004controlled,kang2007entanglement,aono2008dephasing} or to monitoring environments~\cite{sukhorukov2007conditional,ferguson2023measuremnt}
to study the effect of dephasing or quantum measurements on the Kondo effect~\cite{hasegawa2021kondo}. Finally, superconducting circuits are emerging as platform to explore the role of local dissipation in a controlled way~\cite{google_dephasing_abanin,mi2023stable}.

These developments have triggered the interest around a new class of \emph{dissipative quantum impurity models}, where the impurity is both coupled to a quantum bath, i.e. a structured frequency-dependent environment, and exposed to fast Markovian dissipation describing incoherent processes such as particle losses or dephasing, that can often be modelled within a Lindblad master equation~\cite{breuerPetruccione2010}.  The physics of these dissipative quantum impurity models has started only recently to be explored, with a focus on non-interacting chains with localised single particle losses~\cite{Froml2019,damanet2019controlling,visuri2022symmetry,visuri2023nonlinear,visuri2023dc} or pumps~\cite{krapivsky2019free,krapivsky2020free} or local dephasing~\cite{Scarlatella2019,tonielli2019orthogonality,dolgirev2020nongaussian,ferreira2023exact}. Non-Hermitian quantum impurity models, arising from a postselection over quantum trajectories, have also been studied~\cite{Nakagawa2018,
yoshimura2020nonhermitian,stefanini2023orthogonality}. In addition to their intrinsic interest, dissipative quantum impurity models also arise as effective description of open Markovian lattice models in the large connectivity limit, within Dynamical Mean-Field Theory~\cite{scarlatella2021dynamical}.

Despite these recent progresses the physics of dissipative quantum impurities is still largely unexplored, particularly concerning the interplay between local dissipation and strong correlations. This is in part because the range of methods and techniques to solve them efficiently and numerically exactly is rather limited. Several techniques have been developed in the past decade to study the real-time dynamics of unitary quantum impurity models ranging from time-dependent Numerical Renormalization group~\cite{anders2005realtime}, Matrix Product States and their extensions~\cite{heidrich2009realtime,schwarz2018nonequilibrium,Kohn_2022,wauters2023simulations}, or auxiliary master equation approaches~\cite{dorda2014auxiliary,chen2019auxiliary,Chen_2019}. We note a recent development using matrix product state representation in the temporal domain~\cite{strathearn2018efficient,gribben2022exact,jorgensen2019exploiting,thoenniss2023efficient,thoenniss2023nonequilibrium,ng2023realtime,park2024tensor} which is particularly promising.
Diagrammatic Monte Carlo methods, which are the workhorse for imaginary time dynamics, suffer from a severe sign problem which limits in practice their applicability~\cite{schiroPRB2009,schiroFabrizioPRB2009,muhlbacherRabaniPRL2008,Werner_Keldysh_09}, although recent developments have significantly pushed this boundary~\cite{cohen2015taming,corentin2019quantum,macek2020quantum,nunez2022learning,erpenbeck2023quantum}.


In this work we develop a real-time Diagrammatic Monte Carlo (DiagMC) algorithm to tackle dissipative quantum impurity models. The idea is to combine the real-time hybridization expansion algorithm~\cite{schiroPRB2009} with the formalism used to solve Lindblad Markovian problems, often called vectorization or super-fermion representation~\cite{dzhioev2011,HARBOLA2008191,dorda2014auxiliary,Arrigoni2018,werner2023configuration}, in such a way to include local dissipation into the solution of the atomic limit and sample the hybridization expansion in the resulting vectorized Hilbert space. A similar strategy was developed in Refs.~\cite{Scarlatella2019,scarlatella2023selfconsistent} leading to  to a self-consistent diagrammatic theory in the hybridization (Non-Crossing Approximation and its extensions). Here instead we sample all diagrams entering the hybridization expansion using DiagMC. We formulate the algorithm in the most general terms and apply it to the case in which the jump operators are diagonal in the occupation of the impurity, leading to a generalised segment picture~\cite{Gull_RMP11}. As a non trivial application we study the dynamics of an Anderson Impurity Model (AIM) in presence of local dephasing. We show that strong local dissipation helps the convergence of the diagrammatic expansion, reducing the average number of vertex and thus the sign problem, allowing to reach longer time scales than in the usual hybridization expansion algorithm~\cite{schiroPRB2009}. Our results for the charge and spin dynamics of the AIM reveal that the former is strongly slown down by a large local dephasing, a signature of the Zeno effect, while the latter is only partially affected by dissipation. On the other hand, we show that an asymmetric dephasing for the two spin species results in the formation of a metastable state with finite impurity magnetization.

The paper is organized as follows. In Sec.~\ref{sec:dQIM} we introduce the general dissipative quantum impurity model and present a brief recap of the vectorization formalism. In Sec.~\ref{sec:hybexp} we formulate the hybridization expansion in this extended Hilbert space formalism, while in Sec.~\ref{sec:DMC} we describe the diagMC algorithm we developed to sample the hybridization expansion.  Sec.~\ref{sec:results} contains our main results for the Anderson Impurity Model with Dephasing, including an analysis of the algorithm performance, benchmarks in the non-interacting case and the results on charge, spin and entanglement dynamics. Sec.~\ref{sec:conclusions} is devoted to conclusions. Two Appendix complete this work with additional technical details.

\section{Dissipative Quantum Impurity Models}
\label{sec:dQIM}

The aim of this section is to introduce the model and setting we will be focusing throughout this work, namely dissipative quantum impurities and their out of equilibrium dynamics. To this purpose, we consider a small quantum system with a finite number of fermionic degrees of freedom 
 $\{ d_\sigma , d_{\sigma'}^\dagger\} = \delta_{\sigma,\sigma'}$ where the label $\sigma$ may include both spin and orbital degrees of freedom , and described by a local Hamiltonian $H_{I}$, in the present case:
\begin{align}
    H_{I}\left[\{d_\sigma , d_{\sigma'}^\dagger\} \right] = \sum_{\sigma} \epsilon_{\sigma} d_{ \sigma}^\dagger d_{\sigma}   + H_{U} \left[ \{d_\sigma , d_{\sigma'}^\dagger \} \right]
\end{align}
where $H_{U} \left[ \{d_\sigma , d_{\sigma'}^\dagger \} \right]$ contains the many-body interactions, which at this stage are not necessary diagonal in the spin or orbital degrees of freedom.
These quantum levels are coupled to one or more non-interacting baths, i.e described by a free fermions Hamiltonian $H_B=\sum_{\bb{k},\sigma}\varepsilon_{\bb{k}}c^\dagger_{\bb{k},\sigma}c_{\bb{k},\sigma}$ with fermionic bath operators $c_{\bb{k},\sigma} ,c^\dagger_{\bb{k},\sigma} $. In order to simplify we only consider a linear coupling with the bath, described by the Hamiltonian $H_{IB}$:
\begin{align}
    H_{IB} = \sum_{\bb{k},a} \left(V_{\bb{k},\sigma} d_{\sigma}^\dagger c_{\bb{k},\sigma} + h.c \right)
\end{align}
thus, a generic quantum impurity model is described by the following Hamiltonian:
\begin{align}\label{eqn:Hfull}
    H = H_{I} + H_B + H_{IB}
\end{align}
\begin{figure}[t]
	\includegraphics[width=0.40\textwidth]{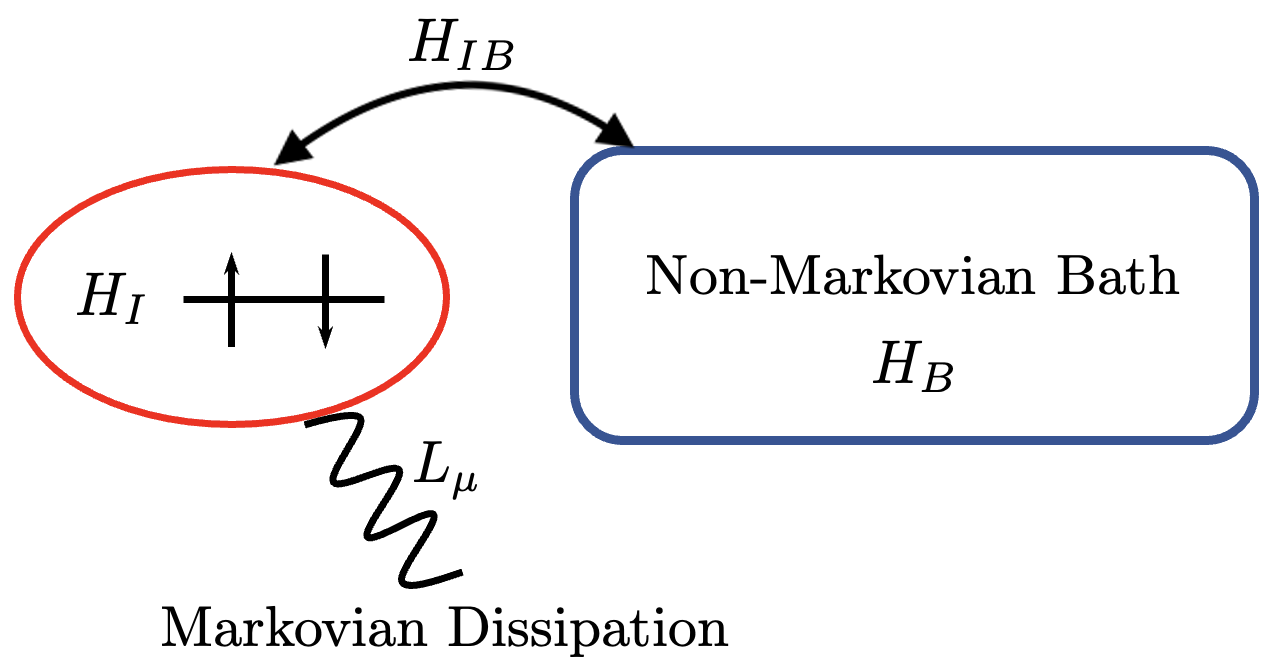}
	\caption{\label{fig0:sketch} Cartoon of the set-up: a dissipative quantum impurity model consisting of a local fermionic level (i.e. an interacting dot with Hamiltonian $H_{I}$) coupled to a fermionic bath $H_B$ through the hybridization $H_{SB}$ and exposed to dissipative Markovian processes with jump operators $L_{\mu},L^{\dagger}_{\mu}$.}
\end{figure}
In addition to the local interactions described by $H_{U}$ we are interested in a situation where the impurity is exposed to local dissipative processes, that we assume to be Markovian and modelled by a Lindblad Master Equation~\cite{breuerPetruccione2010}. This dissipative processes originate from some fast Markovian environment, whose microscopic degrees of freedom are not under our control and so can be traced out from the start. This has to be contrasted with the quantum bath described by the fermions $c_{\bb{k},\sigma} ,c^\dagger_{\bb{k},\sigma} $ which play a key role in the many-body physics of the quantum impurity. As a result of this local dissipation the entire system (quantum bath plus quantum impurity) is described by a density matrix $\rho(t)$ which evolves in time according to the Lindblad equation
\begin{align}\label{eqn:lindblad}
    \partial_t \rho(t) = -i \left[ H, \rho(t) \right] + \sum_{\mu} L_\mu \rho(t) L_\mu^\dagger -\frac{1}{2} \{ L_\mu^\dagger L_\mu , \rho(t) \}
\end{align}
where $H$ is the impurity plus bath Hamiltonian in Eq.~(\ref{eqn:Hfull}) and we have denoted $L_\mu , L_\mu^\dagger $ the jump operators for the impurity system only, 
that is to say they are written only in functions of the operators $d_\sigma , d_\sigma^\dagger$. A sketch of the setup we are considering in this work is provided in Fig.~\ref{fig0:sketch}.

Since we are interested in the non-equilibrium dynamics we want to determine the time evolution of the density matrix $\rho(t)$ starting from an initial configuration given by $\rho(0)$.
For this one, in principle we could prepare our system in a thermal state, then drive the system out of equilibrium, in this case the initial density matrix can be determined by the Boltzmann distribution at a temperature $\beta^{-1}$.
In this paper, we assume to start from a decoupled situation, where the fermionic bath is in thermal equilibrium at temperature $T=0$, while the impurity is prepared in a given initial density matrix $\rho_I(0)$. The initial density matrix for the whole system then factorize:
\begin{align}
    \rho(0) = \rho_I(0) \otimes \rho_B(0)
\end{align}
where $\rho_B(0)$ is a thermal density matrix for the fermions in the bath while $\rho_I(0)$ depends on the initial preparation for the impurity and will be specified later. 
Then we let the entire system evolve under the action of the Lindblad master equation~(\ref{eqn:lindblad}). We note that in principle an initial state with finite impurity-bath correlations could be also implemented within diagrammatic Monte Carlo, by adding a third branch (imaginary-time axis) on the real-time contour, as done in the unitary case~\cite{schiroPRB2009}.

In this work we will be mainly interested in properties of the impurity which can be computed from the reduced impurity density matrix, obtained after tracing out the fermionic degrees of freedom of the quantum bath.

 \subsection{Vectorization and Tilde Space}

In this section we set-up the theoretical framework we will use to study non equilibrium  dynamics in dissipative quantum impurity models, in particular to obtain the hybridization expansion that will be sampled through diagrammatic Monte Carlo. As a first step we discuss how to reformulate the Linbdlad master equation, which is an equation for the density matrix written in terms of a Linbdlad super-operator, in terms of a non-unitary evolution for a vector state which represents a purification of the density matrix and lives in an enlarged Hilbert space. This formalism, sometime referred to as vectorization, third-quantization~\cite{Prosen_2008}, superfermion representation~\cite{dzhioev2011,HARBOLA2008191,dorda2014auxiliary,Arrigoni2018,werner2023configuration} or thermofield~\cite{takahashi96,ojima1981} depending on the communities~\cite{secliphd,mcdonald2023third}, will make the development of the hybridization expansion and of the DiagMC algorithm rather natural as we are going to see in Sec.~\ref{sec:hybexp}.
The advantage of the vectorization formalism is that the superoperator structure usually needed to treat Lindbladian problems and the associated hybridization expansion is now encoded by doubling the local Hilbert space and working with an additional quantum number, similar to an orbital degrees of freedom in conventional diagrammatic Monte Carlo. As a by product the diagrammatic expansion will be formulated on a single real-time contour, rather than on the Keldysh one, the additional label keeping track of the information on whether operators are on the upper/lower branch of the contour.

As a warmup we start describing the vectorization for a single site fermionic problem, which could describe for example the isolated impurity. The Hilbert space is spanned by the orthonormal Fock basis $\vert n\rangle$, with $n=0,1$ and in this space the identity operator is written as
\begin{align}
    I  = \sum_{n} \vert n \rangle  \langle n \vert
\end{align}
In this basis any operator, including the density matrix $\rho$, reads
\begin{align}
    O = \sum_{n,m} O_{n,m}  \vert n \rangle \langle m \vert 
\end{align}
Now, we want to duplicate the physical Hilbert space $\m{H}$ and purify the density matrix. We introduce therefore an auxiliary tilde space $\tilde{\m{H}}$ with orthonormal basis  $\vert \tilde{n}\rangle$, where we can also introduce the identity
\begin{align}
    \tilde{I}  = \sum_{n} \vert \tilde{n} \rangle  \langle \tilde{n} \vert \,.
\end{align}
We can then define fermionic operators in the new Hilbert space, respectively $\{c_{n} ,c^\dagger_n \}$ in the Hilbert space $\m{H}$ and  $\{\tilde{c}_{n} ,\tilde{c}^\dagger_n \}_n$ in $\tilde{\m{H}}$,  satisfying the usual algebra: 
\begin{align}
    \{ c_\alpha , c^\dagger_\beta \} = \delta_{\alpha,\beta} \quad  \{ \tilde{c}_\alpha , \tilde{c}_\beta^\dagger \} =\delta_{\alpha,\beta}
\end{align}
and with all the other anticommutators equal to zero.  The key step is now to vectorize the identity operator, introducing the left vacuum~\cite{dorda2014auxiliary,dzhioev2011} (or vectorized identity)
\begin{align}
    \vert I  \rangle = \sum_{n} \left( -i \right)^{n} \vert n \rangle \otimes | \tilde{n} \rangle
\end{align}
The vectorized identity is particularly useful as it allows to write any operator, in terms of a vector, for example if we can write:
\begin{align}
    \vert O \rangle = O \vert I \rangle = O \otimes \Tilde{I} \vert I\rangle
\end{align}
In particular, the vectorized density matrix reads:
\begin{align}
    \vert \rho \rangle = \rho \vert I \rangle
\end{align}
In the vectorization formalism we can evaluate the average of an operator over the density matrix $\rho(t)$ as
\begin{align}\label{eqn:O_t}
   \langle O(t) \rangle = \Tr{ \rho(t)O} = & \langle I \vert O \vert \rho(t) \rangle 
\end{align}
Since we are interested in the dynamics of the impurity density matrix, we have to write the Lindblad in the Superfermions representation and then write the formal solution of the Lindblad Master equation.

\subsection{Vectorization of the Lindbladian}

We can apply the superfermion formalism to the case of the master equation for a dissipative quantum impurity model, i.e. to Eq.~(\ref{eqn:lindblad}). To this extent we introduce the Hilbert spaces $\m{H}$ and its doubled tilde-version $\tilde{\m{H}}$ and duplicate all the degrees of freedom in the problem, namely the impurity and the bath fermions, and introduce the associated creation/annihilation operators $d_{\sigma},\tilde{d}_\sigma$ and $c_{\bb{k},\sigma},\tilde{c}_{\bb{k},\sigma}$ and their Hermitian conjugate. In terms of these degrees of freedom we can rewrite the Linblad master equation as a non-unitary Schrodinger type of equation~\cite{dzhioev2011,dorda2014auxiliary}
$$
\partial_t\vert\rho\rangle=\mathcal{L}\vert\rho\rangle
$$
where the Lindbladian $\mathcal{L}$ has now two contributions
\begin{equation}\label{eqn:L_vect}
    \m{L} = \m{L}_0 + \m{L}_{IB}
\end{equation}
the first one $\m{L}_0$ is the free Lindbladian for the dissipative impurity and the bath, and the second one $\m{L}_{IB}$ is the coupling term between the two subspaces.  By using the super-fermions rules~\cite{ojima1981,dorda2014auxiliary} ($d_\sigma | I \rangle  = -i \tilde{d}^\dagger_\sigma |I\rangle$ and $d_\sigma^\dagger | I \rangle  = -i \tilde{d}_\sigma |I\rangle$) and since we consider only the dissipation on the impurity degrees of freedom, we can formally write the impurity  Lindbladian $\m{L}_0$ as:
\begin{widetext}
    \begin{equation}
    \m{L}_0 = -i\left(H_I + H_B - \Tilde{H}_I - \Tilde{H}_B \right) + \sum_\mu \left(s_{L_\mu} L_\mu \Tilde{L}_\mu - \frac{1}{2} L_\mu^\dagger L_\mu - \frac{1}{2} \Tilde{L}_\mu^\dagger \Tilde{L}_\mu \right)
\end{equation}
\end{widetext}
where $s_{L_\mu}$ is an extra sign depending on the fermionic ($s_{L_\mu}=-i$) or bosonic ($s_{L_\mu}=1$) nature of the jumps operator. For the second contribution to Eq.~(\ref{eqn:L_vect}), the impurity-bath Lindbladian, we can write it in compact form by introducing the following fields
\begin{align}\label{eqn:phi_psi}
    \Phi_{\sigma} = \sum_{\bb{k}} V_\bb{k} \begin{pmatrix}
        c_{\bb{k},\sigma} \\ 
        \Tilde{c}_{\bb{k},\sigma}^\dagger
    \end{pmatrix} \quad \Psi_\sigma = \begin{pmatrix}
        d_{\sigma} \\ 
        \Tilde{d}_{\sigma}^\dagger
    \end{pmatrix}
\end{align}
which group together the operators living in the space $\m{H}$ and $\tilde{\m{H}}$. Using these fields we can write the system-bath term in a more compact way:
\begin{equation}\label{eqn:L_SB}
    \m{L}_{SB} = -i \sum_{\sigma\alpha} \left( \Bar{\Phi}^{\alpha}_{\sigma} \Psi^{\alpha}_\sigma + \Bar{\Psi}^{\alpha}_\sigma \Phi^{\alpha}_{\sigma} \right)
\end{equation}
where we have introduced a label $\alpha=0,1$ which denotes the Hilbert space $\m{H}$ or $\tilde{\m{H}}$ ($d_\sigma = \Psi_\sigma^{\alpha=0}$ and $\tilde{d}^\dagger_\sigma = \Psi_\sigma^{\alpha=1}$)
At this point we can write the formal solution of the vectorized master equation as 
\begin{equation}\label{eqn:rho_t}
    \vert \rho(t) \rangle = \m{T}_t \exp\left(  \int_0^t \m{L}(s) ds \right) \vert \rho(0) \rangle 
\end{equation}
where we have introduced the time ordering operator $\m{T}_t$ in the Superfermions representation. Unlike the standard Keldysh time-ordering, here the time ordering is defined as:
\begin{align}
    t_{\alpha} > \bar{t}_{\beta} = \left\{ \begin{array}{ll}
    t>\bar{t} \quad \text{if} \quad \alpha = \beta \in \m{H},\Tilde{\m{H}} \\ 
    \alpha \in \m{H} \quad \beta \in \tilde{\m{H}}
    \end{array} \right.
\end{align}
This ordering allows to define a time-ordering operator $\m{T}_t$ such that two operators, $\psi_1$ and $\psi_2$, being $\psi$ a creation or annihilation fermionic operator living in the $\m{H}(\Tilde{\m{H}})$ Hilbert space, anticommute under time-ordering:
\begin{align}
    \m{T}_t \psi_1(t_\alpha) \psi_2(t_\beta) = \left \{ \begin{array}{ll}
    \psi_1(t_\alpha) \psi_2(t_\beta) \quad \text{if} \quad t_\alpha > t_\beta \\
    - \psi_1(t_\alpha) \psi_2(t_\beta) \quad \text{otherwise 
} \end{array} \right.
\end{align}
Eq.~(\ref{eqn:rho_t}) represents the starting point to perform the hybridization expansion, namely an expansion order by order in the system-bath coupling $\mathcal{L}_{IB}$, as we will discuss in the next section.


\section{Hybridization Expansion}\label{sec:hybexp}

In this section we derive for completeness the hybridization expansion in the vectorized formulation of our dissipative quantum impurity model. This type of expansion was first derived for dissipative impurities using the superoperator formalism in Ref.~\cite{Scarlatella2019}.

As in the standard hybridization expansion~\cite{schiroPRB2009} the starting point is to write down the trace of density matrix as a dynamical partition function $Z=\mbox{Tr}[\rho(t)]$. In the vectorized formalism this amount to evaluate $\langle I\vert\rho(t)\rangle$. Using the formal solution of the vectorized master equation, Eq.~(\ref{eqn:rho_t}), that we write in the interaction picture with respect to the free Lindbladian $\m{L}_0$, we obtain
\begin{align}
  \langle I \vert \rho(t) \rangle = \langle I \vert e^{\m{L}_0t}   \m{T}_{t} \exp \left( \int_0^t d\tau \m{L}_{IB}(\tau) \right)|\vert \rho(0) \rangle 
\end{align}

Then, we Taylor expand the time-ordered exponential in power of the impurity-bath hybridization, $\mathcal{L}_{SB}$, 
\begin{align}
       \langle I\vert \rho(t) \rangle = \langle I \vert e^{\m{L}_0t}  \sum_n \frac{1}{n!} \int_0^t \prod_i^n dt_i  \m{T}_t \left[ \m{L}_{IB}(t_1) \cdots \m{L}_{IB}(t_n) \right] \vert \rho(0) \rangle
\end{align}
and take the average over the bath and the impurity degrees of freedom, using the fact that the initial state $\vert\rho(0)\rangle$ is factorized. Since the Lindladian $\m{L}_{SB}$ is bilinear in terms of the bath and impurity operators, it comes directly that only the even terms contribute to the expansion. 
Using Eq.~(\ref{eqn:L_SB}) for the system-bath Lindbladian 
and by factoring the bath operators we can obtain the hyrbdiziation expansion as
\begin{widetext}
\begin{align}\label{eqn:hyb_exp}
    \langle I  \vert \rho(t) \rangle  =
      \sum_n \frac{(-i)^n}{(n!)^2}
     \int_0^t & \prod_{i=1}^{n} dt_i d\Bar{t}_i \sum_{\{\sigma, \Bar{\sigma}\}} \sum_{\{\alpha, \bar{\alpha }\}}  \langle I \vert  \m{T}_t \left[e^{\m{L}_I t}  \Psi^{\alpha_1}_{\sigma_1}(t_1) \Bar{\Psi}^{\Bar{\alpha}_1}_{\Bar{\sigma}_1}(\Bar{t}_1) \cdots \Bar{\Psi}^{\Bar{\alpha}_n}_{\Bar{\sigma}_n}(\Bar{t}_n) \right] \vert \rho_{I}(0) \rangle \text{Det}_{\sigma}\left[ \{\Delta_{\sigma}^{\alpha \bar{\alpha}} \}\right] \text{Det}_{\bar{\sigma}}\left[ \{\Delta_{\bar{\sigma}}^{\alpha \bar{\alpha}} \}\right]\,. 
\end{align}
\end{widetext}
In the expression above the impurity operators are evolved under the local Lindbladian $\m{L}_I$, i.e.
\begin{align}
    \Psi (t) = e^{-\m{L}_I t}  \Psi e^{\m{L}_I t}
\end{align}
For what concerns the bath degrees of freedom, we have used the Wick theorem  since we consider a non-interacting bath, and have introduced the bath hybridization function defined as:
\begin{align}
    \Delta_{\sigma \bar{\sigma}}^{\alpha \bar{\alpha}} (t,\bar{t}) = -i \langle I_{B} \vert \m{T}_t \left[\Bar{\Phi}_{\sigma}^{\alpha}(t) \Phi_{\bar{\sigma}}^{\bar{\alpha}}(\bar{t}) \right] \vert \rho_{B}(0) \rangle
\end{align}
We note that in the case of interest here the hybridization between impurity and bath is diagonal in the spin-index, therefore the hybridization function above can be written as a matrix
\begin{align}
    \bb{\Delta}_\sigma (t,\bar{t}) = \begin{pmatrix}
        \Delta_\sigma^{00} (t,\bar{t}) & \Delta_\sigma^{01} (t,\bar{t}) \\
        \Delta_\sigma^{10} (t,\bar{t}) & \Delta_\sigma^{11} (t,\bar{t})
    \end{pmatrix}
\end{align}
where the different components refer to the structure of the Hilbert space $\m{H}$ or $\tilde{\m{H}}$. We will give explicit expressions for these functions in Appendix A.

\subsection{Trace over the Impurity degrees of freedom}

As we have shown in the previous section, for each order $n$ in the hybridization expansion, the trace over the impurity degrees of freedom involve $2n$ operators evaluated at a time t by the bare impurity Lindbladian $\m{L}_I$. It is therefore quite natural to rewrite all the operator in the diagonal basis of $\m{L}_I$, in order to reduce the computational cost and also to look at the symmetries of the system. So, we denoted $\{ \vert r_\mu \rangle , \vert l_\mu \rangle\}_\mu$ respectively the right and left eigenvectors such that
\begin{align}
    \m{L}_I \vert r_\mu \rangle = \lambda_\mu \vert r_\mu \rangle  \quad \text{and} \quad \langle l_\mu \vert \m{L}_I =\langle l_\mu \vert  \lambda_\mu^*
\end{align}
where $\lambda_\mu$ is the associated eigenvalue, which in the case of a Lindbladian evolution is a complex number, with a real part that can be non-zero. In fact, the imaginary part of the eigenvalues give the coherent part for the dynamics and the real part gives rise to the dissipative dynamics. Moreover, even in a non-unitary dynamics, the set of eigenvectors form an orthonormal basis, with an associate closure relation given by:
\begin{align}
    \m{I} = \sum_\mu \vert r_\mu \rangle \langle l_\mu \vert \quad \text{with} \quad \langle l_\mu \vert r_\nu \rangle = \delta_{\mu,\nu}
\end{align}
by using the orthoganality and the spectral properties of $\m{L}_I$, we can rewrite the local (impurity) evolution operator as:
\begin{align}
    e^{\m{L}_I t} = \sum_{\mu} e^{\lambda_\mu t } \vert r_\mu \rangle \langle l_\mu \vert
\end{align}

Concerning the impurity part of the hybridization expansion, we can insert closure relation in order to rewrite all the operators in the basis of $\m{L}_I$.

\begin{align}
    \Tr_{\text{imp}}[\cdots] &= \langle I \vert \m{T}_t \left[ \Psi^{\alpha_1}_{\sigma_1}(t_1) \Bar{\Psi}^{\Bar{\alpha}_1}_{\Bar{\sigma}_1}(\Bar{t}_1) \cdots \Bar{\Psi}^{\Bar{\alpha}_n}_{\Bar{\sigma}_n}(\Bar{t}_n) \right] \vert \rho_I(0) \rangle \notag \\ &= \langle I \vert  \left( \sum_{\mu,\mu^\prime} \m{A}_{\mu, \mu^\prime } \left( \{ t\} \right) \vert r_\mu \rangle \langle l_{\mu^\prime} \vert \right) \vert \rho_I(0) \rangle 
\end{align}

where we have introduced the matrix $\m{A}$, given by:
\begin{align}
    \m{A} \left(\{ t\}\right) = \left[ e^{\lambda_\mu \left( t-t_1^\prime \right)}\right] \left( \Psi_{\sigma_1}^{\alpha_1}\right)_{\mu , \mu_1}
    \cdots  \left(\Bar{\Psi}_{\sigma_n}^{\alpha_n} \right)_{\mu^{''},\mu^\prime} \left[ e^{\lambda_{\mu^\prime} \left( t-t_1^\prime \right)} \right]
\end{align}
where $\left( \Psi_{\sigma_1^\prime}^{\alpha_1^\prime}\right)_{\mu , \mu^\prime}$ denoted the matrix component of the spinor in the diagonal basis:
\begin{align}
    \left( \Psi_{\sigma_1^\prime}^{\alpha_1^\prime}\right)_{\mu , \mu^\prime} = \langle l_\mu \vert \Psi_{\sigma_1^\prime}^{\alpha_1^\prime} \vert r_{\mu^
    \prime}\rangle
\end{align}
where the sum runs over those sectors which are compatible with the operator sequence.

Note that the evaluation of the trace factor thus involves the multiplication of matrices whose size is equal to the size of the Hilbert space of $H_I$. Since the dimension of the Hilbert space grows exponentially with the number of spin/orbitals, the calculation of the trace factor becomes the computational bottleneck of the simulation, and the matrix formalism is therefore restricted to a relatively small number of spin/orbitals. 
In practice, we can take to account the symmetries of the Lindbladian in order to restrict the diagrams space, and for some specific case we can write a analytic expression for the trace over the impurity degrees of freedom.

\section{Diagrammatic Monte Carlo}\label{sec:DMC}

Diagrammatic Monte Carlo (diagMC) is a numerical algorithm for sampling infinite series of multiples integrals, such as those arising in any perturbative expansion~\cite{VANHOUCKE201095,Gull_RMP11}. Often this expansion admits a diagrammatic representation, even in out-of-equilibrium situations.
One then performs a Monte Carlo sampling of the resulting space of diagrams to evaluate physical quantities.


As it can be immediately read out from Eq.~(\ref{eqn:hyb_exp}) in the previous section, the dynamical evolution of the density matrix can be written as a weighted sum over configuration $\m{C}$
\begin{align}
    \langle I \vert \rho(t) \rangle = \langle I \vert \m{V}(t) \vert \rho(0) \rangle = \sum_{\m{C}} \m{W}(\m{C})
\end{align}
where a given configuration $\m{C}$ contains, for each flavor $\sigma$ , a total of $2k_\sigma$ vertices occurring at times $\{t_i^\sigma , \bar{t}_i^\sigma\}$ with $i=1,\cdots k_\sigma$.
Half of these vertices represent an impurity creation operator $d^\dagger_\sigma$ or $\Tilde{d}^\dagger_\sigma$, and the other half represent an impurity annihilation operator $d_\sigma$ or $\Tilde{d}_\sigma$, both of them being evolved in time with the local Lindbladian $\m{L}_I$. All the operators are stored in such a way to always preserve global time ordering along the contour, a typical configuration reads:
\begin{align}
    \m{C}= \left\{ \begin{array}{ll}
         \sigma= \{\uparrow,\downarrow  \} \\
         k_\sigma = 0,1,\cdots ,\infty \\
         \left(\bar{t}_1^\sigma , \bar{\alpha}_1^\sigma \right) ; \cdots ; \left(\bar{t}_{k_\sigma}^\sigma , \bar{\alpha}_{k_\sigma}^\sigma \right)
         \\ 
         \left(t_1^\sigma , \alpha_1^\sigma \right) ; \cdots ; \left(t_{k_\sigma}^\sigma , \alpha_{k_\sigma}^\sigma \right)
    \end{array}\right.
\end{align}
For each configuration, we defined the Monte Carlo weight directly from the hybridization expansion in Eq.~(\ref{eqn:hyb_exp}), as:
\begin{align}\label{eqn:W_C}
    \m{W}\left[ \m{C}\right] = \text{sign}[\mathcal{C}] \text{Det} [\mathcal{C}]  \Tr_{\text{imp}}  [\mathcal{C}] 
\end{align}
where $sign\left[\m{C}\right]$ includes all the signs (phases) coming from the evolution as well as from the time ordering, while the trace over the impurity degrees of freedom reads 
\begin{figure}[t]
	\includegraphics[width=0.48\textwidth]{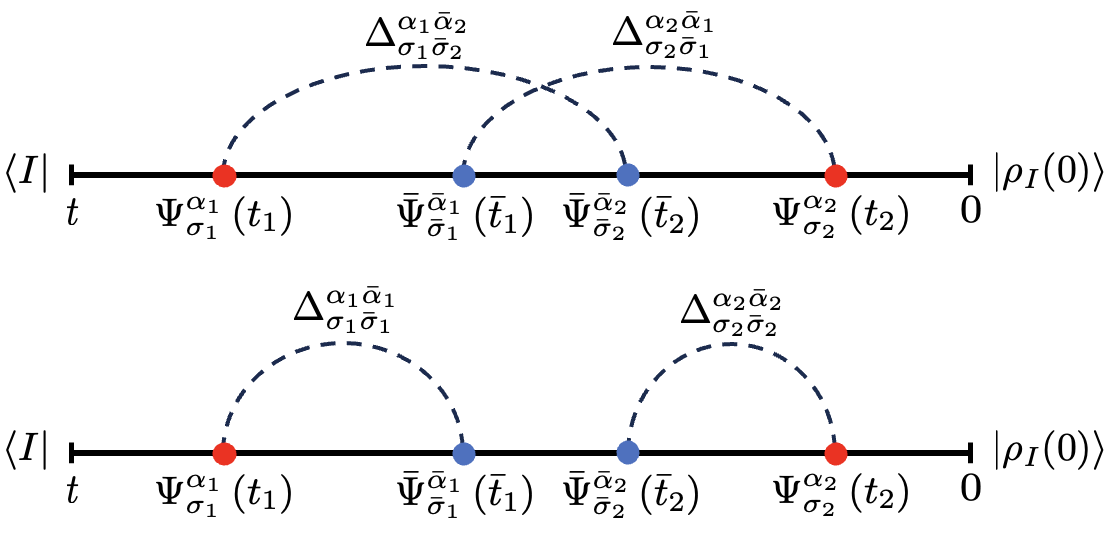}
	\caption{\label{fig0:sketch} An example of configuration $\m{C}$ for the second order of the Hybridization expansion. The diagram in the top(bottom) panel, corresponding to crossing and non-crossing lines, are combined together in a determinant structure. We note the expansion is formulated on a single \emph{collapsed} real-time contour, with an extra index $\alpha$ taking into account whether a given vertex belong to the Hilbert space $\m{H}$ or $\m{\tilde{H}}$. }
\end{figure}
\begin{align}
    \Tr_{\text{imp}}  \left[\m{C} \right] = \langle I \vert D_1(t_1) D_2(t_2) \cdots D_{2n}(t_{2n}) \vert \rho_I(0) \rangle 
\end{align}
where $D$ denoted an impurity operator with some spin/orbital index $\sigma$ and living in a Hilbert space $\m{H}$($\Tilde{\m{H}}$). 
The knowledge of the weight $\m{W}\left[\m{C}\right]$ allows in principle to compute any observable acting on the impurity degrees of freedom. In fact, starting from Eq.~(\ref{eqn:O_t}) we can in principle rewrite the hybridization expansion and obtain
\begin{align}\label{eqn:O_t}
    \langle O(t)\rangle = \frac{\sum_{\m{C}} O(\m{C})\m{W}[\m{C}]}{\sum_{\m{C}} \m{W}[\m{C}]}
\end{align}
where the estimator of local operator has been defined as 
\begin{align}
    \m{O}(\m{C}) = \frac{ \langle I \vert \m{O} D_1(t_1) D_2(t_2) \cdots D_{2n}(t_{2n}) \vert \rho(0) \rangle }{ \langle I \vert D_1(t_1) D_2(t_2) \cdots D_{2n}(t_{2n}) \vert \rho(0) \rangle }
\end{align}
Once the real-time average of a local operator is written like this, it would be natural to sample it using a Monte Carlo method, namely generating a random walk in the configuration space which visit configurations $\m{C}$ with probability $P(\m{C}) = \m{W}[\m{C}] / \sum_{\m{C}^\prime} \m{W}[\m{C}^\prime]$.

One of the challenges of implementing the real-time diagMC is that the weight $\m{W}\left[ \m{C}\right]$ is in general a complex number. In the specific case of the hybridization expansion, the complex value of the weight is due not only to the ”i-factors” coming from the real time evolution but also to the fact the bath part and the contour bath defined previously is a complex function of it's time arguments.
In order to circumvent this problem, we sample the absolute value of the weight $|\m{W}\left[ \m{C}\right]|$, while including the phase of the configuration $\eta(\m{C})$ defined as
\begin{align}
    \eta(\m{C}) = \frac{\m{W}\left[\m{C} \right]}{\vert \m{W}\left[\m{C} \right]\vert}
\end{align}
in the Monte Carlo estimator. In other words, we can rewrite Eq.~(\ref{eqn:O_t}) as 
\begin{align}\label{eqn:O_phase}
    \langle O(t)\rangle = \frac{\sum_{\m{C}} O(\m{C})\eta(\m{C})|\m{W}\left[\m{C} \right]|}{
    \sum_{\m{C}} \eta(\m{C})\vert\m{W}\left[\m{C} \right]\vert}=\frac{\langle O\eta\rangle_{MC}}{\langle \eta\rangle_{MC}}
\end{align}
where we have introduced the Monte Carlo average $\langle X\rangle_{MC}=\sum_{\m{C}}
X(\m{C})P(\m{C})$ with respect to a well defined (positive) probability measure $P(\m{C}) = |\m{W}[\m{C}] |/ \sum_{\m{C}^\prime} |\m{W}[\m{C}^\prime] |$. This approach, despite its simplicity, becomes problematic when the average phase goes to zero, as in this case the accuracy of the algorithm deteriorates as errors become exponentially large with time. As we will see later on, the presence of local dissipation improves the convergence properties of the diagMC algorithm.

\subsection{Metropolis Algorithm}

A standard approach to generate configurations with a given probability $P(\m{C}) = |\m{W}[\m{C}] |/ \sum_{\m{C}^\prime} |\m{W}[\m{C}^\prime] |$ is to build up a Markov chain~\cite{krauth06}, i.e. a stochastic process which describes the evolution of the probability to visit configuration $\m{C}$ after $n$ steps, denoted as $P\left( \m{C}, n \right)$. The way to describe a Markov chain is to introduce the conditional probability $R\left[ \m{C}  \rightarrow \m{C}^\prime \right]$ to be in the configuration $\m{C}^\prime$ at step $n+1$ being in the configuration $\m{C}$ at step $n$. This quantity allows us to define the master equation, the recursive equation that expresses $P\left( \m{C}^\prime, n+1 \right)$ in function of the previous step:
\begin{align}
    P\left(\m{C}^\prime , n+1 \right) = \sum_{\m{C}} R\left[\m{C}\rightarrow  \m{C}^\prime\right] P\left( \m{C},n\right)
\end{align}
In order to reach the desired probability $P\left( \m{C}\right)$, after waiting a proper equilibration time, the matrix $R\left[\m{C}\rightarrow  \m{C}^\prime\right]$ must satisfies two constraints. The first one is the ergodicity of the matrix and the second one is that the matrix must satisfy the detailed balance condition. 
Ergodicity ensures that we can reach any configuration $\m{C}$ from any other configuration $\m{C}^\prime$, after a finite number of steps. This means that all the space of the configuration can be visit during the simulation. While detailed balance means that for any configuration $\m{C}$ and $\m{C}^\prime$, the following relation must be verified
\begin{align}
    R\left[ \m{C} \rightarrow \m{C}^\prime \right] P(\m{C}) = R\left[ \m{C}^\prime \rightarrow \m{C} \right] P(\m{C}^\prime)
\end{align}

where $P(\m{C})$ is the probability distribution we want to sample through the Markov chain. 
One way to generate configurations which satisfies the detailed balance condition is to use the Metropolis Algorithm~\cite{krauth06}. The basic idea is, starting from a initial configuration $\m{C}$ we propose to visit a new configuration $\m{C}^\prime$ with a certain transition probability $T\left( \m{C} \rightarrow \m{C}^\prime \right)$, this probability depends on how we propose the new configuration which in principle it can be independent of the physical system. Then, this new configuration is accepted or rejected according to the probability $A\left( \m{C} \rightarrow \m{C}^\prime \right)$, so in this context the conditional probability $R\left[\m{C} \rightarrow \m{C}^\prime \right]$ to move in the configuration $\m{C}^\prime$ starting from $\m{C}$ is given by:
\begin{align}
    R\left[\m{C} \rightarrow \m{C}^\prime \right] = A\left( \m{C} \rightarrow \m{C}^\prime \right) T\left( \m{C} \rightarrow \m{C}^\prime \right)
\end{align}

Concerning the acceptance probability $\m{A}\left( \m{C} \rightarrow \m{C}^\prime\right)$, the Metropolis algorithm is based on the following relation
\begin{align}
    \m{A}\left( \m{C} \rightarrow \m{C}^\prime\right) = \min \left[ 1, \frac{P\left( \m{C}^\prime \right) T\left( \m{C}^\prime \rightarrow \m{C}\right) }{P\left( \m{C} \right) T\left( \m{C} \rightarrow \m{C}^\prime \right)}\right]
\end{align}
which satisfies the detailed balance condition. While this previous description of the algorithm is generic and model independent, it is interesting to detail how in practice we can compute the acceptance probability and what type of transition probability has to be used,  since these two quantities can strongly affect the performance and the reliability of the Monte Carlo algorithm.

The transition probability $T\left( \m{C} \rightarrow \m{C}^\prime \right)$ is determined according to the types of moves to implement. In the case of interest, we implement two classes of local moves, characterized by their way of exploring the space of configuration.

The first one, allows us to change the number of vertex in a given channel a by unity $\Delta k_\sigma = \pm 1$. These moves amount to add or remove a vertex (one creation and one annihilation fermionic operator) in a  given channel $a$ and at randomly chosen time. In principle, only these two moves are necessary to ensure the ergodicity of the matrix $R$. Indeed, it is obvious that using these two basic updates any configuration can be reached after a finite number of steps, which guarantees the Metropolis algorithm to visit configurations according to the probability $P(\m{C})$. However, although the ergodicity is respected, these two moves do not guarantee the efficiency and the speed-up of the Monte Carlo sampling. Indeed, exploring the space of configurations with a fixed number of vertex is relatively inefficient and requires drastically increasing the number of Monte Carlo steps. For this purpose,it is interesting to implement the second class of moves, which explore the configuration space at a fixed number of vertex in a given channel a ($\Delta k_a = 0$)
such as for example shifting a fermionic operator (annihilation or creation operator). 
In practice, we can also implement other kind of moves, which are more specific, for example some moves in which more than two operators are added/removed/shifted. This types of moves become revelant when dealing with off-diagonal baths or when dealing with two or more particles dissipation process. Global moves are also fundamental in the case of multiorbital dissipation process. In fact, the choice of moves is determined by the structure of the Lindbladian and of the Non-markovian bath.

From the point of view of the computational scaling of the algorithm the key quantity is the acceptance ratio  $\m{A}(\m{C} \rightarrow \m{C}^\prime )$ which needs to be evaluated at each Monte Carlo step. As we can see from the definition of the weight $\m{W}(\m{C})$ in Eq.~(\ref{eqn:W_C}) we have to evaluate the ratio of two determinants and the ratio of the trace over the impurities degrees of freedom. For the ratio of determinants fast updates routines are available~\cite{Gull_RMP11}, which allows us to find a analytical expression and then makes this operation rather efficient, scaling polynomially with the number of vertex. On the other hand concerning the trace over the impurities degrees of freedom,  this usually scales exponentially with the size of the local Hilbert space since one has to rewrite the operators in the basis of local eigenstates of the Lindbladian and store the whole chain of matrix products from left to right (and viceversa). However in some specific case, the symmetries of the Lindbladian allows us to use some segment representation (see next Section) and so find a analytic expression for the trace over the impurity degrees of freedom, this is the case of the impurity models without exchange or hopping terms. 

In the next section, we describe the first application of the diagMC algorithm to the Anderson impurity model in presence of dephasing. We will first discuss its performances, then benchmark it against the exactly solvable dissipative resonant level model and finally present the results in the interacting case.

\section{Results:  Anderson Impurity Model with Dephasing}
\label{sec:results}

In this section we apply our DiagMC algorithm to study the non-equilibrium dynamics of the Anderson Impurity Model (AIM) coupled to a dephasing bath. We consider therefore a single spinful impurity with local Hamiltonian and local jump operator given respectively by
\begin{align}
    H_{I} = \sum_{\sigma} \epsilon_d d_\sigma^\dagger d_\sigma + U n_\uparrow n_\downarrow  \quad \text{and} \quad L_\sigma = \sqrt{\gamma_\sigma} n_\sigma
\end{align}
which both enter the Lindblad master equation given in Eq.~(\ref{eqn:lindblad}). Concerning the fermions describing the non-Markovian bath, we assumed a non-interacting bath coupled to the impurity via an energy-dependant hybridization function $\Gamma\left( \epsilon \right)$ defined as
\begin{align}
    \Gamma(\epsilon) =  \sum_\bb{k} \vert V_\bb{k} \vert^2 \delta \left(\epsilon - \epsilon_\bb{k} \right) =  V^2 \rho(\epsilon)
\end{align}
where $V_\bb{k}$ is assumed independent of the momentum for simplicity and $\rho(\epsilon)$ is the conduction density of state, which at first approach we consider the flat band limit, namely a flat band of width 2W:
\begin{align}
    \rho(\epsilon)  =\rho_0 \Theta \left( \vert \epsilon - W \vert \right)\,.
\end{align}
Although simplistic this state density encodes the main properties of a metallic conduction bath, with a finite bandwidth and a finite weight at the Fermi level. In this case, the hybridization function which describes the coupling between the bath and the impurity becomes energy independent, $\Gamma(\varepsilon)\equiv \Gamma$. In the following we take $\Gamma$ as our unit of energy. Unless stated otherwise we consider the fermionic bath to be in equilibrium at zero temperature.

Let us briefly discuss some notable limit of this model. First, in absence of any dephasing the real-time dynamics of the AIM has been studied in detail with different methods~\cite{anders2005realtime,schiroPRB2009,wauters2023simulations}. Here the spin impurity dynamics is controlled by the emergent Kondo scale $T_K\sim e^{-U/\Gamma}$ while charge dynamics is faster and controlled by higher energy scales, such as $U,\Gamma$. In presence of dephasing but no electron-electron interaction, i.e. $U=0$, the model reduces to a dissipative Resonant Level Model (dRLM) which can be still solved exactly using Keldysh techniques~\cite{ferreira2023exact} (see Appendix~\ref{app:DRLM}). Finally, as we are going to discuss below, in absence of impurity-bath hybridization, $V_k=0$, the local occupation of the impurity remains constant, even though the system acquires a finite lifetime given by the dephasing $\gamma$. In the remaining of this section, we first discuss some aspect of the algorithm in particular the structure of the DiagMC configurations and the performances and convergence properties. Then we present some benchmark results for the dRLM and finally presents our results for the fully interacting Anderson model.

\begin{figure}[t]
    \center     \includegraphics[width=1.0\linewidth]{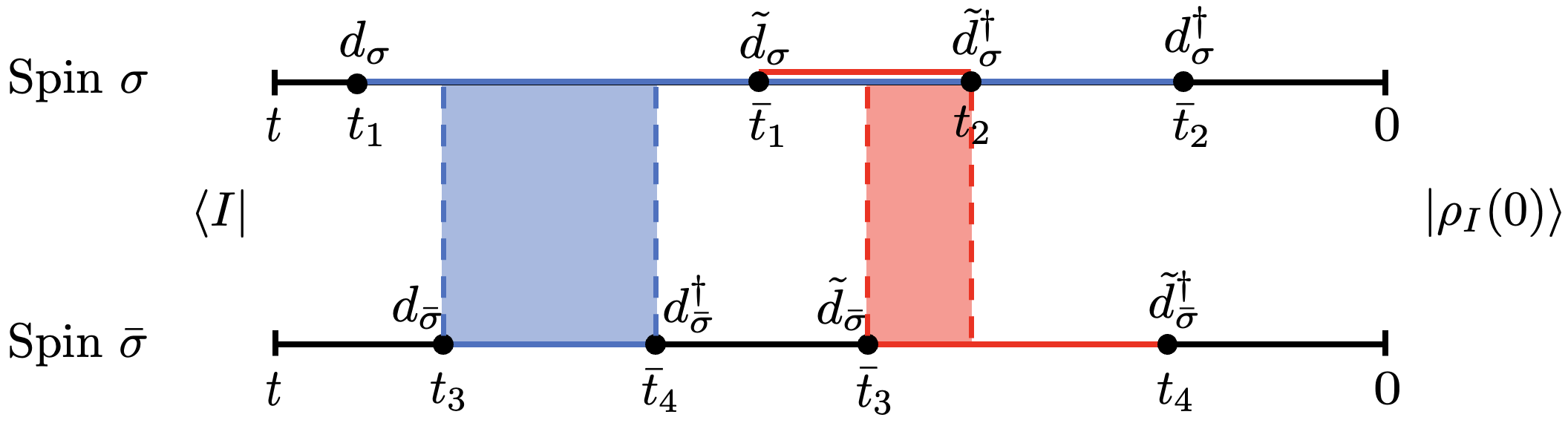}
\caption{\label{fig:segment}Segment representation of the impurity trace in hybridization expansion of the single orbital Anderson model with Dephasing . Upper line: spin up orbital, lower line, spin down orbital. Blue line: orbital in the space $\m{H}$ occupied, Red line: orbital in the space $\Tilde{\m{H}}$ occupied. Shaded areas: regions where both up and down orbitals are filled, so the impurity is doubly occupied. The length of the shaded area enters into an overall weighting factor for the potential energy. }
\end{figure}
\subsection{DiagMC: Generalised Segment Picture}

In Sec.~\ref{sec:DMC} we discussed the general structure of DiagMC configurations, considering all possible vertex types, without taking account the symmetries of the problem. In practice, part of the vertex can induce a zero contribution when we calculate the trace over the impurity degrees of freedom, the symmetries constrains the space of all the configuration to a subspace where we have only the non-zero contribution. 

For open quantum systems described by a Lindbladian one can distinguish between weak and strong symmetries~\cite{albert2014symmetries}. In particular, whenever an operator commutes with the impurity Hamiltonian and with all jump operators than we can associate to it a strong symmetry which reflects in a block diagonal structure of the Lindbladian. 
In the case of the Anderson Impurity with depahsing we have that 
\begin{align}
    \left[ H_{I},n_{\sigma} \right]=\left[ L_{\sigma},n_{\sigma} \right]=\left[ L^{\dagger}_{\sigma},n_{\sigma}\right]=0
\end{align}
In other words we can say that the impurity Lindbladian commutes with the density $n_{\sigma}$. This strong symmetry for the Lindbladian of the impurity means that the evolution through this Lindbladian preserves the number of particles on the impurity. We can therefore perform the trace on the impurity degrees of freedom, since the Lindbladian $\m{L}_I$ commutes with the occupation number operator of each orbital, the evolution operator $e^{\m{L}_I t}$ is diagonal in the Fock space. This allows to simplify the evaluation of the local trace and also to identify in a simple way the configurations with non-zero weight. As in the equilibrium case~\cite{Gull_RMP11}, we can use a segment representation and write a analytical expression for the trace over the impurity degrees of freedom. In this representation, we represent the time evolution of the impurity by collections of segments, which each segments represent time intervals in which an electron with a given spin resides on the impurity.

In Fig.~\ref{fig:segment}, we have illustrated an example of such a segment representation for the Anderson impurity model with dephasing. 

Using this segment representation we can compute in closed form the trace over the impurity configuration that enters the hybridization expansion, see Eq.~(\ref{eqn:hyb_exp}). In particular one can show that this reads
\begin{widetext}
\begin{align}\label{eqn:trace_segm}
    \Tr_{\text{Imp}} \left[ \cdots \right] &=\langle I \vert \m{T}_t \left\{e^{-\m{L}_I t} \Psi_\sigma^{\Bar{\alpha}_1^\sigma}(\Bar{t}_1^\sigma) \Bar{\Psi}_\sigma^{\alpha_1^\sigma}(t_1^\sigma) \cdots \Bar{\Psi}^{\alpha_{k_\sigma}^\sigma}_\sigma (t_{k_\sigma}^\sigma) \right\} \vert \rho_I(0) \rangle  
    = s   e^{i\left( \sum_{\sigma} \epsilon_d \left[ l_\sigma - \Tilde{l}_\sigma  \right] + U \left[ O_{\sigma \bar{\sigma}} - \tilde{O}_{\sigma \bar{\sigma} }\right] \right) -\sum_{\sigma} \gamma_\sigma W_\sigma + \frac{\gamma_\sigma}{2} \left[l_\sigma + \Tilde{l}_\sigma \right]  }
\end{align}
\end{widetext}
where we have introduced the following quantities
\begin{itemize}    \item $l_\sigma$ : total length of segments in spin $\sigma$ and for the Hilbert Space $\m{H}$
    \item $\Tilde{l}_\sigma$ : total length of segments in spin $\sigma$ and for the Hilbert Space $\Tilde{\m{H}}$
    \item $O_{\sigma,\bar{\sigma}}$ :total overlap between segment of flavor $\sigma$ and $\bar{\sigma}$ for the space $\m{H}$

    \item $\Tilde{O}_{\sigma,\bar{\sigma}}$ : total overlap between segment of flavor $\sigma$ and $\bar{\sigma}$ for the space $\Tilde{\m{H}}$

    \item $W_\sigma$: total overlap between segment of same flavor $\sigma$ but living in different Hilbert space.
\end{itemize}
In Eq.~(\ref{eqn:trace_segm})  $s$ is a extra sign, coming from two contributions: the first one is the time ordering operator and the second one from the different permutation of the fermionic operator in order to get the natural ordering of the basis. 
The knowledge of this analytical expression for the local trace greatly simplify the DiagMC algorithm.

\subsection{Performance of the Algorithm }

In order to analyze the performance of DiagMC in presence of dissipation, we will consider two mains quantities, namely the probability distribution of perturbative orders (kinks, or vertex) in the diagrammatic expansion and the average sign of the Monte Carlo weight, both being precise measures of the efficiency of the algorithm and for the determination of error bar. In all this subsection we consider as initial condition an impurity which is initially empty
$\rho_I(0) = | 0 \rangle \langle 0 |$.

\subsubsection{Statistics of Kinks}

\begin{figure}[t]
	\includegraphics[width=0.40\textwidth]{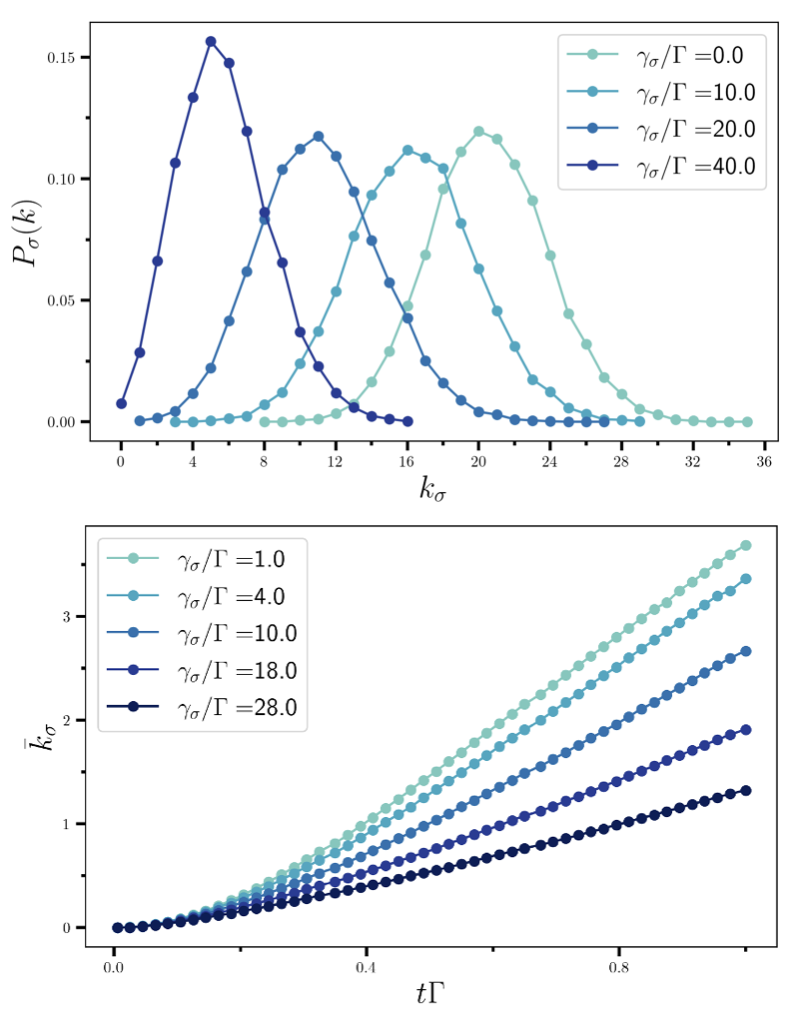}
	\caption{\label{fig4:kinks} Statistic of the kinks (vertex) $k$ sampled during the simulation. Top panel shows the probability distribution of different perturbative orders $k$ in function of the dephasing for a fixed measuring time t. While the bottom panel, is the scaling of the average number of kinks with maximum time t for different values of dephasing $\gamma_\sigma$. the plots are obtained for a empty initial state $\rho_I(0)= \vert 0 \rangle \langle 0 \vert$ and for $T=0$ and $W=2\Gamma$. }
\end{figure}

As we have shown in the previous section,  DiagMC allows to stochastically sample the expansion of the trace of the density matrix in power of the impurity-bath coupling. The main idea of Monte Carlo algorithm is to perform a random walk in the diagrams space. Thus, during the simulation it is natural to verify the stability of the algorithm by looking at the statistics of the different perturbative order, namely the probability distribution to visit a Monte Carlo configuration with $k$ vertex in the spin channel $\sigma$. 
The respective probability is defined as:
\begin{align}
    P_{\sigma}(k) = \frac{\sum_{\m{C}} \vert \m{W}\left( \m{C}\right)\vert \delta\left( k_\sigma \left( \m{C} \right)-k\right)}{\sum_{\m{C}} \vert \m{W}\left( \m{C}\right)\vert }
\end{align}
where $k_\sigma \left( \m{C} \right)$ is the number of vertex with $\sigma$ in the configuration $\m{C}$.  In Fig.~\ref{fig4:kinks}, we have plotted an example of the behaviour of this probability distribution for different values of measuring time $t$ and dephasing $\gamma_\sigma$. As in the unitary case, all histograms of $P_\sigma(k)$ are peaked around an average value $\bar{k}$, with an exponentially small probability for higher perturbative order. However, Fig.~\ref{fig4:kinks} confirms that all orders contributed and are included, so diagMC calculation is an unbiased result which does not truncate at any finite perturbative order the hybridization expansion but rather perform an exact resummation of all the perturbative orders. Importantly, we note in Fig.~\ref{fig4:kinks} that the effect of dephasing is to shift the hystogram towards the low diagram-order sector. This means that for a fixed measuring time $t$ the hybridization expansion converges faster, i.e. with a smaller number of diagrams, in presence of dissipation than in the purely unitary case. We note that a similar effect occurs in the imaginary-time hybridization expansion algorithm~\cite{Gull_RMP11}, upon increasing the local interaction and it is one of the reasons of its success. 
We can understand this decreasing of the perturbative order by looking at eigenvalues of the local lindbladian $\m{L}_I$ which can be written in general as

\begin{align}
    \lambda = \m{R}e(\lambda) + i \m{I}m(\lambda)
\end{align}

For the Anderson impurity Model with dephasing the associated eigenvectors have the following form $\{ \vert n , \Tilde{m}\rangle \}_{n,m}$, i.e. they are diagonal in Fock space of both Hilbert spaces $\m{H},\m{\tilde{H}}$. In the segment representation, only the segment $\m{U}_{[t , \bar{t}]}$ with the same number of particle in the space $\m{H}$ and $\Tilde{\m{H}}$ for a given spin $\sigma$ are not affected by the dissipation. For the other segments where we have a state of the form $|n,\Tilde{m}\rangle_{n \ne m}$, the trace over the impurity degrees of freedom gives:
\begin{align}
    \Tr_{\text{Imp}} \left[ \cdots\right]_{\m{U}_{[t , \bar{t}]}} \propto e^{\m{R}e\left( \lambda \right) (\bar{t} -t)} e^{i \m{I}m\left( \lambda\right)(\bar{t} -t)}
\end{align}
Since $\m{R}e\left( \lambda \right) <0$ ,  the dissipation then decrease the probability of sampling the vertex with a non-zero real part eigenvalue. In this respect, the effect of the dissipation is to constraint the sampling to a subspace of diagrams. In particular, in the strong dissipative regime only the state with $\m{R}e\left( \lambda\right) = 0$ contribute to the dynamics, we can then write an effective model by projecting the Lindbladian onto its states.

In order to quantify the impact of dephasing on the statistics of diagrams order, it is interesting to look at the average perturbative order $\bar{k}_\sigma$. In the bottom panel of Fig.~(\ref{fig4:kinks}),  we plot $\bar{k}_\sigma$ as a function of time $t$ for different value of the dephasing $\gamma_\sigma$ and for a initial empty impurity state.  We note an almost linear scaling with time with a slope which, as expected, decreases as the value of dephasing is increased, i.e. the effect of the markovian dissipation is to reduce the number of kinks and so the scaling with the time. In fact, since in the strong dissipative regime the space of diagrams are reduced to a subspace, it can then be interesting to modify the probabilities of sampling in the algorithm of Metropolis in order to favour the diagrams with a non-zero probability of sampling. 
Thus, to summarize the scaling of the average number of diagrams for our real-time DiagMC reads
\begin{align}
    \bar{k}_\sigma = C t
\end{align}
with $C$ a constant which depends of $\gamma_\sigma$, but which is independent of other local energy scales. Both in the unitary case and in presence of markovian dissipation, the coefficient $C$ strongly depends on the bandwidth $W$ and $\Gamma$. Note that even with the dissipation accessing long time scale in the regime $W\gg \Gamma$, becomes difficult with this approach.
Overall the results of this section shows that Markovian dissipation such as dephasing is beneficial for the convergence properties of DiagMC and can help reach longer time scales compared to the unitary case.

\begin{figure}[!t]
	\includegraphics[width=0.40\textwidth]{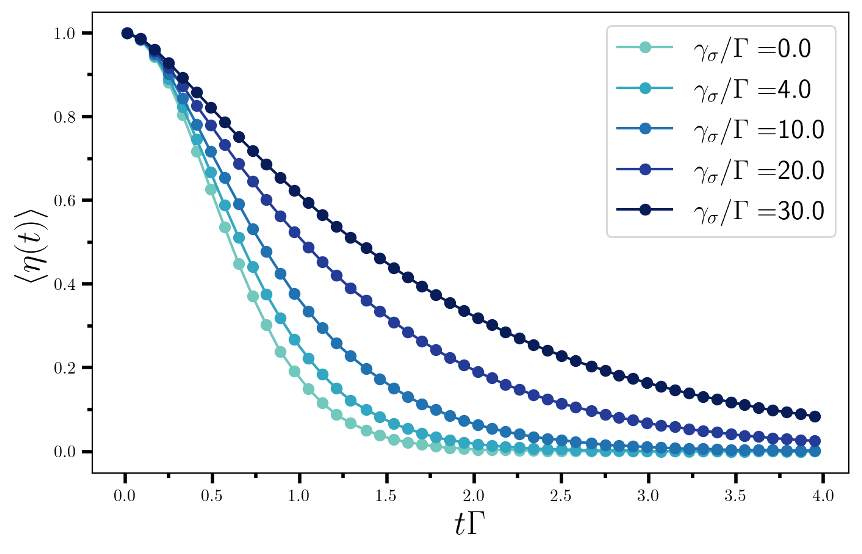}
	\caption{\label{fig4:Phase} Average Phase as a function of time t for different dephasing $\gamma_\sigma$ in the regime $W = 2 \Gamma$, the dephasing being the same for each spin channel. We clearly see an exponential decay on a scale of time all the larger as the dissipation is important.   }
\end{figure}

\subsubsection{Average Sign}

Another important quantity to monitor during the simulation is the average phase of the Monte Carlo configurations. Indeed, the relation between the physical quantities and the MC phase is given by Eq.~(\ref{eqn:O_phase}), a vanishing average sign turns into very large error bars on Monte Carlo averages that makes the simulation unstable and then restricts the regimes accessible by diagMC. In the real-time diagMC, the average phase of the MonteCarlo configurations is defined according to the complex nature of the MC weights,
\begin{align}
    \langle \eta(t) \rangle  = \frac{ \sum_{\m{C}} \eta\left( \m{C }\right) \vert \m{W}(\m{C})\vert }{\sum_{\m{C}}\vert \m{W}(\m{C})\vert }
\end{align}
In Fig.~(\ref{fig4:Phase}) we plot the average phase as a function of time, for different values of the dissipation. We see that, consistently with the decrease of the average perturbative order, the average sign decays to zero in a slower fashion in presence of strong dissipation. This result, which is one of the important one of this work, implies that longer time scales can be reached within our diagMC algorithm at fixed computational resources as compared to the purely unitary evolution algorithm.

\subsection{Benchmark: Dissipative Resonant Level Model}

\begin{figure}[!t]
	\includegraphics[width=0.40\textwidth]{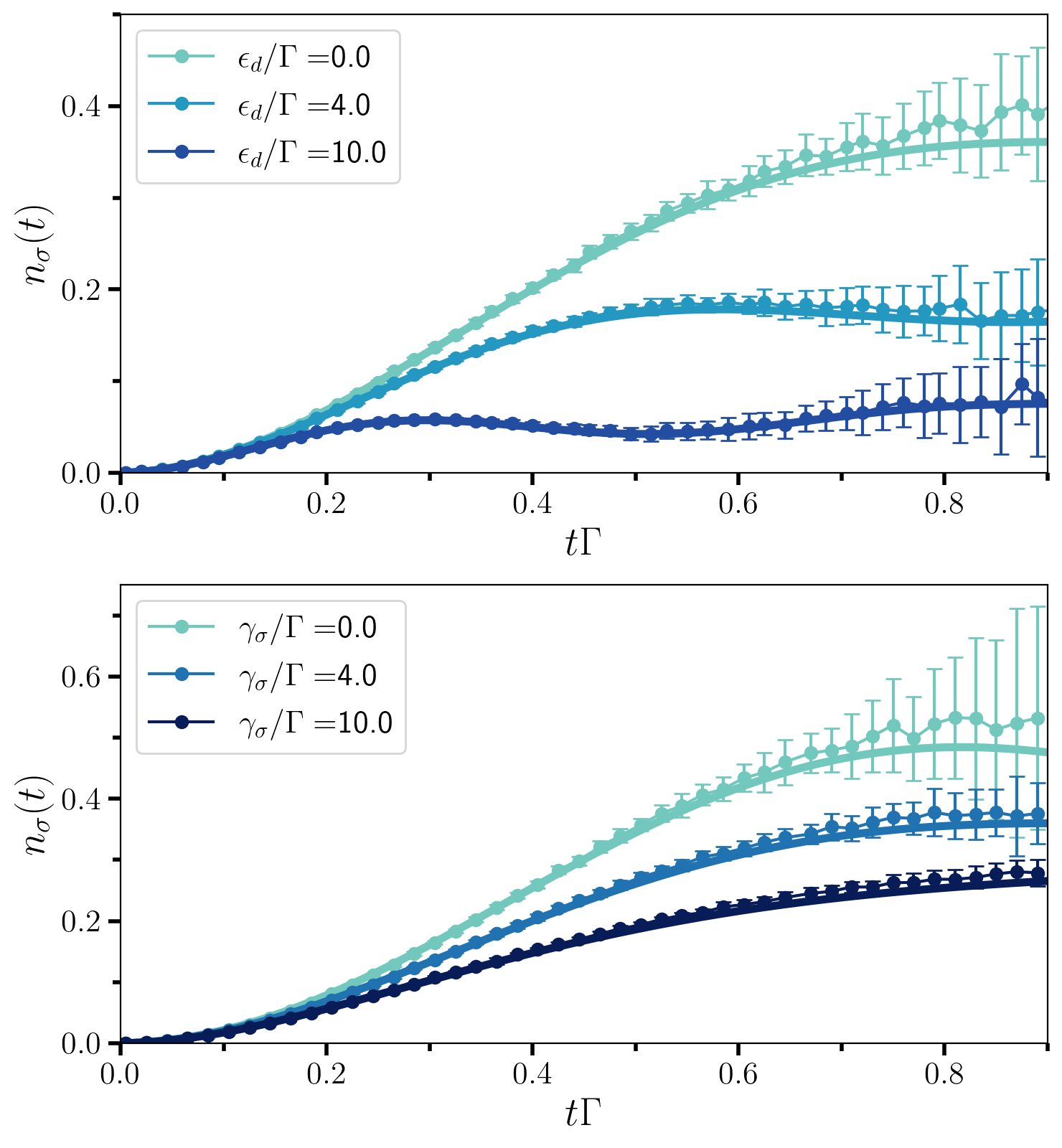}
	\caption{\label{fig4:benchmark} Dynamics of the Dissipative Resonant level model for differents values of dephasing (bottom) and of the impurity energy level $\varepsilon_d$ (top). 
	The solid line correspond to the exact solution obtained by the standard Keldysh methods. All the results are obtained in the regime $W = 2 \Gamma$.}
\end{figure}

We start by considering the non interacting case with $U=0$, the so called dissipative Resonant Level Model (dRLM), which allows for an exact solution in the unitary and dissipative case by using standard Keldysh techniques (see Appendix~\ref{app:DRLM} and Ref.~\cite{ferreira2023exact}). As a result this model can be used in order to benchmark the diagMC algorithm and also in order to understand the effect of the Markovian dissipation. We consider for concreteness the case of symmetric spin dephasing $\gamma_\uparrow = \gamma_\downarrow$ and start from an empty initial state of the impurity $\rho_I(0) = | 0 \rangle \langle 0 |$.

In Fig.~\ref{fig4:benchmark} we plot the real-time dynamics of the impurity density $n_\sigma(t)$ for different values of dephasing (bottom panel) and impurity energy level (top panel). We note that spin symmetry is preserved through the time evolution, therefore $n_{\sigma}(t)=n_{\uparrow}(t)=n_{\downarrow}(t)$.  The comparison between the DiagMC results and the exact solution shows an excellent agreement at short times, with the Keldysh  results remaining well within the error bars at long time scales where the sign problem becomes more severe. The agreement is particularly good for large dephasing (see right panel) where as discussed our algorithm is more efficient.
Overall we see that the effect of a finite energy level introduces oscillations in the dynamics of the impurity, which are nevertheless well captured by DiagMC. 


\subsection{Charge and Spin Dynamics of the Dissipative AIM}

\begin{figure}[!t]
    \includegraphics[width=0.40\textwidth]{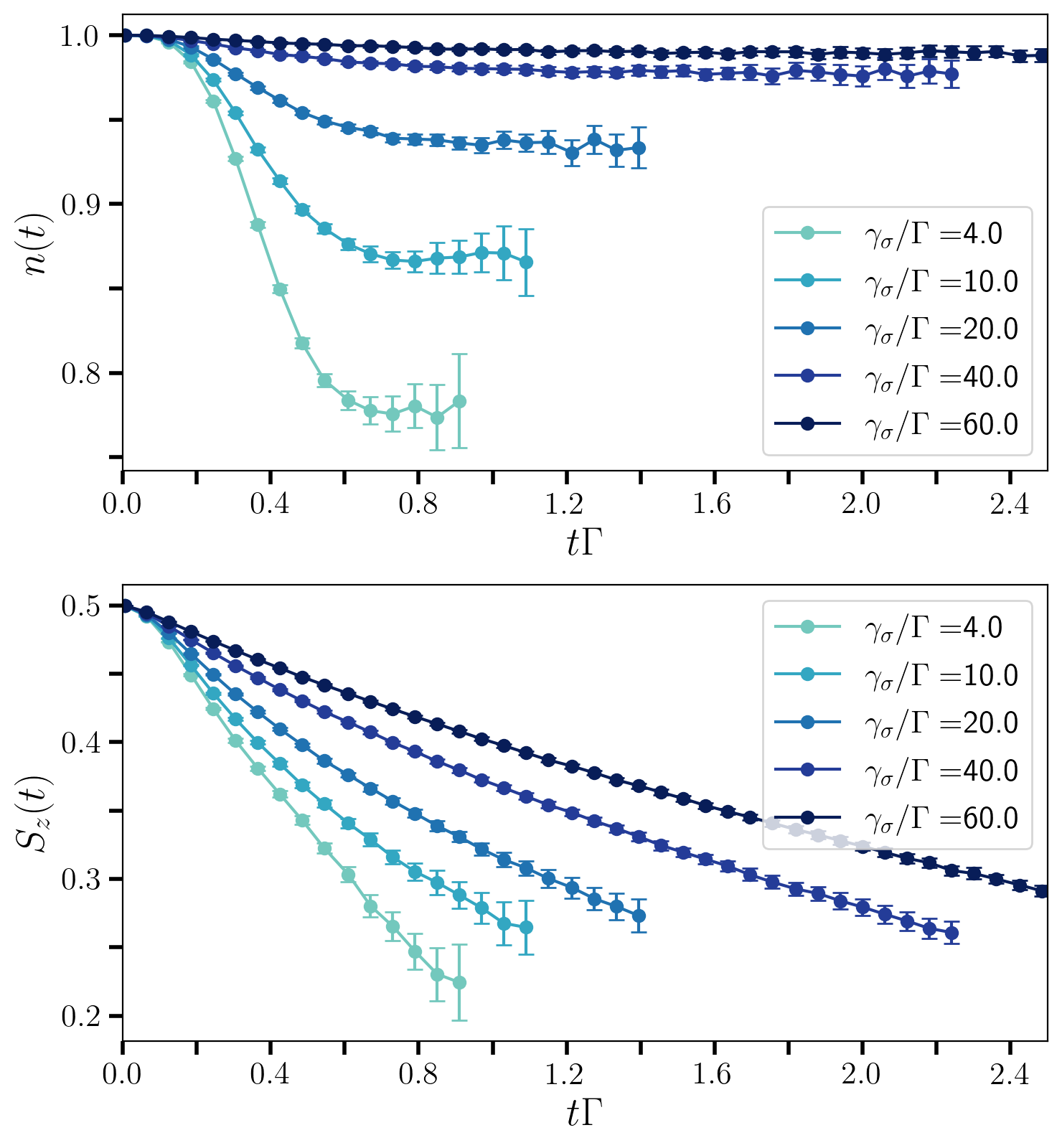}
    \caption{Charge and Spin impurity dynamics in the strong interaction limit, $U/\Gamma=10$, for different values of the dephasing.}
    \label{fig5:dephasing}
\end{figure}

We  now move to the interacting Anderson Impurity with dephasing and discuss the dynamics of charge and spin impurity as a function of different system parameters. We consider an initial condition with a single occupied, spin up impurity fermion, for simplicity, and discuss the role of the initial condition later on. Throughout this section we take $\varepsilon_d=0$.

\subsubsection{Effect of Dephasing}

We start discussing the dynamics in the strong interacting regime, $U=10 \Gamma$. In Figure~\ref{fig5:dephasing} we plot the dynamics of the impurity density and impurity spin
as a function of time for increasing value of the dephasing. In absence of dephasing, i.e. within the unitary AIM, we expect the initially polarised spin to hybridize with the bath and decay and also the charge on the dot to delocalize in the bath until an equilibrium value is reached (note that here we are not at particle-hole symmetry even for $\gamma_{\sigma}=0$, since $\varepsilon_d\neq -U/2$). 
In presence of dephasing this remains true, however we observe immediately  an interesting and counter-intuitive effect, namely upon increasing the dephasing rate the charge dynamics slow down significantly (see top panel) and the system remains frozen close to the initial state. This effect is particularly pronounced for the charge sector but is also visible on the spin dynamics (bottom panel): the initially prepared polarised spin decays in time with a slower rate in presence of a large dissipation. We interpret this result as a signature of the Zeno effect~\cite{misra1977a,garcia-ripoll2009,Froml2019,krapivsky2019free,Chaudhari_2022,secli2022steady}, where strong monitoring of a dot population leads to a freezing of the state. 
We note (not shown) that this behavior emerges also for moderate values of the local interaction $U\sim \Gamma$, suggesting its origin comes from a many-body effect due to the interplay between impurity-bath hybridization and strong dephasing. This result
is also in line with what discussed in the previous section, namely that dephasing reduces the average number of diagrams sampled, i.e. makes the system close to the atomic limit. 


\begin{figure}[!t] 
	\includegraphics[width=0.40\textwidth]{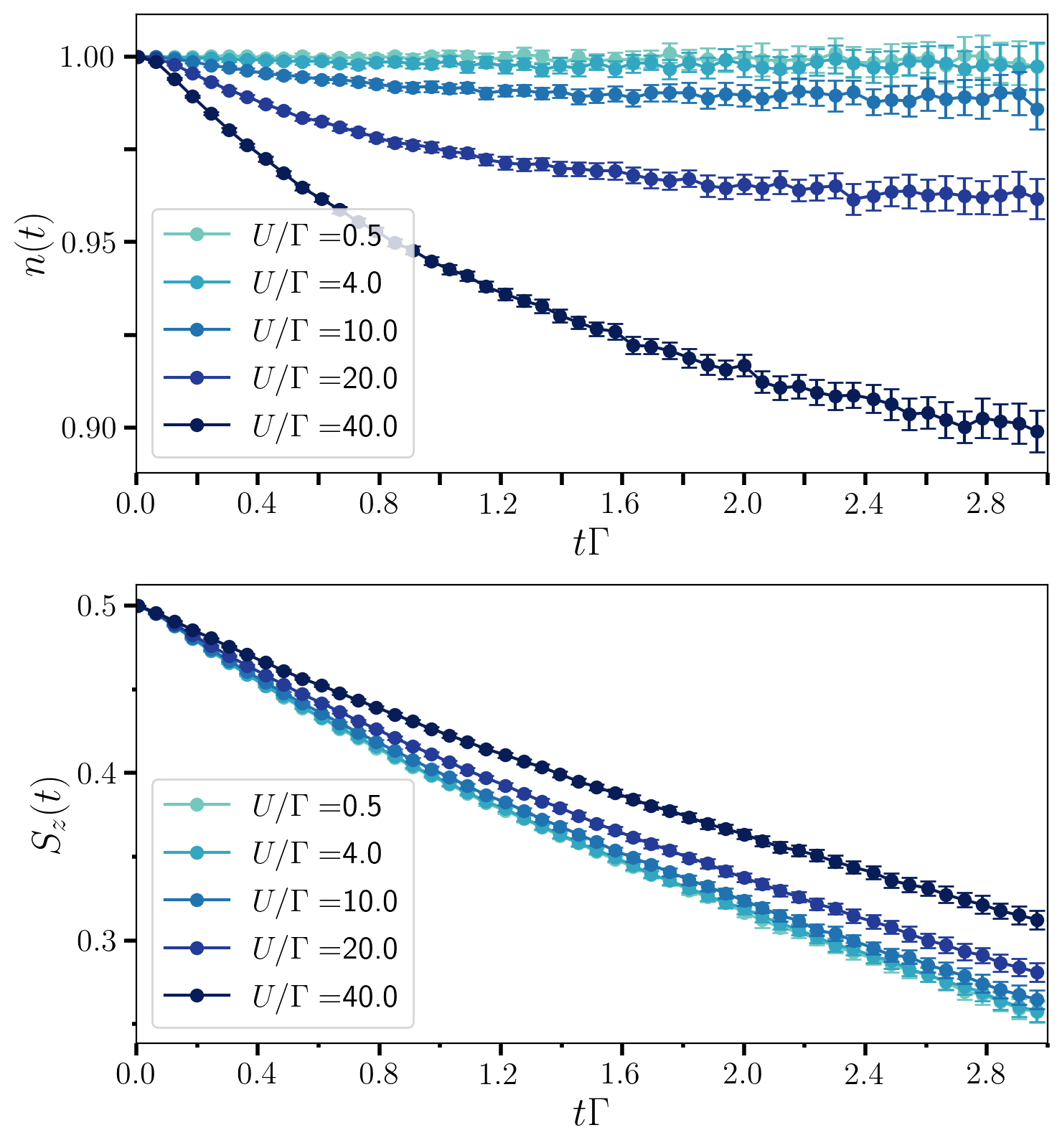}
    \caption{Charge and Spin impurity dynamics in the strong dephasing limit, $\gamma_{\sigma}/\Gamma=60$, for different values of the interaction. }
    \label{fig6:interaction}
\end{figure}

\subsubsection{Strong Dephasing Limit and Role of Interaction}

We now consider the regime of strong dephasing $\gamma_{\sigma}=60 \Gamma$ and study the charge and spin dynamics for different values of interaction	 $U/\Gamma$. In Fig.~\ref{fig6:interaction} we plot again the impurity density and the impurity spin starting from spin-up polarised state. We first of all note how in this regime our DiagMC algorithm is able to reach time scales of order $t\Gamma\sim 3$, while retaining very small error bars. This substantial increase with respect to the basic version of the hybridization algorithm~\cite{schiroPRB2009}, which is usually limited to $t\Gamma<1$, is due to the role played by the dephasing.
From these results we see clearly that increasing the interaction has the effect of slow down the dynamics of the impurity spin. We can understand this behavior from what is known about the unitary Anderson Impurity model, in particular a slow down of the impurity spin dynamics is a signature of the onset of the Kondo effect. An interesting effect is observed however in the impurity density which remains almost constant for weak interaction while start decaying for large $U$. We can understand this effect as the coupling with the fermionic bath induce some residual losses on the impurity, whose charge would otherwise be constant due to the Zeno effect and which however displays a slow decay.

\begin{figure}[!t]
    \includegraphics[width=0.40\textwidth]{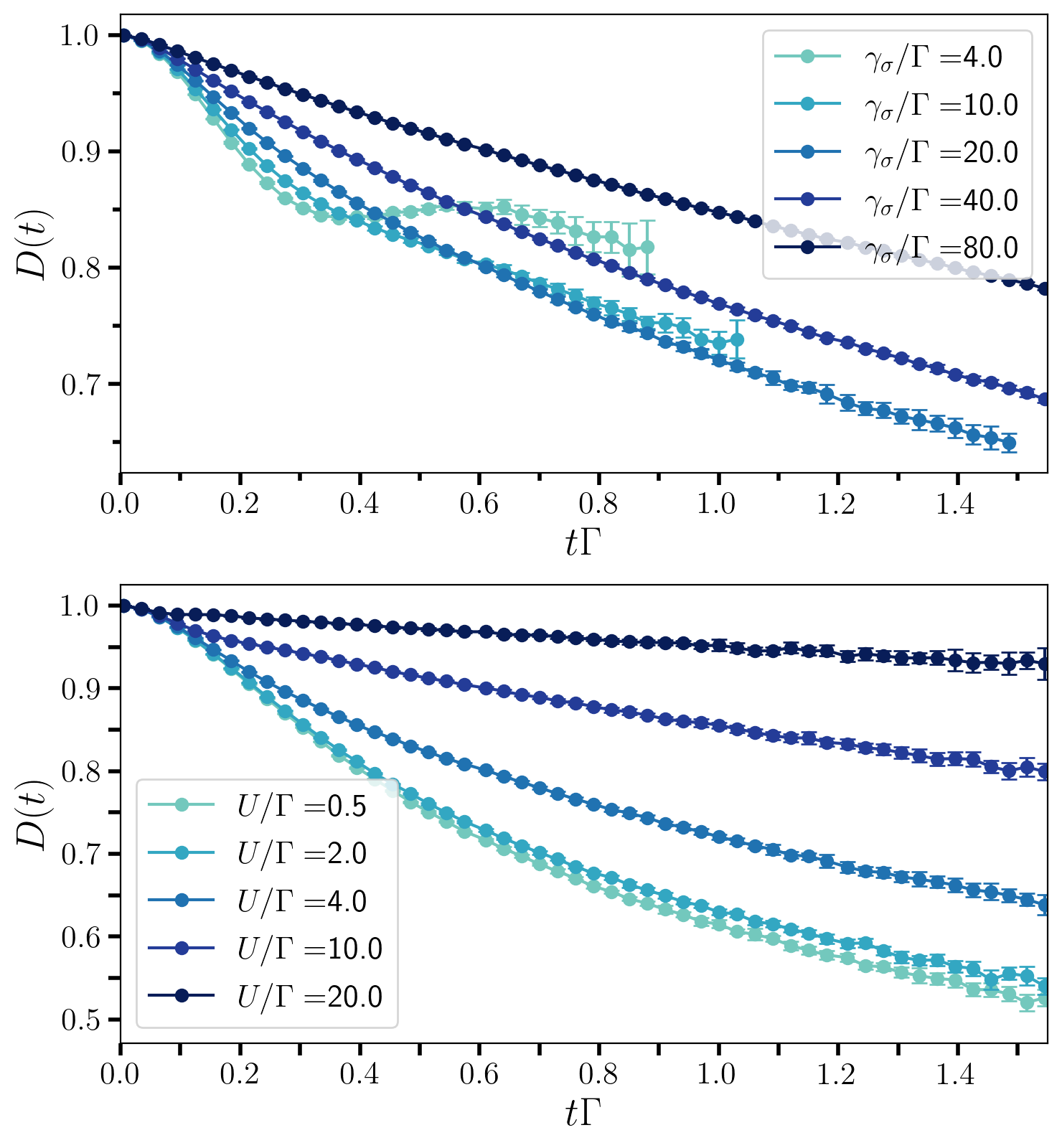}	
	\caption{Doublon dynamics starting from a full impurity,  for different values of dephasing (top panel)	at fixed interaction $U/\Gamma=10$ and at fixed dephasing (bottom panel)$\gamma_{\sigma}/\Gamma=20$ and different values of interaction.
	 }\label{fig7:doublon}
\end{figure}

\subsubsection{Dynamics of Doublons}

We now move our attention to the dynamics of doublons and discuss how this is affected by the presence of dephasing. In particular we consider an initial state of the impurity containing a doubly occupied site and study the time evolution after a quench of the bath coupling, in presence of dephasing. In this case we have therefore to modify the initial condition, which implies some differences in the algorithm as discussed previously. In Fig.~\ref{fig7:doublon} we plot the dynamics of doublon fraction at fixed interaction changing the dephasing (top panel) and fixed dephasing while changing the interaction (bottom panel). In both cases the initially prepared doublon decay with time, with a decay rate that increases with both interaction and dephasing. The first effect is the well known result related to the lifetime of a doublon in the strong interacting regime. The second one can be again be interpreted as the onset of the Zeno effect. We note however that, as compared to the total density (see Fig.~\ref{fig5:dephasing}) which remains practically constant for large dephasing here the doublon fraction still decays with time.



\begin{figure*}[!t]
    \includegraphics[width=0.85\textwidth]{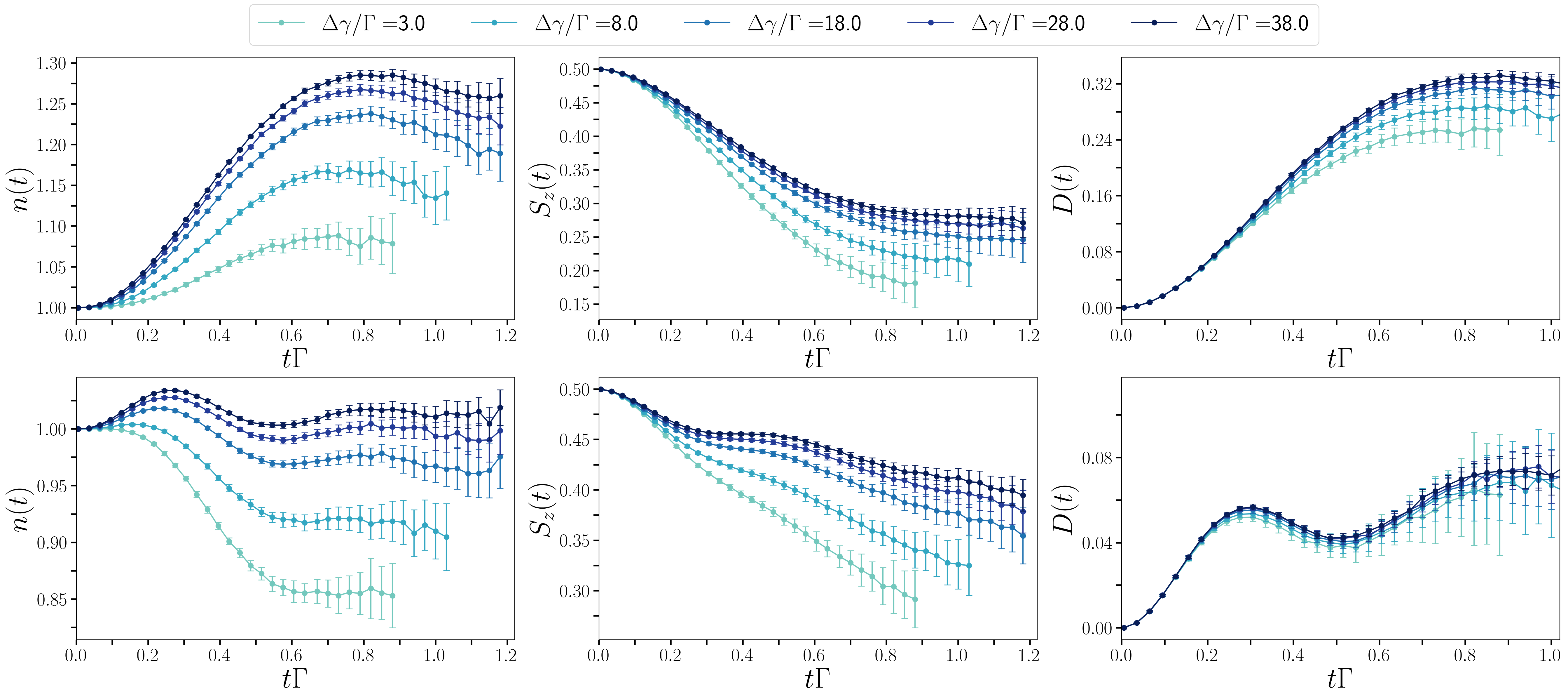}
    \caption{Charge dynamics of impurity density (left panels), Spin dynamics (middle panels) and doublon fraction (right panels) in the asymmetric dephasing case, with $\gamma_{\downarrow}/\Gamma=4$ and $U/\Gamma=0.5,10$ from top to bottom panels. }
	\label{fig7:asymmetric_charge}
\end{figure*}

\subsection{Asymmetric Dephasing}

Until now we have considered the case in which the dephasing acts equally on the two spin species of impurity fermions. We now discuss the case of spin-dependent dephasing $\gamma_{\uparrow}\neq\gamma_{\downarrow}$. Specifically we fix $\gamma_{\downarrow}/\Gamma=4$ and change the value of $\gamma_{\uparrow}$. In Fig.~\ref{fig7:asymmetric_charge} we plot the dynamics of charge, spin and doublon fraction upon increasing $\gamma_{\uparrow}$ at weak (top) and strong (bottom) interaction $U/\Gamma$. 

We first focus on the charge dynamics of the impurity (top left panel). At weak interactions $U=0.5\Gamma$ we observe a non-monotonous dynamics for the impurity density which increases at short times, reaches a maximum and then decay. For large asymmetry in the dephasing, i.e. when only one of the two spin species is strongly dissipative, this result in an increase of particle density, which is otherwise absent in the Zeno phase for symmetric dephasing. The maximum in the impurity density seems to be controlled by the interaction and indeed moves towards short times and smaller values upon increasing $U/\Gamma$ (bottom left panel). 
The dynamics of doublons (top/bottom right panels) on the other hand is much less affected by the asymmetry in the dephasing. We see for small interactions a large production of doublons, while upon increasing $U/\Gamma$ we see the emergence of coherent oscillations. Finally, the spin dynamics (top/bottom central panel) shows a rather interesting effect, namely that increasing the dephasing rate for the up spin results in a slow down of the dynamics at short time with the formation of a well defined magnetization plateau for very large $\gamma_{\uparrow}$. 

At longer time scales the dynamics seems to escape from this plateau and continue decaying towards zero magnetization. For weak interactions on the other hand there is no sign of the plateau at short times, yet the dynamics seems to reach a steady state where the impurity is still polarised. This can be understood since the asymmetric dephasing breaks the spin-rotation symmetry of the Anderson impurity model.

\subsection{Dynamics of Entanglement Entropy}

\begin{figure}[!t]
\includegraphics[width=0.40\textwidth]{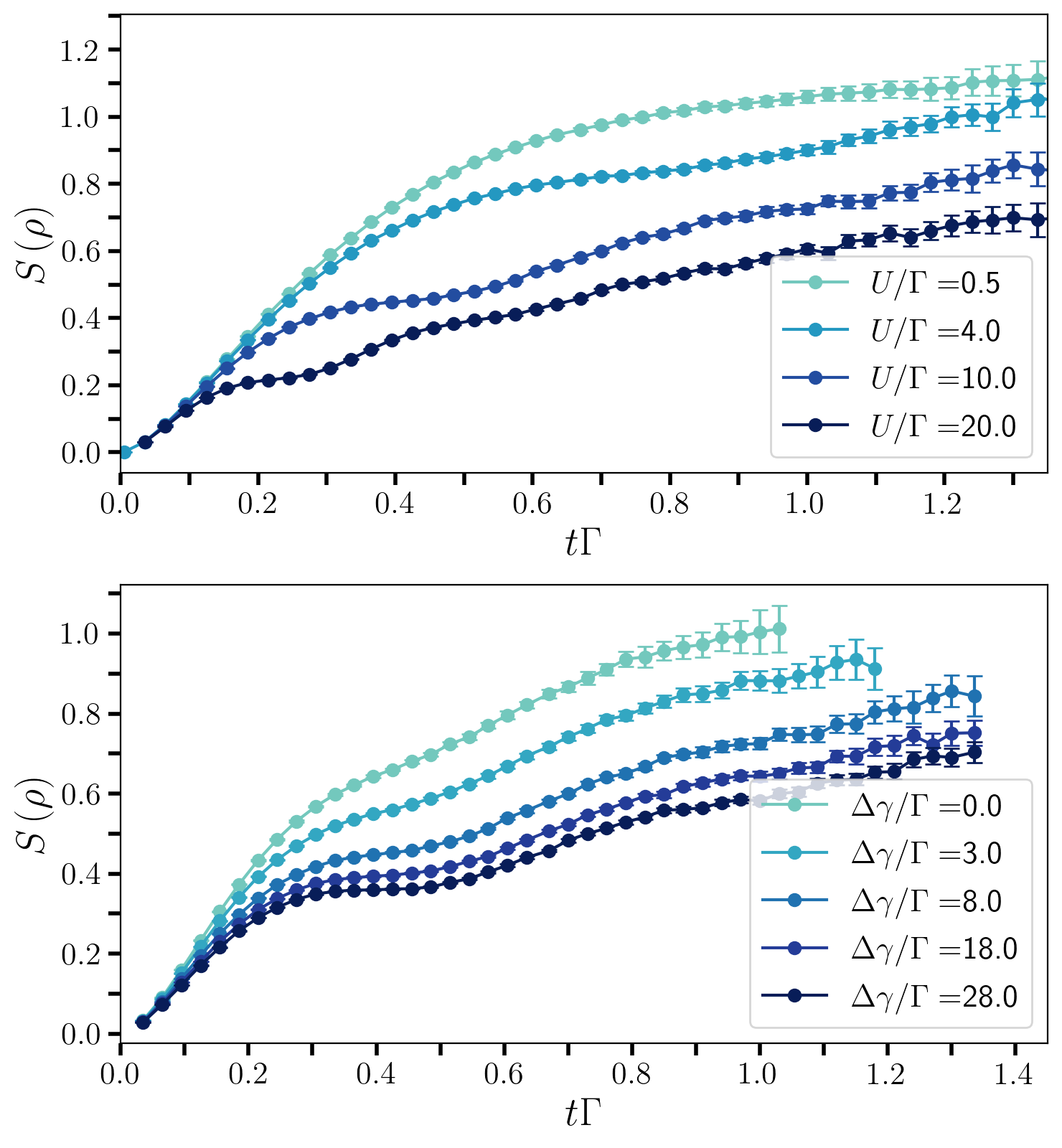}	
	\caption{Dynamics of Entanglement Entropy for different values of the interaction (at fixed dephasing $\gamma_{\sigma}/\Gamma=40$(top) and different values of dephasing asymmetry (bottom)	(at fixed interaction $U/\Gamma=10$).}\label{fig9:entanglement}
\end{figure}

In addition to the charge and spin dynamics we can compute the dynamics of the impurity entropy, which corresponds to the entanglement entropy after tracing out the fermionic bath 
We emphasize therefore that the state of the system is mixed to begin with, due to the dephasing, therefore the entropy of entanglement also takes contribution from the thermal entropy. To compute the entanglement entropy we reconstruct the impurity density matrix 
$$
\rho(t)=\sum_{ab}\rho_{ab}(t)\vert a,b\rangle\langle a,b\vert
$$
by sampling each individual matrix element $\rho_{ab}(t)$ and reconstruct the entropy from $S(\rho)=-\mbox{Tr}\left(\rho\mbox{log}\rho\right)$. In Fig.~\ref{fig9:entanglement} we plot the dynamics of the entanglement entropy for different values of the interaction at fixed large dephasing (top panel). We see that strong correlations on the impurity slows down the growth of entropy at short time. Similar effect is obtained by tuning the dephasing asymmetry at fixed interaction (bottom panel), where we see signatures of the magnetization plateau observed in the spin dynamics shown in Fig.~(\ref{fig7:asymmetric_charge}).

\section{Conclusion}
\label{sec:conclusions}

In this work we have extended the Diagrammatic Monte Carlo hybridization expansion algorithm to study the real-time dynamics of dissipative quantum impurity models, where the impurity is coupled to local Markovian dissipative processes (involving different impurity degrees of freedom) as well as to a fermionic bath. We formulate the hybridization expansion using the vectorization (or thermofield) formalism in which all the degrees of freedom are doubled to account for the correct structure of the density matrix (analog to upper and lower Keldysh contour in the conventional diagMC~\cite{schiroFabrizioPRB2009}).  In this picture the local Markovian dissipation is incorporated as dissipative interaction for the impurity degrees of freedom, thus entering the atomic limit around which the hybridization expansion is performed. With respect to the standard diagMC the main difference arises due to the fact that the theory is formulated on a single (collapsed) real-time contour where each impurity operator carries an extra quantum number (for the duplicated tilde Hilbert space). While our results are fairly general, independent on the specific form of the local Hamiltonian and local dissipator, we apply our algorithm to the Anderson impurity model with local dephasing. From the point of view of the algorithm we show that dissipation helps the convergence of the diagMC and alleviates the sign problem, thus allowing to reach longer time scales than for the unitary case. After benchmarking our method with an exactly solvable case we discuss how dephasing affects charge and spin dynamics of the Anderson impurity. We further discuss the case of asymmetric dephasing between spin up and spin down, which gives rise to an interesting dynamics for the impurity magnetization showing metastable plateau. In the future our algorithm can be further extended, implementing for examples the inchworm algorithm~\cite{cohen2015taming} and can be used as impurity solver for Dynamical Mean-Field Theory~\cite{scarlatella2021dynamical}.

\begin{acknowledgements}
We acknowledge computational resources on the Coll\'ege de France IPH cluster. This project has received funding from the European Research Council (ERC) under the European Union’s Horizon 2020 research and innovation programme (Grant agreement No. 101002955 — CONQUER).
\end{acknowledgements}

\appendix

\section{Structure of the Hybridization Function}

In the Hybridization expansion representation, the bath is completely characterized by the hybridization functions $\Delta_{\sigma,\bar{\sigma}}^{\alpha \bar{\alpha}}$, this function encodes the effect of the bath on the impurity degrees of freedom. In the case of the Anderson model, the coupling between the bath and the impurity degrees of freedom do not hybridize the spin channel, so we only have to consider the diagonal hybridization functions in spin, which are defined by:
\begin{align}
    \Delta_\sigma^{\alpha \bar{\alpha}}(\tau,\bar{\tau}) = -i \langle I_B \vert \m{T}_t \left[ \bar{\Phi}^\alpha_\sigma (\tau) \Phi_\sigma^{\bar{\alpha}}(\bar{\tau})\right] \vert \rho_{B}(0)\rangle
\end{align}
where both time arguments $\tau$ and $\bar{\tau}$ live on a single-real time contour and the operators
$\bar{\Phi}^\alpha_\sigma,\Phi_\sigma^{\bar{\alpha}}$ have been defined in the main text, see Eq.~(\ref{eqn:phi_psi}). To evaluate the hybridization function one needs to compute the Green's function of the fermion in the bath
\begin{align}
    G_{\bb{k},\sigma}^{\alpha , \bar{\alpha}}(\tau, \bar{\tau}) = -i \langle I_B \vert \m{T}_\tau \left[ \Phi^\alpha_{\bb{k},\sigma}(\tau) \bar{\Phi}^{\bar{\alpha}}_{\bb{k},\sigma}(\bar{\tau}) \right] \vert \rho_{B}(0)\rangle
\end{align}
where the average is taken overt the initial density matrix of the bath  $ \rho_{B}(0)$
\begin{align}
    \rho_{0,B} = \frac{e^{-\beta H_B}}{Z}
\end{align}
with bath Hamiltonian $H_B$ given by
\begin{align}
    H_B = \sum_{\bb{k},\sigma} \epsilon_{\bb{k},\sigma }  c_{\bb{k},\sigma}^\dagger c_{\bb{k},\sigma}
\end{align}
The time ordering operator $\m{T}_t$ entering the definition of the Green's function orders the operators according to their time and the Hilbert space they belong to.

For the time evolution of the bath operators (creation and annihilation operators) it is defined as usual with the Lindbladian of the Bath $\m{L}_B = -i \left( H_B -\tilde{H}_B\right) $:
\begin{align}
    \Phi_\sigma (t) = e^{-\m{L}_B t} \Phi_\sigma  e^{\m{L}_B t}
\end{align}
Depending on the position of the time arguments  $\tau$ and $\bar{\tau}$ along the single contour and their Hilbert space label $\alpha$, the hybridization function acquires a matrix structure 
\begin{align}
    \vec{\Delta}_\sigma (\tau,\bar{\tau}) = \begin{pmatrix}
        \Delta_\sigma^{00} (\tau,\bar{\tau}) & \Delta_\sigma^{01} (\tau,\bar{\tau}) \\
        \Delta_\sigma^{10} (\tau,\bar{\tau}) & \Delta_\sigma^{11} (\tau,\bar{\tau})
    \end{pmatrix}
\end{align}
From this we recognize a certain similarity with the Keldysh structure of the hybridization function in the conventional diagMC~\cite{schiroPRB2009}, as we will discuss more in detail below.
Moreover, we note that since we consider a time independent quantum impurity model, with a bath which is in thermal equilibrium, all the components of the hybridization function only depend on the time differences $\tau - \bar{\tau}$. Below we give the explicit expressions for the hybridization function entering the matrix representation above.

\subsection{Diagonal Sector}

We first consider the case when both operators $\Phi / \bar{\Phi}$  live on the same Hilbert Space $\m{H}/\Tilde{\m{H}}$. In this case, the contour time ordering $\m{T}_t$ acts as the real time-ordering operator,
\begin{align}
    \m{T}_t \left[  \Phi^\alpha(\tau) \Phi^\alpha(\bar{\tau})\right] = \left\{\begin{array}{ll}
         \Phi^\alpha(\tau) \Phi^\alpha(\bar{\tau}) & \tau>\bar{\tau} \\
         - \Phi^\alpha(\bar{\tau}) \Phi^\alpha(\tau)  & Else
    \end{array} \right.
\end{align}
where the two operators are living in the same Hilbert space $\m{H}/ \tilde{\m{H}}$. By using the expression of the spinor described in the main text, the bath Green's function can be expressed
as
\begin{align}
    G_{\bb{k},\sigma}^{0 0}(\tau, \bar{\tau}) = &-i \Theta(\tau - \Bar{\tau}) \langle c_{\bb{k},\sigma}(\tau) c^\dagger_{\bb{k},\sigma}(\bar{\tau})  \rangle \notag \\ &+ i \Theta(\bar{\tau} - \tau) \langle c^\dagger_{\bb{k},\sigma}(\bar{\tau})c_{\bb{k},\sigma}(\tau)\rangle 
\end{align}
where we recognize the usual Keldysh time-ordered Green function defines as $G^{\m{C}}_{\bb{k},\sigma} (\tau , \bar{\tau})= -i \langle \m{T}_t \left[ c_{\bb{k},\sigma}(\tau) c^\dagger_{\bb{k},\sigma}(\bar{\tau}) \right] \rangle$. As for the second diagonal component where the two operators are living in the $\tilde{\m{H}}$ Hilbert space, 
\begin{align}
    G_{\bb{k},\sigma}^{1 1}(\tau, \bar{\tau}) =& -i \Theta(\tau - \Bar{\tau}) \langle \tilde{c}^\dagger_{\bb{k},\sigma}(\tau) \tilde{c}_{\bb{k},\sigma}(\bar{\tau})  \rangle \notag \\ & + i \Theta(\bar{\tau} - \tau) \langle \tilde{c}_{\bb{k},\sigma}(\bar{\tau}) \tilde{c}^\dagger_{\bb{k},\sigma}(\tau)\rangle 
\end{align}
annihilation and creation operators in the tilde space can be expressed in terms of the physical operators by using the Superfermion relation,
\begin{align}
    &c_{\bb{k},\sigma} \vert I_B \rangle = -i \tilde{c}_{\bb{k},\sigma}^\dagger \vert I_B \rangle \notag \\ &
    c_{\bb{k},\sigma}^\dagger \vert I_B \rangle = -i \Tilde{c}_{\bb{k}\sigma} \vert I_B \rangle
\end{align}
which lead to the following relation
\begin{align}
    G_{\bb{k}\sigma}^{11} (\tau, \Bar{\tau}) = - G^{\tilde{\m{C}}}_{\bb{k}\sigma} (\tau ,\bar{\tau})
\end{align}
where $G^{\tilde{\m{C}}}_{\bb{k}\sigma} (\tau ,\bar{\tau})$ is the antitime-ordered Green function for the bath degrees of freedom.   

\subsection{Mixed Sector}

We now consider the case in which the two operators are living in different Hilbert Space,  this one correspond to the off diagonal ($\alpha \ne \bar{\alpha}$) component of the Hybridization function. Contrarily to the previous case, in the mixed sector the time ordering operator acts as,
\begin{align}
    &\m{T}_t \left[\Phi(\tau) \tilde{\Phi}(\Bar{\tau}) \right] = \Phi(\tau) \tilde{\Phi}(\Bar{\tau}) \notag \\ & \m{T}_t \left[ \tilde{\Phi}(\Bar{\tau}) \Phi(\tau) \right] =  - \Phi(\tau) \tilde{\Phi}(\Bar{\tau}) 
\end{align}
In the same spirit of the diagonal component of the Hybridization function, we can write the Green's function in the Keldysh formalism as:
\begin{align}
    i G_{\bb{k},\sigma}^{01}(\tau,\bar{\tau}) = G_{\bb{k}\sigma}^{<}(\tau,\bar{\tau}) \quad \text{and} \quad i G_{\bb{k},\sigma}^{10}(\tau,\bar{\tau}) = G_{\bb{k}\sigma}^{>}(\tau,\bar{\tau}) 
\end{align}
where $G^{<(>)}_{\bb{k}\sigma}$ are the lesser (greater) Green's functions.
With regard to the hybridization functions we obtain the standard result used also in the Keldysh formalism
\begin{align}
    \Delta_\sigma^{01}(\tau,\bar{\tau}) = \int d\epsilon \left(1-n_F(\epsilon) \right)\Gamma_\sigma(\epsilon) e^{- i \epsilon (\tau -\bar{\tau})}
\end{align}
and 
\begin{align}
    \Delta_\sigma^{10}(\tau,\bar{\tau}) = -\int d\epsilon n_F(\epsilon) \Gamma_\sigma(\epsilon) e^{- i \epsilon (\tau -\bar{\tau})}
\end{align}
where $n_F(\epsilon)$ is the Fermi distribution and $\Gamma(\epsilon)$ the energy-dependent hybridization for the channel $\sigma$, given by 
\begin{align}
    \Gamma_\sigma(\epsilon) = \sum_{\bb{k}} \vert V_{\bb{k}\sigma} \vert^2 \delta(\epsilon - \epsilon_{\bb{k}\sigma})
\end{align}
Finally, we obtain the two diagonal component $\Delta^{00}_\sigma/\Delta^{1}_\sigma$ of the hybridization function, which reduce to the off-diagonal ones depending, namely
\begin{align}
    \Delta^{00}_\sigma(\tau,\bar{\tau}) = i \Theta(\tau -\bar{\tau}) \Delta_\sigma^{10}(\tau,\bar{\tau}) + i \Theta( \bar{\tau}-\tau ) \Delta_\sigma^{01}(\tau,\bar{\tau})
\end{align}
and 
\begin{align}
    \Delta^{11}_\sigma(\tau,\bar{\tau}) = -i \Theta(\tau -\bar{\tau}) \Delta_\sigma^{01}(\tau,\bar{\tau}) - i \Theta( \bar{\tau}-\tau ) \Delta_\sigma^{10}(\tau,\bar{\tau})
\end{align}

\section{Dissipative Resonant-Level Model}\label{app:DRLM}

In this Appendix we briefly discuss the Keldysh solution of the dissipative Resonant Level Model (dRLM), that we use to benchmark the diagMC algorithm. This corresponds to the $U=0$ limit of the Anderson Impurity discussed in the main text. In absence of interaction the Hamiltonian is quadratic in all the fermionic degrees of freedom. The dephasing on the other hand introduces a dissipative interacting vertex, which however does not prevent to compute exactly certain quantities, in particular the single particle Green's functions. There are two ways to proceed to obtain the exact dynamics of the model. The first one is to look at the stochastic version of the Lindblad Master equation, corresponding to a unitary unravelling~\cite{ferreira2023exact}, in which the problem remains quadratic and averages over the noise can be taken exactly. The second one we follow in this paper is to write down the Dyson equation for the Green's function, starting from the Lindbladian. In particular we define the contour-time ordered Green's function
\begin{align}
    D_\sigma (s,s^\prime) = -i \langle \m{T}_C d_\sigma (\tau) d_\sigma(\tau^\prime) \rangle
\end{align}
where $\m{T}_C$ is the standard Keldysh time ordering operator. Even in the presence of the Markovian dissipation the retarded component of this Green's function satisfies a closed equation of motion which reads in frequency domain:
\begin{align}
    D^R_\sigma(\omega) = \frac{D^R_{0,\sigma}(\omega)}{1 - \Sigma^R_\sigma(\omega) D^R_\sigma(\omega) }
\end{align}
where $\Sigma^R_\sigma(\omega)$ is the retarded self energy for the spin channel $\sigma$. In fact, since we are interesting in the simplest resonant level model without any coupling between the spin channel, we can treat each channel of spin independently. For a given spin channel the retarded self energy reads:
\begin{align}
    \Sigma_\sigma^R(\omega) = -i \frac{\gamma_\sigma}{2} + \sum_{\bb{k}} |V_\bb{k}|^2 G_{0,\sigma\bb{k}}^R(\omega)
\end{align}
The first contribution to the self energy is the dephasing term, which is frequency independent and does not couple to the bath degrees of freedom. The second contribution is just the usual bath hybridization contribution, where $G^\sigma_{0,\bb{k}}(s,s^\prime) = -i \langle \m{T}_C c_{\bb{k}\sigma}(s) c^\dagger_{\bb{k}\sigma}(s^\prime) \rangle_0 $ denotes the bare bath Green’s function. By using the Langreth rules~\cite{kamenev_2011}, we can write the lesser Green's function $D^<_\sigma(s,s^\prime) = i \langle d_\sigma^\dagger(s^\prime) d_\sigma(s) \rangle$ as:
\begin{align}
    D^{<}=\left(1+D^r \Sigma^r\right) D_0^{<}\left(1+\Sigma^a D^a\right)+D^r \Sigma^{<} D^a
\end{align}
where the constraint given by the initial condition is encoded in $D_0^{<}$. Concerning the lesser self-energy $\Sigma^<$, as in the retarded case, we can decompose it into two contributions:
\begin{align}
     \Sigma^< (\tau,\tau^\prime) =  \Sigma^<_{B} (\tau,\tau^\prime) +  \Sigma^<_{\text{Deph}} (\tau,\tau^\prime)
\end{align}
the bath hybridization contribution $\Sigma^<_{B}$ and dephasing part $\Sigma^<_{Deph}$ given by
\begin{align}
     \Sigma^<_{\text{Deph}} (\tau,\tau^\prime) =  \gamma_\sigma D^{<}(\tau,\tau) \delta(\tau-\tau^\prime)
\end{align}
the instantaneous nature of dephasing self energy is due to the fact that in the Lindblad master equation we assume a Markovian environment with no memory and with a typical relaxation time that is negligible compared to the other relaxation times of the system.
Solving the Dyson equation for the retarded Green's function and then for the lesser component we can directly compute the dynamics of the impurity density, $n_{\sigma}(t)=
-iD^{<}_{\sigma}(t,t)$ which we use to benchmark the diagMC algorithm in the main text.


\begin{thebibliography}{80}%
\makeatletter
\providecommand \@ifxundefined [1]{%
 \@ifx{#1\undefined}
}%
\providecommand \@ifnum [1]{%
 \ifnum #1\expandafter \@firstoftwo
 \else \expandafter \@secondoftwo
 \fi
}%
\providecommand \@ifx [1]{%
 \ifx #1\expandafter \@firstoftwo
 \else \expandafter \@secondoftwo
 \fi
}%
\providecommand \natexlab [1]{#1}%
\providecommand \enquote  [1]{``#1''}%
\providecommand \bibnamefont  [1]{#1}%
\providecommand \bibfnamefont [1]{#1}%
\providecommand \citenamefont [1]{#1}%
\providecommand \href@noop [0]{\@secondoftwo}%
\providecommand \href [0]{\begingroup \@sanitize@url \@href}%
\providecommand \@href[1]{\@@startlink{#1}\@@href}%
\providecommand \@@href[1]{\endgroup#1\@@endlink}%
\providecommand \@sanitize@url [0]{\catcode `\\12\catcode `\$12\catcode
  `\&12\catcode `\#12\catcode `\^12\catcode `\_12\catcode `\%12\relax}%
\providecommand \@@startlink[1]{}%
\providecommand \@@endlink[0]{}%
\providecommand \url  [0]{\begingroup\@sanitize@url \@url }%
\providecommand \@url [1]{\endgroup\@href {#1}{\urlprefix }}%
\providecommand \urlprefix  [0]{URL }%
\providecommand \Eprint [0]{\href }%
\providecommand \doibase [0]{https://doi.org/}%
\providecommand \selectlanguage [0]{\@gobble}%
\providecommand \bibinfo  [0]{\@secondoftwo}%
\providecommand \bibfield  [0]{\@secondoftwo}%
\providecommand \translation [1]{[#1]}%
\providecommand \BibitemOpen [0]{}%
\providecommand \bibitemStop [0]{}%
\providecommand \bibitemNoStop [0]{.\EOS\space}%
\providecommand \EOS [0]{\spacefactor3000\relax}%
\providecommand \BibitemShut  [1]{\csname bibitem#1\endcsname}%
\let\auto@bib@innerbib\@empty
\bibitem [{\citenamefont {Leggett}\ \emph {et~al.}(1987)\citenamefont
  {Leggett}, \citenamefont {Chakravarty}, \citenamefont {Dorsey}, \citenamefont
  {Fisher}, \citenamefont {Garg},\ and\ \citenamefont
  {Zwerger}}]{leggett1987dynamics}%
  \BibitemOpen
  \bibfield  {author} {\bibinfo {author} {\bibfnamefont {A.~J.}\ \bibnamefont
  {Leggett}}, \bibinfo {author} {\bibfnamefont {S.}~\bibnamefont
  {Chakravarty}}, \bibinfo {author} {\bibfnamefont {A.~T.}\ \bibnamefont
  {Dorsey}}, \bibinfo {author} {\bibfnamefont {M.~P.~A.}\ \bibnamefont
  {Fisher}}, \bibinfo {author} {\bibfnamefont {A.}~\bibnamefont {Garg}},\ and\
  \bibinfo {author} {\bibfnamefont {W.}~\bibnamefont {Zwerger}},\ }\bibfield
  {title} {\bibinfo {title} {Dynamics of the dissipative two-state system},\
  }\href {https://doi.org/10.1103/RevModPhys.59.1} {\bibfield  {journal}
  {\bibinfo  {journal} {Rev. Mod. Phys.}\ }\textbf {\bibinfo {volume} {59}},\
  \bibinfo {pages} {1} (\bibinfo {year} {1987})}\BibitemShut {NoStop}%
\bibitem [{\citenamefont {Pustilnik}\ and\ \citenamefont
  {Glazman}(2004)}]{Pustilnik_2004}%
  \BibitemOpen
  \bibfield  {author} {\bibinfo {author} {\bibfnamefont {M.}~\bibnamefont
  {Pustilnik}}\ and\ \bibinfo {author} {\bibfnamefont {L.}~\bibnamefont
  {Glazman}},\ }\bibfield  {title} {\bibinfo {title} {Kondo effect in quantum
  dots},\ }\href {https://doi.org/10.1088/0953-8984/16/16/R01} {\bibfield
  {journal} {\bibinfo  {journal} {Journal of Physics: Condensed Matter}\
  }\textbf {\bibinfo {volume} {16}},\ \bibinfo {pages} {R513} (\bibinfo {year}
  {2004})}\BibitemShut {NoStop}%
\bibitem [{\citenamefont {Haroche}\ and\ \citenamefont
  {Raimond}(2006)}]{Haroche:993568}%
  \BibitemOpen
  \bibfield  {author} {\bibinfo {author} {\bibfnamefont {S.}~\bibnamefont
  {Haroche}}\ and\ \bibinfo {author} {\bibfnamefont {J.~M.}\ \bibnamefont
  {Raimond}},\ }\href
  {https://doi.org/10.1093/acprof:oso/9780198509141.001.0001} {\emph {\bibinfo
  {title} {{Exploring the Quantum: Atoms, Cavities, and Photons}}}}\ (\bibinfo
  {publisher} {Oxford Univ. Press},\ \bibinfo {address} {Oxford},\ \bibinfo
  {year} {2006})\BibitemShut {NoStop}%
\bibitem [{\citenamefont {Breuer}\ and\ \citenamefont
  {Petruccione}(2007)}]{breuerPetruccione2010}%
  \BibitemOpen
  \bibfield  {author} {\bibinfo {author} {\bibfnamefont {H.~P.}\ \bibnamefont
  {Breuer}}\ and\ \bibinfo {author} {\bibfnamefont {F.}~\bibnamefont
  {Petruccione}},\ }\href
  {https://doi.org/10.1093/acprof:oso/9780199213900.001.0001} {\emph {\bibinfo
  {title} {The {{Theory}} of {{Open Quantum Systems}}}}},\ \bibinfo {edition}
  {1st}\ ed.,\ Vol.\ \bibinfo {volume} {9780199213}\ (\bibinfo  {publisher}
  {{OUP Oxford}},\ \bibinfo {year} {2007})\BibitemShut {NoStop}%
\bibitem [{\citenamefont {Weiss}(2022)}]{weissquantumdissipative}%
  \BibitemOpen
  \bibfield  {author} {\bibinfo {author} {\bibfnamefont {U.}~\bibnamefont
  {Weiss}},\ }\href@noop {} {\emph {\bibinfo {title} {Quantum Dissipative
  Systems}}},\ \bibinfo {edition} {4th}\ ed.\ (\bibinfo  {publisher} {World
  Scientific},\ \bibinfo {address} {New Jersey},\ \bibinfo {year}
  {2022})\BibitemShut {NoStop}%
\bibitem [{\citenamefont {Mi}\ \emph {et~al.}(2022)\citenamefont {Mi},
  \citenamefont {Sonner}, \citenamefont {Niu}, \citenamefont {Lee},
  \citenamefont {Foxen}, \citenamefont {Acharya}, \citenamefont {Aleiner},
  \citenamefont {Andersen}, \citenamefont {Arute}, \citenamefont {Arya},
  \citenamefont {Asfaw}, \citenamefont {Atalaya}, \citenamefont {Bardin},
  \citenamefont {Basso}, \citenamefont {Bengtsson}, \citenamefont {Bortoli},
  \citenamefont {Bourassa}, \citenamefont {Brill}, \citenamefont {Broughton},
  \citenamefont {Buckley}, \citenamefont {Buell}, \citenamefont {Burkett},
  \citenamefont {Bushnell}, \citenamefont {Chen}, \citenamefont {Chiaro},
  \citenamefont {Collins}, \citenamefont {Conner}, \citenamefont {Courtney},
  \citenamefont {Crook}, \citenamefont {Debroy}, \citenamefont {Demura},
  \citenamefont {Dunsworth}, \citenamefont {Eppens}, \citenamefont {Erickson},
  \citenamefont {Faoro}, \citenamefont {Farhi}, \citenamefont {Fatemi},
  \citenamefont {Flores}, \citenamefont {Forati}, \citenamefont {Fowler},
  \citenamefont {Giang}, \citenamefont {Gidney}, \citenamefont {Gilboa},
  \citenamefont {Giustina}, \citenamefont {Dau}, \citenamefont {Gross},
  \citenamefont {Habegger}, \citenamefont {Harrigan}, \citenamefont {Hoffmann},
  \citenamefont {Hong}, \citenamefont {Huang}, \citenamefont {Huff},
  \citenamefont {Huggins}, \citenamefont {Ioffe}, \citenamefont {Isakov},
  \citenamefont {Iveland}, \citenamefont {Jeffrey}, \citenamefont {Jiang},
  \citenamefont {Jones}, \citenamefont {Kafri}, \citenamefont {Kechedzhi},
  \citenamefont {Khattar}, \citenamefont {Kim}, \citenamefont {Kitaev},
  \citenamefont {Klimov}, \citenamefont {Klots}, \citenamefont {Korotkov},
  \citenamefont {Kostritsa}, \citenamefont {Kreikebaum}, \citenamefont
  {Landhuis}, \citenamefont {Laptev}, \citenamefont {Lau}, \citenamefont {Lee},
  \citenamefont {Laws}, \citenamefont {Liu}, \citenamefont {Locharla},
  \citenamefont {Martin}, \citenamefont {McClean}, \citenamefont {McEwen},
  \citenamefont {Costa}, \citenamefont {Miao}, \citenamefont {Mohseni},
  \citenamefont {Montazeri}, \citenamefont {Morvan}, \citenamefont {Mount},
  \citenamefont {Mruczkiewicz}, \citenamefont {Naaman}, \citenamefont {Neeley},
  \citenamefont {Neill}, \citenamefont {Newman}, \citenamefont {O’Brien},
  \citenamefont {Opremcak}, \citenamefont {Petukhov}, \citenamefont {Potter},
  \citenamefont {Quintana}, \citenamefont {Rubin}, \citenamefont {Saei},
  \citenamefont {Sank}, \citenamefont {Sankaragomathi}, \citenamefont
  {Satzinger}, \citenamefont {Schuster}, \citenamefont {Shearn}, \citenamefont
  {Shvarts}, \citenamefont {Strain}, \citenamefont {Su}, \citenamefont
  {Szalay}, \citenamefont {Vidal}, \citenamefont {Villalonga}, \citenamefont
  {Vollgraff-Heidweiller}, \citenamefont {White}, \citenamefont {Yao},
  \citenamefont {Yeh}, \citenamefont {Yoo}, \citenamefont {Zalcman},
  \citenamefont {Zhang}, \citenamefont {Zhu}, \citenamefont {Neven},
  \citenamefont {Bacon}, \citenamefont {Hilton}, \citenamefont {Lucero},
  \citenamefont {Babbush}, \citenamefont {Boixo}, \citenamefont {Megrant},
  \citenamefont {Chen}, \citenamefont {Kelly}, \citenamefont {Smelyanskiy},
  \citenamefont {Abanin},\ and\ \citenamefont
  {Roushan}}]{google_dephasing_abanin}%
  \BibitemOpen
  \bibfield  {author} {\bibinfo {author} {\bibfnamefont {X.}~\bibnamefont
  {Mi}}, \bibinfo {author} {\bibfnamefont {M.}~\bibnamefont {Sonner}}, \bibinfo
  {author} {\bibfnamefont {M.~Y.}\ \bibnamefont {Niu}}, \bibinfo {author}
  {\bibfnamefont {K.~W.}\ \bibnamefont {Lee}}, \bibinfo {author} {\bibfnamefont
  {B.}~\bibnamefont {Foxen}}, \bibinfo {author} {\bibfnamefont
  {R.}~\bibnamefont {Acharya}}, \bibinfo {author} {\bibfnamefont
  {I.}~\bibnamefont {Aleiner}}, \bibinfo {author} {\bibfnamefont {T.~I.}\
  \bibnamefont {Andersen}}, \bibinfo {author} {\bibfnamefont {F.}~\bibnamefont
  {Arute}}, \bibinfo {author} {\bibfnamefont {K.}~\bibnamefont {Arya}},
  \bibinfo {author} {\bibfnamefont {A.}~\bibnamefont {Asfaw}}, \bibinfo
  {author} {\bibfnamefont {J.}~\bibnamefont {Atalaya}}, \bibinfo {author}
  {\bibfnamefont {J.~C.}\ \bibnamefont {Bardin}}, \bibinfo {author}
  {\bibfnamefont {J.}~\bibnamefont {Basso}}, \bibinfo {author} {\bibfnamefont
  {A.}~\bibnamefont {Bengtsson}}, \bibinfo {author} {\bibfnamefont
  {G.}~\bibnamefont {Bortoli}}, \bibinfo {author} {\bibfnamefont
  {A.}~\bibnamefont {Bourassa}}, \bibinfo {author} {\bibfnamefont
  {L.}~\bibnamefont {Brill}}, \bibinfo {author} {\bibfnamefont
  {M.}~\bibnamefont {Broughton}}, \bibinfo {author} {\bibfnamefont {B.~B.}\
  \bibnamefont {Buckley}}, \bibinfo {author} {\bibfnamefont {D.~A.}\
  \bibnamefont {Buell}}, \bibinfo {author} {\bibfnamefont {B.}~\bibnamefont
  {Burkett}}, \bibinfo {author} {\bibfnamefont {N.}~\bibnamefont {Bushnell}},
  \bibinfo {author} {\bibfnamefont {Z.}~\bibnamefont {Chen}}, \bibinfo {author}
  {\bibfnamefont {B.}~\bibnamefont {Chiaro}}, \bibinfo {author} {\bibfnamefont
  {R.}~\bibnamefont {Collins}}, \bibinfo {author} {\bibfnamefont
  {P.}~\bibnamefont {Conner}}, \bibinfo {author} {\bibfnamefont
  {W.}~\bibnamefont {Courtney}}, \bibinfo {author} {\bibfnamefont {A.~L.}\
  \bibnamefont {Crook}}, \bibinfo {author} {\bibfnamefont {D.~M.}\ \bibnamefont
  {Debroy}}, \bibinfo {author} {\bibfnamefont {S.}~\bibnamefont {Demura}},
  \bibinfo {author} {\bibfnamefont {A.}~\bibnamefont {Dunsworth}}, \bibinfo
  {author} {\bibfnamefont {D.}~\bibnamefont {Eppens}}, \bibinfo {author}
  {\bibfnamefont {C.}~\bibnamefont {Erickson}}, \bibinfo {author}
  {\bibfnamefont {L.}~\bibnamefont {Faoro}}, \bibinfo {author} {\bibfnamefont
  {E.}~\bibnamefont {Farhi}}, \bibinfo {author} {\bibfnamefont
  {R.}~\bibnamefont {Fatemi}}, \bibinfo {author} {\bibfnamefont
  {L.}~\bibnamefont {Flores}}, \bibinfo {author} {\bibfnamefont
  {E.}~\bibnamefont {Forati}}, \bibinfo {author} {\bibfnamefont {A.~G.}\
  \bibnamefont {Fowler}}, \bibinfo {author} {\bibfnamefont {W.}~\bibnamefont
  {Giang}}, \bibinfo {author} {\bibfnamefont {C.}~\bibnamefont {Gidney}},
  \bibinfo {author} {\bibfnamefont {D.}~\bibnamefont {Gilboa}}, \bibinfo
  {author} {\bibfnamefont {M.}~\bibnamefont {Giustina}}, \bibinfo {author}
  {\bibfnamefont {A.~G.}\ \bibnamefont {Dau}}, \bibinfo {author} {\bibfnamefont
  {J.~A.}\ \bibnamefont {Gross}}, \bibinfo {author} {\bibfnamefont
  {S.}~\bibnamefont {Habegger}}, \bibinfo {author} {\bibfnamefont {M.~P.}\
  \bibnamefont {Harrigan}}, \bibinfo {author} {\bibfnamefont {M.}~\bibnamefont
  {Hoffmann}}, \bibinfo {author} {\bibfnamefont {S.}~\bibnamefont {Hong}},
  \bibinfo {author} {\bibfnamefont {T.}~\bibnamefont {Huang}}, \bibinfo
  {author} {\bibfnamefont {A.}~\bibnamefont {Huff}}, \bibinfo {author}
  {\bibfnamefont {W.~J.}\ \bibnamefont {Huggins}}, \bibinfo {author}
  {\bibfnamefont {L.~B.}\ \bibnamefont {Ioffe}}, \bibinfo {author}
  {\bibfnamefont {S.~V.}\ \bibnamefont {Isakov}}, \bibinfo {author}
  {\bibfnamefont {J.}~\bibnamefont {Iveland}}, \bibinfo {author} {\bibfnamefont
  {E.}~\bibnamefont {Jeffrey}}, \bibinfo {author} {\bibfnamefont
  {Z.}~\bibnamefont {Jiang}}, \bibinfo {author} {\bibfnamefont
  {C.}~\bibnamefont {Jones}}, \bibinfo {author} {\bibfnamefont
  {D.}~\bibnamefont {Kafri}}, \bibinfo {author} {\bibfnamefont
  {K.}~\bibnamefont {Kechedzhi}}, \bibinfo {author} {\bibfnamefont
  {T.}~\bibnamefont {Khattar}}, \bibinfo {author} {\bibfnamefont
  {S.}~\bibnamefont {Kim}}, \bibinfo {author} {\bibfnamefont {A.~Y.}\
  \bibnamefont {Kitaev}}, \bibinfo {author} {\bibfnamefont {P.~V.}\
  \bibnamefont {Klimov}}, \bibinfo {author} {\bibfnamefont {A.~R.}\
  \bibnamefont {Klots}}, \bibinfo {author} {\bibfnamefont {A.~N.}\ \bibnamefont
  {Korotkov}}, \bibinfo {author} {\bibfnamefont {F.}~\bibnamefont {Kostritsa}},
  \bibinfo {author} {\bibfnamefont {J.~M.}\ \bibnamefont {Kreikebaum}},
  \bibinfo {author} {\bibfnamefont {D.}~\bibnamefont {Landhuis}}, \bibinfo
  {author} {\bibfnamefont {P.}~\bibnamefont {Laptev}}, \bibinfo {author}
  {\bibfnamefont {K.-M.}\ \bibnamefont {Lau}}, \bibinfo {author} {\bibfnamefont
  {J.}~\bibnamefont {Lee}}, \bibinfo {author} {\bibfnamefont {L.}~\bibnamefont
  {Laws}}, \bibinfo {author} {\bibfnamefont {W.}~\bibnamefont {Liu}}, \bibinfo
  {author} {\bibfnamefont {A.}~\bibnamefont {Locharla}}, \bibinfo {author}
  {\bibfnamefont {O.}~\bibnamefont {Martin}}, \bibinfo {author} {\bibfnamefont
  {J.~R.}\ \bibnamefont {McClean}}, \bibinfo {author} {\bibfnamefont
  {M.}~\bibnamefont {McEwen}}, \bibinfo {author} {\bibfnamefont {B.~M.}\
  \bibnamefont {Costa}}, \bibinfo {author} {\bibfnamefont {K.~C.}\ \bibnamefont
  {Miao}}, \bibinfo {author} {\bibfnamefont {M.}~\bibnamefont {Mohseni}},
  \bibinfo {author} {\bibfnamefont {S.}~\bibnamefont {Montazeri}}, \bibinfo
  {author} {\bibfnamefont {A.}~\bibnamefont {Morvan}}, \bibinfo {author}
  {\bibfnamefont {E.}~\bibnamefont {Mount}}, \bibinfo {author} {\bibfnamefont
  {W.}~\bibnamefont {Mruczkiewicz}}, \bibinfo {author} {\bibfnamefont
  {O.}~\bibnamefont {Naaman}}, \bibinfo {author} {\bibfnamefont
  {M.}~\bibnamefont {Neeley}}, \bibinfo {author} {\bibfnamefont
  {C.}~\bibnamefont {Neill}}, \bibinfo {author} {\bibfnamefont
  {M.}~\bibnamefont {Newman}}, \bibinfo {author} {\bibfnamefont {T.~E.}\
  \bibnamefont {O’Brien}}, \bibinfo {author} {\bibfnamefont {A.}~\bibnamefont
  {Opremcak}}, \bibinfo {author} {\bibfnamefont {A.}~\bibnamefont {Petukhov}},
  \bibinfo {author} {\bibfnamefont {R.}~\bibnamefont {Potter}}, \bibinfo
  {author} {\bibfnamefont {C.}~\bibnamefont {Quintana}}, \bibinfo {author}
  {\bibfnamefont {N.~C.}\ \bibnamefont {Rubin}}, \bibinfo {author}
  {\bibfnamefont {N.}~\bibnamefont {Saei}}, \bibinfo {author} {\bibfnamefont
  {D.}~\bibnamefont {Sank}}, \bibinfo {author} {\bibfnamefont {K.}~\bibnamefont
  {Sankaragomathi}}, \bibinfo {author} {\bibfnamefont {K.~J.}\ \bibnamefont
  {Satzinger}}, \bibinfo {author} {\bibfnamefont {C.}~\bibnamefont {Schuster}},
  \bibinfo {author} {\bibfnamefont {M.~J.}\ \bibnamefont {Shearn}}, \bibinfo
  {author} {\bibfnamefont {V.}~\bibnamefont {Shvarts}}, \bibinfo {author}
  {\bibfnamefont {D.}~\bibnamefont {Strain}}, \bibinfo {author} {\bibfnamefont
  {Y.}~\bibnamefont {Su}}, \bibinfo {author} {\bibfnamefont {M.}~\bibnamefont
  {Szalay}}, \bibinfo {author} {\bibfnamefont {G.}~\bibnamefont {Vidal}},
  \bibinfo {author} {\bibfnamefont {B.}~\bibnamefont {Villalonga}}, \bibinfo
  {author} {\bibfnamefont {C.}~\bibnamefont {Vollgraff-Heidweiller}}, \bibinfo
  {author} {\bibfnamefont {T.}~\bibnamefont {White}}, \bibinfo {author}
  {\bibfnamefont {Z.}~\bibnamefont {Yao}}, \bibinfo {author} {\bibfnamefont
  {P.}~\bibnamefont {Yeh}}, \bibinfo {author} {\bibfnamefont {J.}~\bibnamefont
  {Yoo}}, \bibinfo {author} {\bibfnamefont {A.}~\bibnamefont {Zalcman}},
  \bibinfo {author} {\bibfnamefont {Y.}~\bibnamefont {Zhang}}, \bibinfo
  {author} {\bibfnamefont {N.}~\bibnamefont {Zhu}}, \bibinfo {author}
  {\bibfnamefont {H.}~\bibnamefont {Neven}}, \bibinfo {author} {\bibfnamefont
  {D.}~\bibnamefont {Bacon}}, \bibinfo {author} {\bibfnamefont
  {J.}~\bibnamefont {Hilton}}, \bibinfo {author} {\bibfnamefont
  {E.}~\bibnamefont {Lucero}}, \bibinfo {author} {\bibfnamefont
  {R.}~\bibnamefont {Babbush}}, \bibinfo {author} {\bibfnamefont
  {S.}~\bibnamefont {Boixo}}, \bibinfo {author} {\bibfnamefont
  {A.}~\bibnamefont {Megrant}}, \bibinfo {author} {\bibfnamefont
  {Y.}~\bibnamefont {Chen}}, \bibinfo {author} {\bibfnamefont {J.}~\bibnamefont
  {Kelly}}, \bibinfo {author} {\bibfnamefont {V.}~\bibnamefont {Smelyanskiy}},
  \bibinfo {author} {\bibfnamefont {D.~A.}\ \bibnamefont {Abanin}},\ and\
  \bibinfo {author} {\bibfnamefont {P.}~\bibnamefont {Roushan}},\ }\bibfield
  {title} {\bibinfo {title} {Noise-resilient edge modes on a chain of
  superconducting qubits},\ }\href {https://doi.org/10.1126/science.abq5769}
  {\bibfield  {journal} {\bibinfo  {journal} {Science}\ }\textbf {\bibinfo
  {volume} {378}},\ \bibinfo {pages} {785} (\bibinfo {year} {2022})},\ \Eprint
  {https://arxiv.org/abs/https://www.science.org/doi/pdf/10.1126/science.abq5769}
  {https://www.science.org/doi/pdf/10.1126/science.abq5769} \BibitemShut
  {NoStop}%
\bibitem [{\citenamefont {Lebrat}\ \emph {et~al.}(2019)\citenamefont {Lebrat},
  \citenamefont {H\"ausler}, \citenamefont {Fabritius}, \citenamefont
  {Husmann}, \citenamefont {Corman},\ and\ \citenamefont
  {Esslinger}}]{lebrat2019quantized}%
  \BibitemOpen
  \bibfield  {author} {\bibinfo {author} {\bibfnamefont {M.}~\bibnamefont
  {Lebrat}}, \bibinfo {author} {\bibfnamefont {S.}~\bibnamefont {H\"ausler}},
  \bibinfo {author} {\bibfnamefont {P.}~\bibnamefont {Fabritius}}, \bibinfo
  {author} {\bibfnamefont {D.}~\bibnamefont {Husmann}}, \bibinfo {author}
  {\bibfnamefont {L.}~\bibnamefont {Corman}},\ and\ \bibinfo {author}
  {\bibfnamefont {T.}~\bibnamefont {Esslinger}},\ }\bibfield  {title} {\bibinfo
  {title} {Quantized conductance through a spin-selective atomic point
  contact},\ }\href {https://doi.org/10.1103/PhysRevLett.123.193605} {\bibfield
   {journal} {\bibinfo  {journal} {Phys. Rev. Lett.}\ }\textbf {\bibinfo
  {volume} {123}},\ \bibinfo {pages} {193605} (\bibinfo {year}
  {2019})}\BibitemShut {NoStop}%
\bibitem [{\citenamefont {Corman}\ \emph {et~al.}(2019)\citenamefont {Corman},
  \citenamefont {Fabritius}, \citenamefont {H\"ausler}, \citenamefont {Mohan},
  \citenamefont {Dogra}, \citenamefont {Husmann}, \citenamefont {Lebrat},\ and\
  \citenamefont {Esslinger}}]{corman2019quantized}%
  \BibitemOpen
  \bibfield  {author} {\bibinfo {author} {\bibfnamefont {L.}~\bibnamefont
  {Corman}}, \bibinfo {author} {\bibfnamefont {P.}~\bibnamefont {Fabritius}},
  \bibinfo {author} {\bibfnamefont {S.}~\bibnamefont {H\"ausler}}, \bibinfo
  {author} {\bibfnamefont {J.}~\bibnamefont {Mohan}}, \bibinfo {author}
  {\bibfnamefont {L.~H.}\ \bibnamefont {Dogra}}, \bibinfo {author}
  {\bibfnamefont {D.}~\bibnamefont {Husmann}}, \bibinfo {author} {\bibfnamefont
  {M.}~\bibnamefont {Lebrat}},\ and\ \bibinfo {author} {\bibfnamefont
  {T.}~\bibnamefont {Esslinger}},\ }\bibfield  {title} {\bibinfo {title}
  {Quantized conductance through a dissipative atomic point contact},\ }\href
  {https://doi.org/10.1103/PhysRevA.100.053605} {\bibfield  {journal} {\bibinfo
   {journal} {Phys. Rev. A}\ }\textbf {\bibinfo {volume} {100}},\ \bibinfo
  {pages} {053605} (\bibinfo {year} {2019})}\BibitemShut {NoStop}%
\bibitem [{\citenamefont {Huang}\ \emph {et~al.}(2023)\citenamefont {Huang},
  \citenamefont {Mohan}, \citenamefont {Visuri}, \citenamefont {Fabritius},
  \citenamefont {Talebi}, \citenamefont {Wili}, \citenamefont {Uchino},
  \citenamefont {Giamarchi},\ and\ \citenamefont
  {Esslinger}}]{huang2023superfluid}%
  \BibitemOpen
  \bibfield  {author} {\bibinfo {author} {\bibfnamefont {M.-Z.}\ \bibnamefont
  {Huang}}, \bibinfo {author} {\bibfnamefont {J.}~\bibnamefont {Mohan}},
  \bibinfo {author} {\bibfnamefont {A.-M.}\ \bibnamefont {Visuri}}, \bibinfo
  {author} {\bibfnamefont {P.}~\bibnamefont {Fabritius}}, \bibinfo {author}
  {\bibfnamefont {M.}~\bibnamefont {Talebi}}, \bibinfo {author} {\bibfnamefont
  {S.}~\bibnamefont {Wili}}, \bibinfo {author} {\bibfnamefont {S.}~\bibnamefont
  {Uchino}}, \bibinfo {author} {\bibfnamefont {T.}~\bibnamefont {Giamarchi}},\
  and\ \bibinfo {author} {\bibfnamefont {T.}~\bibnamefont {Esslinger}},\
  }\bibfield  {title} {\bibinfo {title} {Superfluid signatures in a dissipative
  quantum point contact},\ }\href
  {https://doi.org/10.1103/PhysRevLett.130.200404} {\bibfield  {journal}
  {\bibinfo  {journal} {Phys. Rev. Lett.}\ }\textbf {\bibinfo {volume} {130}},\
  \bibinfo {pages} {200404} (\bibinfo {year} {2023})}\BibitemShut {NoStop}%
\bibitem [{\citenamefont {Riegger}\ \emph {et~al.}(2018)\citenamefont
  {Riegger}, \citenamefont {Darkwah~Oppong}, \citenamefont {H\"ofer},
  \citenamefont {Fernandes}, \citenamefont {Bloch},\ and\ \citenamefont
  {F\"olling}}]{riegger2018localized}%
  \BibitemOpen
  \bibfield  {author} {\bibinfo {author} {\bibfnamefont {L.}~\bibnamefont
  {Riegger}}, \bibinfo {author} {\bibfnamefont {N.}~\bibnamefont
  {Darkwah~Oppong}}, \bibinfo {author} {\bibfnamefont {M.}~\bibnamefont
  {H\"ofer}}, \bibinfo {author} {\bibfnamefont {D.~R.}\ \bibnamefont
  {Fernandes}}, \bibinfo {author} {\bibfnamefont {I.}~\bibnamefont {Bloch}},\
  and\ \bibinfo {author} {\bibfnamefont {S.}~\bibnamefont {F\"olling}},\
  }\bibfield  {title} {\bibinfo {title} {Localized magnetic moments with
  tunable spin exchange in a gas of ultracold fermions},\ }\href
  {https://doi.org/10.1103/PhysRevLett.120.143601} {\bibfield  {journal}
  {\bibinfo  {journal} {Phys. Rev. Lett.}\ }\textbf {\bibinfo {volume} {120}},\
  \bibinfo {pages} {143601} (\bibinfo {year} {2018})}\BibitemShut {NoStop}%
\bibitem [{\citenamefont {Zhang}\ \emph {et~al.}(2020)\citenamefont {Zhang},
  \citenamefont {Cheng}, \citenamefont {Zhang},\ and\ \citenamefont
  {Zhai}}]{zhang2020controlling}%
  \BibitemOpen
  \bibfield  {author} {\bibinfo {author} {\bibfnamefont {R.}~\bibnamefont
  {Zhang}}, \bibinfo {author} {\bibfnamefont {Y.}~\bibnamefont {Cheng}},
  \bibinfo {author} {\bibfnamefont {P.}~\bibnamefont {Zhang}},\ and\ \bibinfo
  {author} {\bibfnamefont {H.}~\bibnamefont {Zhai}},\ }\bibfield  {title}
  {\bibinfo {title} {Controlling the interaction of ultracold alkaline-earth
  atoms},\ }\href {https://doi.org/10.1038/s42254-020-0157-9} {\bibfield
  {journal} {\bibinfo  {journal} {Nature Reviews Physics}\ }\textbf {\bibinfo
  {volume} {2}},\ \bibinfo {pages} {213} (\bibinfo {year} {2020})}\BibitemShut
  {NoStop}%
\bibitem [{\citenamefont {Gerbier}\ and\ \citenamefont
  {Castin}(2010)}]{gerbier2010heating}%
  \BibitemOpen
  \bibfield  {author} {\bibinfo {author} {\bibfnamefont {F.}~\bibnamefont
  {Gerbier}}\ and\ \bibinfo {author} {\bibfnamefont {Y.}~\bibnamefont
  {Castin}},\ }\bibfield  {title} {\bibinfo {title} {Heating rates for an atom
  in a far-detuned optical lattice},\ }\href
  {https://doi.org/10.1103/PhysRevA.82.013615} {\bibfield  {journal} {\bibinfo
  {journal} {Phys. Rev. A}\ }\textbf {\bibinfo {volume} {82}},\ \bibinfo
  {pages} {013615} (\bibinfo {year} {2010})}\BibitemShut {NoStop}%
\bibitem [{\citenamefont {Bouganne}\ \emph {et~al.}(2020)\citenamefont
  {Bouganne}, \citenamefont {Bosch~Aguilera}, \citenamefont {Ghermaoui},
  \citenamefont {Beugnon},\ and\ \citenamefont {Gerbier}}]{bouganne2020}%
  \BibitemOpen
  \bibfield  {author} {\bibinfo {author} {\bibfnamefont {R.}~\bibnamefont
  {Bouganne}}, \bibinfo {author} {\bibfnamefont {M.}~\bibnamefont
  {Bosch~Aguilera}}, \bibinfo {author} {\bibfnamefont {A.}~\bibnamefont
  {Ghermaoui}}, \bibinfo {author} {\bibfnamefont {J.}~\bibnamefont {Beugnon}},\
  and\ \bibinfo {author} {\bibfnamefont {F.}~\bibnamefont {Gerbier}},\
  }\bibfield  {title} {\bibinfo {title} {Anomalous decay of coherence in a
  dissipative many-body system},\ }\href
  {https://doi.org/10.1038/s41567-019-0678-2} {\bibfield  {journal} {\bibinfo
  {journal} {Nature Physics}\ }\textbf {\bibinfo {volume} {16}},\ \bibinfo
  {pages} {21} (\bibinfo {year} {2020})}\BibitemShut {NoStop}%
\bibitem [{\citenamefont {{Garc{\'i}a-Ripoll}}\ \emph
  {et~al.}(2009)\citenamefont {{Garc{\'i}a-Ripoll}}, \citenamefont {D{\"u}rr},
  \citenamefont {Syassen}, \citenamefont {Bauer}, \citenamefont {Lettner},
  \citenamefont {Rempe},\ and\ \citenamefont {Cirac}}]{garcia-ripoll2009}%
  \BibitemOpen
  \bibfield  {author} {\bibinfo {author} {\bibfnamefont {J.~J.}\ \bibnamefont
  {{Garc{\'i}a-Ripoll}}}, \bibinfo {author} {\bibfnamefont {S.}~\bibnamefont
  {D{\"u}rr}}, \bibinfo {author} {\bibfnamefont {N.}~\bibnamefont {Syassen}},
  \bibinfo {author} {\bibfnamefont {D.~M.}\ \bibnamefont {Bauer}}, \bibinfo
  {author} {\bibfnamefont {M.}~\bibnamefont {Lettner}}, \bibinfo {author}
  {\bibfnamefont {G.}~\bibnamefont {Rempe}},\ and\ \bibinfo {author}
  {\bibfnamefont {J.~I.}\ \bibnamefont {Cirac}},\ }\bibfield  {title} {\bibinfo
  {title} {Dissipation-induced hard-core boson gas in an optical lattice},\
  }\href@noop {} {\bibfield  {journal} {\bibinfo  {journal} {New Journal of
  Physics}\ }\textbf {\bibinfo {volume} {11}},\ \bibinfo {pages} {013053}
  (\bibinfo {year} {2009})}\BibitemShut {NoStop}%
\bibitem [{\citenamefont {Tomita}\ \emph {et~al.}(2017)\citenamefont {Tomita},
  \citenamefont {Nakajima}, \citenamefont {Danshita}, \citenamefont {Takasu},\
  and\ \citenamefont {Takahashi}}]{TomitaEtAlScienceAdv17}%
  \BibitemOpen
  \bibfield  {author} {\bibinfo {author} {\bibfnamefont {T.}~\bibnamefont
  {Tomita}}, \bibinfo {author} {\bibfnamefont {S.}~\bibnamefont {Nakajima}},
  \bibinfo {author} {\bibfnamefont {I.}~\bibnamefont {Danshita}}, \bibinfo
  {author} {\bibfnamefont {Y.}~\bibnamefont {Takasu}},\ and\ \bibinfo {author}
  {\bibfnamefont {Y.}~\bibnamefont {Takahashi}},\ }\bibfield  {title} {\bibinfo
  {title} {Observation of the mott insulator to superfluid crossover of a
  driven-dissipative bose-hubbard system},\ }\bibfield  {journal} {\bibinfo
  {journal} {Science Advances}\ }\textbf {\bibinfo {volume} {3}},\ \href
  {https://doi.org/10.1126/sciadv.1701513} {10.1126/sciadv.1701513} (\bibinfo
  {year} {2017})\BibitemShut {NoStop}%
\bibitem [{\citenamefont {Honda}\ \emph {et~al.}(2022)\citenamefont {Honda},
  \citenamefont {Taie}, \citenamefont {Takasu}, \citenamefont {Nishizawa},
  \citenamefont {Nakagawa},\ and\ \citenamefont
  {Takahashi}}]{honda2022observation}%
  \BibitemOpen
  \bibfield  {author} {\bibinfo {author} {\bibfnamefont {K.}~\bibnamefont
  {Honda}}, \bibinfo {author} {\bibfnamefont {S.}~\bibnamefont {Taie}},
  \bibinfo {author} {\bibfnamefont {Y.}~\bibnamefont {Takasu}}, \bibinfo
  {author} {\bibfnamefont {N.}~\bibnamefont {Nishizawa}}, \bibinfo {author}
  {\bibfnamefont {M.}~\bibnamefont {Nakagawa}},\ and\ \bibinfo {author}
  {\bibfnamefont {Y.}~\bibnamefont {Takahashi}},\ }\href
  {https://doi.org/10.48550/ARXIV.2205.13162} {\bibinfo {title} {Observation of
  the sign reversal of the magnetic correlation in a driven-dissipative
  fermi-hubbard system}} (\bibinfo {year} {2022})\BibitemShut {NoStop}%
\bibitem [{\citenamefont {Avinun-Kalish}\ \emph {et~al.}(2004)\citenamefont
  {Avinun-Kalish}, \citenamefont {Heiblum}, \citenamefont {Silva},
  \citenamefont {Mahalu},\ and\ \citenamefont
  {Umansky}}]{avinun2004controlled}%
  \BibitemOpen
  \bibfield  {author} {\bibinfo {author} {\bibfnamefont {M.}~\bibnamefont
  {Avinun-Kalish}}, \bibinfo {author} {\bibfnamefont {M.}~\bibnamefont
  {Heiblum}}, \bibinfo {author} {\bibfnamefont {A.}~\bibnamefont {Silva}},
  \bibinfo {author} {\bibfnamefont {D.}~\bibnamefont {Mahalu}},\ and\ \bibinfo
  {author} {\bibfnamefont {V.}~\bibnamefont {Umansky}},\ }\bibfield  {title}
  {\bibinfo {title} {Controlled dephasing of a quantum dot in the kondo
  regime},\ }\href {https://doi.org/10.1103/PhysRevLett.92.156801} {\bibfield
  {journal} {\bibinfo  {journal} {Phys. Rev. Lett.}\ }\textbf {\bibinfo
  {volume} {92}},\ \bibinfo {pages} {156801} (\bibinfo {year}
  {2004})}\BibitemShut {NoStop}%
\bibitem [{\citenamefont {Kang}\ and\ \citenamefont
  {Khym}(2007)}]{kang2007entanglement}%
  \BibitemOpen
  \bibfield  {author} {\bibinfo {author} {\bibfnamefont {K.}~\bibnamefont
  {Kang}}\ and\ \bibinfo {author} {\bibfnamefont {G.~L.}\ \bibnamefont
  {Khym}},\ }\bibfield  {title} {\bibinfo {title} {Entanglement, measurement,
  and conditional evolution of the kondo singlet interacting with a mesoscopic
  detector},\ }\href {https://doi.org/10.1088/1367-2630/9/5/121} {\bibfield
  {journal} {\bibinfo  {journal} {New Journal of Physics}\ }\textbf {\bibinfo
  {volume} {9}},\ \bibinfo {pages} {121} (\bibinfo {year} {2007})}\BibitemShut
  {NoStop}%
\bibitem [{\citenamefont {Aono}(2008)}]{aono2008dephasing}%
  \BibitemOpen
  \bibfield  {author} {\bibinfo {author} {\bibfnamefont {T.}~\bibnamefont
  {Aono}},\ }\bibfield  {title} {\bibinfo {title} {Dephasing in a quantum dot
  coupled to a quantum point contact},\ }\href
  {https://doi.org/10.1103/PhysRevB.77.081303} {\bibfield  {journal} {\bibinfo
  {journal} {Phys. Rev. B}\ }\textbf {\bibinfo {volume} {77}},\ \bibinfo
  {pages} {081303} (\bibinfo {year} {2008})}\BibitemShut {NoStop}%
\bibitem [{\citenamefont {Sukhorukov}\ \emph {et~al.}(2007)\citenamefont
  {Sukhorukov}, \citenamefont {Jordan}, \citenamefont {Gustavsson},
  \citenamefont {Leturcq}, \citenamefont {Ihn},\ and\ \citenamefont
  {Ensslin}}]{sukhorukov2007conditional}%
  \BibitemOpen
  \bibfield  {author} {\bibinfo {author} {\bibfnamefont {E.~V.}\ \bibnamefont
  {Sukhorukov}}, \bibinfo {author} {\bibfnamefont {A.~N.}\ \bibnamefont
  {Jordan}}, \bibinfo {author} {\bibfnamefont {S.}~\bibnamefont {Gustavsson}},
  \bibinfo {author} {\bibfnamefont {R.}~\bibnamefont {Leturcq}}, \bibinfo
  {author} {\bibfnamefont {T.}~\bibnamefont {Ihn}},\ and\ \bibinfo {author}
  {\bibfnamefont {K.}~\bibnamefont {Ensslin}},\ }\bibfield  {title} {\bibinfo
  {title} {Conditional statistics of electron transport in interacting
  nanoscale conductors},\ }\href {https://doi.org/10.1038/nphys564} {\bibfield
  {journal} {\bibinfo  {journal} {Nature Physics}\ }\textbf {\bibinfo {volume}
  {3}},\ \bibinfo {pages} {243} (\bibinfo {year} {2007})}\BibitemShut {NoStop}%
\bibitem [{\citenamefont {Ferguson}\ \emph {et~al.}(2023)\citenamefont
  {Ferguson}, \citenamefont {Camenzind}, \citenamefont {M\"uller},
  \citenamefont {Biesinger}, \citenamefont {Scheller}, \citenamefont
  {Braunecker}, \citenamefont {Zumb\"uhl},\ and\ \citenamefont
  {Zilberberg}}]{ferguson2023measuremnt}%
  \BibitemOpen
  \bibfield  {author} {\bibinfo {author} {\bibfnamefont {M.~S.}\ \bibnamefont
  {Ferguson}}, \bibinfo {author} {\bibfnamefont {L.~C.}\ \bibnamefont
  {Camenzind}}, \bibinfo {author} {\bibfnamefont {C.}~\bibnamefont {M\"uller}},
  \bibinfo {author} {\bibfnamefont {D.~E.~F.}\ \bibnamefont {Biesinger}},
  \bibinfo {author} {\bibfnamefont {C.~P.}\ \bibnamefont {Scheller}}, \bibinfo
  {author} {\bibfnamefont {B.}~\bibnamefont {Braunecker}}, \bibinfo {author}
  {\bibfnamefont {D.~M.}\ \bibnamefont {Zumb\"uhl}},\ and\ \bibinfo {author}
  {\bibfnamefont {O.}~\bibnamefont {Zilberberg}},\ }\bibfield  {title}
  {\bibinfo {title} {Measurement-induced population switching},\ }\href
  {https://doi.org/10.1103/PhysRevResearch.5.023028} {\bibfield  {journal}
  {\bibinfo  {journal} {Phys. Rev. Res.}\ }\textbf {\bibinfo {volume} {5}},\
  \bibinfo {pages} {023028} (\bibinfo {year} {2023})}\BibitemShut {NoStop}%
\bibitem [{\citenamefont {Hasegawa}\ \emph {et~al.}(2021)\citenamefont
  {Hasegawa}, \citenamefont {Nakagawa},\ and\ \citenamefont
  {Saito}}]{hasegawa2021kondo}%
  \BibitemOpen
  \bibfield  {author} {\bibinfo {author} {\bibfnamefont {M.}~\bibnamefont
  {Hasegawa}}, \bibinfo {author} {\bibfnamefont {M.}~\bibnamefont {Nakagawa}},\
  and\ \bibinfo {author} {\bibfnamefont {K.}~\bibnamefont {Saito}},\
  }\href@noop {} {\bibinfo {title} {Kondo effect in a quantum dot under
  continuous quantum measurement}} (\bibinfo {year} {2021}),\ \Eprint
  {https://arxiv.org/abs/2111.07771} {arXiv:2111.07771 [cond-mat.mes-hall]}
  \BibitemShut {NoStop}%
\bibitem [{\citenamefont {Mi}\ \emph {et~al.}(2023)\citenamefont {Mi},
  \citenamefont {Michailidis}, \citenamefont {Shabani}, \citenamefont {Miao},
  \citenamefont {Klimov}, \citenamefont {Lloyd}, \citenamefont {Rosenberg},
  \citenamefont {Acharya}, \citenamefont {Aleiner}, \citenamefont {Andersen},
  \citenamefont {Ansmann}, \citenamefont {Arute}, \citenamefont {Arya},
  \citenamefont {Asfaw}, \citenamefont {Atalaya}, \citenamefont {Bardin},
  \citenamefont {Bengtsson}, \citenamefont {Bortoli}, \citenamefont {Bourassa},
  \citenamefont {Bovaird}, \citenamefont {Brill}, \citenamefont {Broughton},
  \citenamefont {Buckley}, \citenamefont {Buell}, \citenamefont {Burger},
  \citenamefont {Burkett}, \citenamefont {Bushnell}, \citenamefont {Chen},
  \citenamefont {Chiaro}, \citenamefont {Chik}, \citenamefont {Chou},
  \citenamefont {Cogan}, \citenamefont {Collins}, \citenamefont {Conner},
  \citenamefont {Courtney}, \citenamefont {Crook}, \citenamefont {Curtin},
  \citenamefont {Dau}, \citenamefont {Debroy}, \citenamefont {Barba},
  \citenamefont {Demura}, \citenamefont {Paolo}, \citenamefont {Drozdov},
  \citenamefont {Dunsworth}, \citenamefont {Erickson}, \citenamefont {Faoro},
  \citenamefont {Farhi}, \citenamefont {Fatemi}, \citenamefont {Ferreira},
  \citenamefont {Forati}, \citenamefont {Fowler}, \citenamefont {Foxen},
  \citenamefont {Genois}, \citenamefont {Giang}, \citenamefont {Gidney},
  \citenamefont {Gilboa}, \citenamefont {Giustina}, \citenamefont {Gosula},
  \citenamefont {Gross}, \citenamefont {Habegger}, \citenamefont {Hamilton},
  \citenamefont {Hansen}, \citenamefont {Harrigan}, \citenamefont {Harrington},
  \citenamefont {Heu}, \citenamefont {Hoffmann}, \citenamefont {Hong},
  \citenamefont {Huang}, \citenamefont {Huff}, \citenamefont {Huggins},
  \citenamefont {Ioffe}, \citenamefont {Isakov}, \citenamefont {Iveland},
  \citenamefont {Jeffrey}, \citenamefont {Jiang}, \citenamefont {Jones},
  \citenamefont {Juhas}, \citenamefont {Kafri}, \citenamefont {Kechedzhi},
  \citenamefont {Khattar}, \citenamefont {Khezri}, \citenamefont {Kieferova},
  \citenamefont {Kim}, \citenamefont {Kitaev}, \citenamefont {Klots},
  \citenamefont {Korotkov}, \citenamefont {Kostritsa}, \citenamefont
  {Kreikebaum}, \citenamefont {Landhuis}, \citenamefont {Laptev}, \citenamefont
  {Lau}, \citenamefont {Laws}, \citenamefont {Lee}, \citenamefont {Lee},
  \citenamefont {Lensky}, \citenamefont {Lester}, \citenamefont {Lill},
  \citenamefont {Liu}, \citenamefont {Locharla}, \citenamefont {Malone},
  \citenamefont {Martin}, \citenamefont {McClean}, \citenamefont {McEwen},
  \citenamefont {Mieszala}, \citenamefont {Montazeri}, \citenamefont {Morvan},
  \citenamefont {Movassagh}, \citenamefont {Mruczkiewicz}, \citenamefont
  {Neeley}, \citenamefont {Neill}, \citenamefont {Nersisyan}, \citenamefont
  {Newman}, \citenamefont {Ng}, \citenamefont {Nguyen}, \citenamefont {Nguyen},
  \citenamefont {Niu}, \citenamefont {OBrien}, \citenamefont {Opremcak},
  \citenamefont {Petukhov}, \citenamefont {Potter}, \citenamefont {Pryadko},
  \citenamefont {Quintana}, \citenamefont {Rocque}, \citenamefont {Rubin},
  \citenamefont {Saei}, \citenamefont {Sank}, \citenamefont {Sankaragomathi},
  \citenamefont {Satzinger}, \citenamefont {Schurkus}, \citenamefont
  {Schuster}, \citenamefont {Shearn}, \citenamefont {Shorter}, \citenamefont
  {Shutty}, \citenamefont {Shvarts}, \citenamefont {Skruzny}, \citenamefont
  {Smith}, \citenamefont {Somma}, \citenamefont {Sterling}, \citenamefont
  {Strain}, \citenamefont {Szalay}, \citenamefont {Torres}, \citenamefont
  {Vidal}, \citenamefont {Villalonga}, \citenamefont {Heidweiller},
  \citenamefont {White}, \citenamefont {Woo}, \citenamefont {Xing},
  \citenamefont {Yao}, \citenamefont {Yeh}, \citenamefont {Yoo}, \citenamefont
  {Young}, \citenamefont {Zalcman}, \citenamefont {Zhang}, \citenamefont {Zhu},
  \citenamefont {Zobrist}, \citenamefont {Neven}, \citenamefont {Babbush},
  \citenamefont {Bacon}, \citenamefont {Boixo}, \citenamefont {Hilton},
  \citenamefont {Lucero}, \citenamefont {Megrant}, \citenamefont {Kelly},
  \citenamefont {Chen}, \citenamefont {Roushan}, \citenamefont {Smelyanskiy},\
  and\ \citenamefont {Abanin}}]{mi2023stable}%
  \BibitemOpen
  \bibfield  {author} {\bibinfo {author} {\bibfnamefont {X.}~\bibnamefont
  {Mi}}, \bibinfo {author} {\bibfnamefont {A.~A.}\ \bibnamefont {Michailidis}},
  \bibinfo {author} {\bibfnamefont {S.}~\bibnamefont {Shabani}}, \bibinfo
  {author} {\bibfnamefont {K.~C.}\ \bibnamefont {Miao}}, \bibinfo {author}
  {\bibfnamefont {P.~V.}\ \bibnamefont {Klimov}}, \bibinfo {author}
  {\bibfnamefont {J.}~\bibnamefont {Lloyd}}, \bibinfo {author} {\bibfnamefont
  {E.}~\bibnamefont {Rosenberg}}, \bibinfo {author} {\bibfnamefont
  {R.}~\bibnamefont {Acharya}}, \bibinfo {author} {\bibfnamefont
  {I.}~\bibnamefont {Aleiner}}, \bibinfo {author} {\bibfnamefont {T.~I.}\
  \bibnamefont {Andersen}}, \bibinfo {author} {\bibfnamefont {M.}~\bibnamefont
  {Ansmann}}, \bibinfo {author} {\bibfnamefont {F.}~\bibnamefont {Arute}},
  \bibinfo {author} {\bibfnamefont {K.}~\bibnamefont {Arya}}, \bibinfo {author}
  {\bibfnamefont {A.}~\bibnamefont {Asfaw}}, \bibinfo {author} {\bibfnamefont
  {J.}~\bibnamefont {Atalaya}}, \bibinfo {author} {\bibfnamefont {J.~C.}\
  \bibnamefont {Bardin}}, \bibinfo {author} {\bibfnamefont {A.}~\bibnamefont
  {Bengtsson}}, \bibinfo {author} {\bibfnamefont {G.}~\bibnamefont {Bortoli}},
  \bibinfo {author} {\bibfnamefont {A.}~\bibnamefont {Bourassa}}, \bibinfo
  {author} {\bibfnamefont {J.}~\bibnamefont {Bovaird}}, \bibinfo {author}
  {\bibfnamefont {L.}~\bibnamefont {Brill}}, \bibinfo {author} {\bibfnamefont
  {M.}~\bibnamefont {Broughton}}, \bibinfo {author} {\bibfnamefont {B.~B.}\
  \bibnamefont {Buckley}}, \bibinfo {author} {\bibfnamefont {D.~A.}\
  \bibnamefont {Buell}}, \bibinfo {author} {\bibfnamefont {T.}~\bibnamefont
  {Burger}}, \bibinfo {author} {\bibfnamefont {B.}~\bibnamefont {Burkett}},
  \bibinfo {author} {\bibfnamefont {N.}~\bibnamefont {Bushnell}}, \bibinfo
  {author} {\bibfnamefont {Z.}~\bibnamefont {Chen}}, \bibinfo {author}
  {\bibfnamefont {B.}~\bibnamefont {Chiaro}}, \bibinfo {author} {\bibfnamefont
  {D.}~\bibnamefont {Chik}}, \bibinfo {author} {\bibfnamefont {C.}~\bibnamefont
  {Chou}}, \bibinfo {author} {\bibfnamefont {J.}~\bibnamefont {Cogan}},
  \bibinfo {author} {\bibfnamefont {R.}~\bibnamefont {Collins}}, \bibinfo
  {author} {\bibfnamefont {P.}~\bibnamefont {Conner}}, \bibinfo {author}
  {\bibfnamefont {W.}~\bibnamefont {Courtney}}, \bibinfo {author}
  {\bibfnamefont {A.~L.}\ \bibnamefont {Crook}}, \bibinfo {author}
  {\bibfnamefont {B.}~\bibnamefont {Curtin}}, \bibinfo {author} {\bibfnamefont
  {A.~G.}\ \bibnamefont {Dau}}, \bibinfo {author} {\bibfnamefont {D.~M.}\
  \bibnamefont {Debroy}}, \bibinfo {author} {\bibfnamefont {A.~D.~T.}\
  \bibnamefont {Barba}}, \bibinfo {author} {\bibfnamefont {S.}~\bibnamefont
  {Demura}}, \bibinfo {author} {\bibfnamefont {A.~D.}\ \bibnamefont {Paolo}},
  \bibinfo {author} {\bibfnamefont {I.~K.}\ \bibnamefont {Drozdov}}, \bibinfo
  {author} {\bibfnamefont {A.}~\bibnamefont {Dunsworth}}, \bibinfo {author}
  {\bibfnamefont {C.}~\bibnamefont {Erickson}}, \bibinfo {author}
  {\bibfnamefont {L.}~\bibnamefont {Faoro}}, \bibinfo {author} {\bibfnamefont
  {E.}~\bibnamefont {Farhi}}, \bibinfo {author} {\bibfnamefont
  {R.}~\bibnamefont {Fatemi}}, \bibinfo {author} {\bibfnamefont {V.~S.}\
  \bibnamefont {Ferreira}}, \bibinfo {author} {\bibfnamefont {L.~F. B.~E.}\
  \bibnamefont {Forati}}, \bibinfo {author} {\bibfnamefont {A.~G.}\
  \bibnamefont {Fowler}}, \bibinfo {author} {\bibfnamefont {B.}~\bibnamefont
  {Foxen}}, \bibinfo {author} {\bibfnamefont {E.}~\bibnamefont {Genois}},
  \bibinfo {author} {\bibfnamefont {W.}~\bibnamefont {Giang}}, \bibinfo
  {author} {\bibfnamefont {C.}~\bibnamefont {Gidney}}, \bibinfo {author}
  {\bibfnamefont {D.}~\bibnamefont {Gilboa}}, \bibinfo {author} {\bibfnamefont
  {M.}~\bibnamefont {Giustina}}, \bibinfo {author} {\bibfnamefont
  {R.}~\bibnamefont {Gosula}}, \bibinfo {author} {\bibfnamefont {J.~A.}\
  \bibnamefont {Gross}}, \bibinfo {author} {\bibfnamefont {S.}~\bibnamefont
  {Habegger}}, \bibinfo {author} {\bibfnamefont {M.~C.}\ \bibnamefont
  {Hamilton}}, \bibinfo {author} {\bibfnamefont {M.}~\bibnamefont {Hansen}},
  \bibinfo {author} {\bibfnamefont {M.~P.}\ \bibnamefont {Harrigan}}, \bibinfo
  {author} {\bibfnamefont {S.~D.}\ \bibnamefont {Harrington}}, \bibinfo
  {author} {\bibfnamefont {P.}~\bibnamefont {Heu}}, \bibinfo {author}
  {\bibfnamefont {M.~R.}\ \bibnamefont {Hoffmann}}, \bibinfo {author}
  {\bibfnamefont {S.}~\bibnamefont {Hong}}, \bibinfo {author} {\bibfnamefont
  {T.}~\bibnamefont {Huang}}, \bibinfo {author} {\bibfnamefont
  {A.}~\bibnamefont {Huff}}, \bibinfo {author} {\bibfnamefont {W.~J.}\
  \bibnamefont {Huggins}}, \bibinfo {author} {\bibfnamefont {L.~B.}\
  \bibnamefont {Ioffe}}, \bibinfo {author} {\bibfnamefont {S.~V.}\ \bibnamefont
  {Isakov}}, \bibinfo {author} {\bibfnamefont {J.}~\bibnamefont {Iveland}},
  \bibinfo {author} {\bibfnamefont {E.}~\bibnamefont {Jeffrey}}, \bibinfo
  {author} {\bibfnamefont {Z.}~\bibnamefont {Jiang}}, \bibinfo {author}
  {\bibfnamefont {C.}~\bibnamefont {Jones}}, \bibinfo {author} {\bibfnamefont
  {P.}~\bibnamefont {Juhas}}, \bibinfo {author} {\bibfnamefont
  {D.}~\bibnamefont {Kafri}}, \bibinfo {author} {\bibfnamefont
  {K.}~\bibnamefont {Kechedzhi}}, \bibinfo {author} {\bibfnamefont
  {T.}~\bibnamefont {Khattar}}, \bibinfo {author} {\bibfnamefont
  {M.}~\bibnamefont {Khezri}}, \bibinfo {author} {\bibfnamefont
  {M.}~\bibnamefont {Kieferova}}, \bibinfo {author} {\bibfnamefont
  {S.}~\bibnamefont {Kim}}, \bibinfo {author} {\bibfnamefont {A.}~\bibnamefont
  {Kitaev}}, \bibinfo {author} {\bibfnamefont {A.~R.}\ \bibnamefont {Klots}},
  \bibinfo {author} {\bibfnamefont {A.~N.}\ \bibnamefont {Korotkov}}, \bibinfo
  {author} {\bibfnamefont {F.}~\bibnamefont {Kostritsa}}, \bibinfo {author}
  {\bibfnamefont {J.~M.}\ \bibnamefont {Kreikebaum}}, \bibinfo {author}
  {\bibfnamefont {D.}~\bibnamefont {Landhuis}}, \bibinfo {author}
  {\bibfnamefont {P.}~\bibnamefont {Laptev}}, \bibinfo {author} {\bibfnamefont
  {K.~M.}\ \bibnamefont {Lau}}, \bibinfo {author} {\bibfnamefont
  {L.}~\bibnamefont {Laws}}, \bibinfo {author} {\bibfnamefont {J.}~\bibnamefont
  {Lee}}, \bibinfo {author} {\bibfnamefont {K.~W.}\ \bibnamefont {Lee}},
  \bibinfo {author} {\bibfnamefont {Y.~D.}\ \bibnamefont {Lensky}}, \bibinfo
  {author} {\bibfnamefont {B.~J.}\ \bibnamefont {Lester}}, \bibinfo {author}
  {\bibfnamefont {A.~T.}\ \bibnamefont {Lill}}, \bibinfo {author}
  {\bibfnamefont {W.}~\bibnamefont {Liu}}, \bibinfo {author} {\bibfnamefont
  {A.}~\bibnamefont {Locharla}}, \bibinfo {author} {\bibfnamefont {F.~D.}\
  \bibnamefont {Malone}}, \bibinfo {author} {\bibfnamefont {O.}~\bibnamefont
  {Martin}}, \bibinfo {author} {\bibfnamefont {J.~R.}\ \bibnamefont {McClean}},
  \bibinfo {author} {\bibfnamefont {M.}~\bibnamefont {McEwen}}, \bibinfo
  {author} {\bibfnamefont {A.}~\bibnamefont {Mieszala}}, \bibinfo {author}
  {\bibfnamefont {S.}~\bibnamefont {Montazeri}}, \bibinfo {author}
  {\bibfnamefont {A.}~\bibnamefont {Morvan}}, \bibinfo {author} {\bibfnamefont
  {R.}~\bibnamefont {Movassagh}}, \bibinfo {author} {\bibfnamefont
  {W.}~\bibnamefont {Mruczkiewicz}}, \bibinfo {author} {\bibfnamefont
  {M.}~\bibnamefont {Neeley}}, \bibinfo {author} {\bibfnamefont
  {C.}~\bibnamefont {Neill}}, \bibinfo {author} {\bibfnamefont
  {A.}~\bibnamefont {Nersisyan}}, \bibinfo {author} {\bibfnamefont
  {M.}~\bibnamefont {Newman}}, \bibinfo {author} {\bibfnamefont {J.~H.}\
  \bibnamefont {Ng}}, \bibinfo {author} {\bibfnamefont {A.}~\bibnamefont
  {Nguyen}}, \bibinfo {author} {\bibfnamefont {M.}~\bibnamefont {Nguyen}},
  \bibinfo {author} {\bibfnamefont {M.~Y.}\ \bibnamefont {Niu}}, \bibinfo
  {author} {\bibfnamefont {T.~E.}\ \bibnamefont {OBrien}}, \bibinfo {author}
  {\bibfnamefont {A.}~\bibnamefont {Opremcak}}, \bibinfo {author}
  {\bibfnamefont {A.}~\bibnamefont {Petukhov}}, \bibinfo {author}
  {\bibfnamefont {R.}~\bibnamefont {Potter}}, \bibinfo {author} {\bibfnamefont
  {L.~P.}\ \bibnamefont {Pryadko}}, \bibinfo {author} {\bibfnamefont
  {C.}~\bibnamefont {Quintana}}, \bibinfo {author} {\bibfnamefont
  {C.}~\bibnamefont {Rocque}}, \bibinfo {author} {\bibfnamefont {N.~C.}\
  \bibnamefont {Rubin}}, \bibinfo {author} {\bibfnamefont {N.}~\bibnamefont
  {Saei}}, \bibinfo {author} {\bibfnamefont {D.}~\bibnamefont {Sank}}, \bibinfo
  {author} {\bibfnamefont {K.}~\bibnamefont {Sankaragomathi}}, \bibinfo
  {author} {\bibfnamefont {K.~J.}\ \bibnamefont {Satzinger}}, \bibinfo {author}
  {\bibfnamefont {H.~F.}\ \bibnamefont {Schurkus}}, \bibinfo {author}
  {\bibfnamefont {C.}~\bibnamefont {Schuster}}, \bibinfo {author}
  {\bibfnamefont {M.~J.}\ \bibnamefont {Shearn}}, \bibinfo {author}
  {\bibfnamefont {A.}~\bibnamefont {Shorter}}, \bibinfo {author} {\bibfnamefont
  {N.}~\bibnamefont {Shutty}}, \bibinfo {author} {\bibfnamefont
  {V.}~\bibnamefont {Shvarts}}, \bibinfo {author} {\bibfnamefont
  {J.}~\bibnamefont {Skruzny}}, \bibinfo {author} {\bibfnamefont {W.~C.}\
  \bibnamefont {Smith}}, \bibinfo {author} {\bibfnamefont {R.}~\bibnamefont
  {Somma}}, \bibinfo {author} {\bibfnamefont {G.}~\bibnamefont {Sterling}},
  \bibinfo {author} {\bibfnamefont {D.}~\bibnamefont {Strain}}, \bibinfo
  {author} {\bibfnamefont {M.}~\bibnamefont {Szalay}}, \bibinfo {author}
  {\bibfnamefont {A.}~\bibnamefont {Torres}}, \bibinfo {author} {\bibfnamefont
  {G.}~\bibnamefont {Vidal}}, \bibinfo {author} {\bibfnamefont
  {B.}~\bibnamefont {Villalonga}}, \bibinfo {author} {\bibfnamefont {C.~V.}\
  \bibnamefont {Heidweiller}}, \bibinfo {author} {\bibfnamefont
  {T.}~\bibnamefont {White}}, \bibinfo {author} {\bibfnamefont {B.~W.~K.}\
  \bibnamefont {Woo}}, \bibinfo {author} {\bibfnamefont {C.}~\bibnamefont
  {Xing}}, \bibinfo {author} {\bibfnamefont {Z.~J.}\ \bibnamefont {Yao}},
  \bibinfo {author} {\bibfnamefont {P.}~\bibnamefont {Yeh}}, \bibinfo {author}
  {\bibfnamefont {J.}~\bibnamefont {Yoo}}, \bibinfo {author} {\bibfnamefont
  {G.}~\bibnamefont {Young}}, \bibinfo {author} {\bibfnamefont
  {A.}~\bibnamefont {Zalcman}}, \bibinfo {author} {\bibfnamefont
  {Y.}~\bibnamefont {Zhang}}, \bibinfo {author} {\bibfnamefont
  {N.}~\bibnamefont {Zhu}}, \bibinfo {author} {\bibfnamefont {N.}~\bibnamefont
  {Zobrist}}, \bibinfo {author} {\bibfnamefont {H.}~\bibnamefont {Neven}},
  \bibinfo {author} {\bibfnamefont {R.}~\bibnamefont {Babbush}}, \bibinfo
  {author} {\bibfnamefont {D.}~\bibnamefont {Bacon}}, \bibinfo {author}
  {\bibfnamefont {S.}~\bibnamefont {Boixo}}, \bibinfo {author} {\bibfnamefont
  {J.}~\bibnamefont {Hilton}}, \bibinfo {author} {\bibfnamefont
  {E.}~\bibnamefont {Lucero}}, \bibinfo {author} {\bibfnamefont
  {A.}~\bibnamefont {Megrant}}, \bibinfo {author} {\bibfnamefont
  {J.}~\bibnamefont {Kelly}}, \bibinfo {author} {\bibfnamefont
  {Y.}~\bibnamefont {Chen}}, \bibinfo {author} {\bibfnamefont {P.}~\bibnamefont
  {Roushan}}, \bibinfo {author} {\bibfnamefont {V.}~\bibnamefont
  {Smelyanskiy}},\ and\ \bibinfo {author} {\bibfnamefont {D.~A.}\ \bibnamefont
  {Abanin}},\ }\href@noop {} {\bibinfo {title} {Stable quantum-correlated many
  body states via engineered dissipation}} (\bibinfo {year} {2023}),\ \Eprint
  {https://arxiv.org/abs/2304.13878} {arXiv:2304.13878 [quant-ph]} \BibitemShut
  {NoStop}%
\bibitem [{\citenamefont {Fr{\"o}ml}\ \emph {et~al.}(2019)\citenamefont
  {Fr{\"o}ml}, \citenamefont {Chiocchetta}, \citenamefont {Kollath},\ and\
  \citenamefont {Diehl}}]{Froml2019}%
  \BibitemOpen
  \bibfield  {author} {\bibinfo {author} {\bibfnamefont {H.}~\bibnamefont
  {Fr{\"o}ml}}, \bibinfo {author} {\bibfnamefont {A.}~\bibnamefont
  {Chiocchetta}}, \bibinfo {author} {\bibfnamefont {C.}~\bibnamefont
  {Kollath}},\ and\ \bibinfo {author} {\bibfnamefont {S.}~\bibnamefont
  {Diehl}},\ }\bibfield  {title} {\bibinfo {title} {Fluctuation-{{Induced
  Quantum Zeno Effect}}},\ }\href
  {https://doi.org/10.1103/PhysRevLett.122.040402} {\bibfield  {journal}
  {\bibinfo  {journal} {Physical Review Letters}\ }\textbf {\bibinfo {volume}
  {122}},\ \bibinfo {pages} {040402} (\bibinfo {year} {2019})}\BibitemShut
  {NoStop}%
\bibitem [{\citenamefont {Damanet}\ \emph {et~al.}(2019)\citenamefont
  {Damanet}, \citenamefont {Mascarenhas}, \citenamefont {Pekker},\ and\
  \citenamefont {Daley}}]{damanet2019controlling}%
  \BibitemOpen
  \bibfield  {author} {\bibinfo {author} {\bibfnamefont {F.~m.~c.}\
  \bibnamefont {Damanet}}, \bibinfo {author} {\bibfnamefont {E.}~\bibnamefont
  {Mascarenhas}}, \bibinfo {author} {\bibfnamefont {D.}~\bibnamefont
  {Pekker}},\ and\ \bibinfo {author} {\bibfnamefont {A.~J.}\ \bibnamefont
  {Daley}},\ }\bibfield  {title} {\bibinfo {title} {Controlling quantum
  transport via dissipation engineering},\ }\href
  {https://doi.org/10.1103/PhysRevLett.123.180402} {\bibfield  {journal}
  {\bibinfo  {journal} {Phys. Rev. Lett.}\ }\textbf {\bibinfo {volume} {123}},\
  \bibinfo {pages} {180402} (\bibinfo {year} {2019})}\BibitemShut {NoStop}%
\bibitem [{\citenamefont {Visuri}\ \emph {et~al.}(2022)\citenamefont {Visuri},
  \citenamefont {Giamarchi},\ and\ \citenamefont
  {Kollath}}]{visuri2022symmetry}%
  \BibitemOpen
  \bibfield  {author} {\bibinfo {author} {\bibfnamefont {A.-M.}\ \bibnamefont
  {Visuri}}, \bibinfo {author} {\bibfnamefont {T.}~\bibnamefont {Giamarchi}},\
  and\ \bibinfo {author} {\bibfnamefont {C.}~\bibnamefont {Kollath}},\
  }\bibfield  {title} {\bibinfo {title} {Symmetry-protected transport through a
  lattice with a local particle loss},\ }\href
  {https://doi.org/10.1103/PhysRevLett.129.056802} {\bibfield  {journal}
  {\bibinfo  {journal} {Phys. Rev. Lett.}\ }\textbf {\bibinfo {volume} {129}},\
  \bibinfo {pages} {056802} (\bibinfo {year} {2022})}\BibitemShut {NoStop}%
\bibitem [{\citenamefont {Visuri}\ \emph {et~al.}(2023)\citenamefont {Visuri},
  \citenamefont {Giamarchi},\ and\ \citenamefont
  {Kollath}}]{visuri2023nonlinear}%
  \BibitemOpen
  \bibfield  {author} {\bibinfo {author} {\bibfnamefont {A.-M.}\ \bibnamefont
  {Visuri}}, \bibinfo {author} {\bibfnamefont {T.}~\bibnamefont {Giamarchi}},\
  and\ \bibinfo {author} {\bibfnamefont {C.}~\bibnamefont {Kollath}},\
  }\bibfield  {title} {\bibinfo {title} {Nonlinear transport in the presence of
  a local dissipation},\ }\href
  {https://doi.org/10.1103/PhysRevResearch.5.013195} {\bibfield  {journal}
  {\bibinfo  {journal} {Phys. Rev. Res.}\ }\textbf {\bibinfo {volume} {5}},\
  \bibinfo {pages} {013195} (\bibinfo {year} {2023})}\BibitemShut {NoStop}%
\bibitem [{\citenamefont {{Visuri}}\ \emph {et~al.}(2023)\citenamefont
  {{Visuri}}, \citenamefont {{Mohan}}, \citenamefont {{Uchino}}, \citenamefont
  {{Huang}}, \citenamefont {{Esslinger}},\ and\ \citenamefont
  {{Giamarchi}}}]{visuri2023dc}%
  \BibitemOpen
  \bibfield  {author} {\bibinfo {author} {\bibfnamefont {A.-M.}\ \bibnamefont
  {{Visuri}}}, \bibinfo {author} {\bibfnamefont {J.}~\bibnamefont {{Mohan}}},
  \bibinfo {author} {\bibfnamefont {S.}~\bibnamefont {{Uchino}}}, \bibinfo
  {author} {\bibfnamefont {M.-Z.}\ \bibnamefont {{Huang}}}, \bibinfo {author}
  {\bibfnamefont {T.}~\bibnamefont {{Esslinger}}},\ and\ \bibinfo {author}
  {\bibfnamefont {T.}~\bibnamefont {{Giamarchi}}},\ }\bibfield  {title}
  {\bibinfo {title} {{DC transport in a dissipative superconducting quantum
  point contact}},\ }\href {https://doi.org/10.48550/arXiv.2304.00928}
  {\bibfield  {journal} {\bibinfo  {journal} {arXiv e-prints}\ ,\ \bibinfo
  {eid} {arXiv:2304.00928}} (\bibinfo {year} {2023})},\ \Eprint
  {https://arxiv.org/abs/2304.00928} {arXiv:2304.00928 [cond-mat.quant-gas]}
  \BibitemShut {NoStop}%
\bibitem [{\citenamefont {Krapivsky}\ \emph {et~al.}(2019)\citenamefont
  {Krapivsky}, \citenamefont {Mallick},\ and\ \citenamefont
  {Sels}}]{krapivsky2019free}%
  \BibitemOpen
  \bibfield  {author} {\bibinfo {author} {\bibfnamefont {P.~L.}\ \bibnamefont
  {Krapivsky}}, \bibinfo {author} {\bibfnamefont {K.}~\bibnamefont {Mallick}},\
  and\ \bibinfo {author} {\bibfnamefont {D.}~\bibnamefont {Sels}},\ }\bibfield
  {title} {\bibinfo {title} {Free fermions with a localized source},\ }\href
  {https://doi.org/10.1088/1742-5468/ab4e8e} {\bibfield  {journal} {\bibinfo
  {journal} {Journal of Statistical Mechanics: Theory and Experiment}\ }\textbf
  {\bibinfo {volume} {2019}},\ \bibinfo {pages} {113108} (\bibinfo {year}
  {2019})}\BibitemShut {NoStop}%
\bibitem [{\citenamefont {Krapivsky}\ \emph {et~al.}(2020)\citenamefont
  {Krapivsky}, \citenamefont {Mallick},\ and\ \citenamefont
  {Sels}}]{krapivsky2020free}%
  \BibitemOpen
  \bibfield  {author} {\bibinfo {author} {\bibfnamefont {P.~L.}\ \bibnamefont
  {Krapivsky}}, \bibinfo {author} {\bibfnamefont {K.}~\bibnamefont {Mallick}},\
  and\ \bibinfo {author} {\bibfnamefont {D.}~\bibnamefont {Sels}},\ }\bibfield
  {title} {\bibinfo {title} {Free bosons with a localized source},\ }\href
  {https://doi.org/10.1088/1742-5468/ab8118} {\bibfield  {journal} {\bibinfo
  {journal} {Journal of Statistical Mechanics: Theory and Experiment}\ }\textbf
  {\bibinfo {volume} {2020}},\ \bibinfo {pages} {063101} (\bibinfo {year}
  {2020})}\BibitemShut {NoStop}%
\bibitem [{\citenamefont {Schiro}\ and\ \citenamefont
  {Scarlatella}(2019)}]{Scarlatella2019}%
  \BibitemOpen
  \bibfield  {author} {\bibinfo {author} {\bibfnamefont {M.}~\bibnamefont
  {Schiro}}\ and\ \bibinfo {author} {\bibfnamefont {O.}~\bibnamefont
  {Scarlatella}},\ }\bibfield  {title} {\bibinfo {title} {Quantum impurity
  models coupled to {{Markovian}} and non-{{Markovian}} baths},\ }\href
  {https://doi.org/10.1063/1.5100157} {\bibfield  {journal} {\bibinfo
  {journal} {The Journal of Chemical Physics}\ }\textbf {\bibinfo {volume}
  {151}},\ \bibinfo {pages} {044102} (\bibinfo {year} {2019})}\BibitemShut
  {NoStop}%
\bibitem [{\citenamefont {Tonielli}\ \emph {et~al.}(2019)\citenamefont
  {Tonielli}, \citenamefont {Fazio}, \citenamefont {Diehl},\ and\ \citenamefont
  {Marino}}]{tonielli2019orthogonality}%
  \BibitemOpen
  \bibfield  {author} {\bibinfo {author} {\bibfnamefont {F.}~\bibnamefont
  {Tonielli}}, \bibinfo {author} {\bibfnamefont {R.}~\bibnamefont {Fazio}},
  \bibinfo {author} {\bibfnamefont {S.}~\bibnamefont {Diehl}},\ and\ \bibinfo
  {author} {\bibfnamefont {J.}~\bibnamefont {Marino}},\ }\bibfield  {title}
  {\bibinfo {title} {Orthogonality catastrophe in dissipative quantum many-body
  systems},\ }\href {https://doi.org/10.1103/PhysRevLett.122.040604} {\bibfield
   {journal} {\bibinfo  {journal} {Phys. Rev. Lett.}\ }\textbf {\bibinfo
  {volume} {122}},\ \bibinfo {pages} {040604} (\bibinfo {year}
  {2019})}\BibitemShut {NoStop}%
\bibitem [{\citenamefont {Dolgirev}\ \emph {et~al.}(2020)\citenamefont
  {Dolgirev}, \citenamefont {Marino}, \citenamefont {Sels},\ and\ \citenamefont
  {Demler}}]{dolgirev2020nongaussian}%
  \BibitemOpen
  \bibfield  {author} {\bibinfo {author} {\bibfnamefont {P.~E.}\ \bibnamefont
  {Dolgirev}}, \bibinfo {author} {\bibfnamefont {J.}~\bibnamefont {Marino}},
  \bibinfo {author} {\bibfnamefont {D.}~\bibnamefont {Sels}},\ and\ \bibinfo
  {author} {\bibfnamefont {E.}~\bibnamefont {Demler}},\ }\bibfield  {title}
  {\bibinfo {title} {Non-gaussian correlations imprinted by local dephasing in
  fermionic wires},\ }\href {https://doi.org/10.1103/PhysRevB.102.100301}
  {\bibfield  {journal} {\bibinfo  {journal} {Phys. Rev. B}\ }\textbf {\bibinfo
  {volume} {102}},\ \bibinfo {pages} {100301} (\bibinfo {year}
  {2020})}\BibitemShut {NoStop}%
\bibitem [{\citenamefont {Ferreira}\ \emph {et~al.}(2023)\citenamefont
  {Ferreira}, \citenamefont {Jin}, \citenamefont {Mannhart}, \citenamefont
  {Giamarchi},\ and\ \citenamefont {Filippone}}]{ferreira2023exact}%
  \BibitemOpen
  \bibfield  {author} {\bibinfo {author} {\bibfnamefont {J.}~\bibnamefont
  {Ferreira}}, \bibinfo {author} {\bibfnamefont {T.}~\bibnamefont {Jin}},
  \bibinfo {author} {\bibfnamefont {J.}~\bibnamefont {Mannhart}}, \bibinfo
  {author} {\bibfnamefont {T.}~\bibnamefont {Giamarchi}},\ and\ \bibinfo
  {author} {\bibfnamefont {M.}~\bibnamefont {Filippone}},\ }\href@noop {}
  {\bibinfo {title} {Exact description of transport and non-reciprocity in
  monitored quantum devices}} (\bibinfo {year} {2023}),\ \Eprint
  {https://arxiv.org/abs/2306.16452} {arXiv:2306.16452 [quant-ph]} \BibitemShut
  {NoStop}%
\bibitem [{\citenamefont {Nakagawa}\ \emph {et~al.}(2018)\citenamefont
  {Nakagawa}, \citenamefont {Kawakami},\ and\ \citenamefont
  {Ueda}}]{Nakagawa2018}%
  \BibitemOpen
  \bibfield  {author} {\bibinfo {author} {\bibfnamefont {M.}~\bibnamefont
  {Nakagawa}}, \bibinfo {author} {\bibfnamefont {N.}~\bibnamefont {Kawakami}},\
  and\ \bibinfo {author} {\bibfnamefont {M.}~\bibnamefont {Ueda}},\ }\bibfield
  {title} {\bibinfo {title} {Non-{{Hermitian Kondo Effect}} in {{Ultracold
  Alkaline}}-{{Earth Atoms}}},\ }\href
  {https://doi.org/10.1103/PhysRevLett.121.203001} {\bibfield  {journal}
  {\bibinfo  {journal} {Physical Review Letters}\ }\textbf {\bibinfo {volume}
  {121}},\ \bibinfo {pages} {203001} (\bibinfo {year} {2018})}\BibitemShut
  {NoStop}%
\bibitem [{\citenamefont {Yoshimura}\ \emph {et~al.}(2020)\citenamefont
  {Yoshimura}, \citenamefont {Bidzhiev},\ and\ \citenamefont
  {Saleur}}]{yoshimura2020nonhermitian}%
  \BibitemOpen
  \bibfield  {author} {\bibinfo {author} {\bibfnamefont {T.}~\bibnamefont
  {Yoshimura}}, \bibinfo {author} {\bibfnamefont {K.}~\bibnamefont
  {Bidzhiev}},\ and\ \bibinfo {author} {\bibfnamefont {H.}~\bibnamefont
  {Saleur}},\ }\bibfield  {title} {\bibinfo {title} {Non-hermitian quantum
  impurity systems in and out of equilibrium: Noninteracting case},\ }\href
  {https://doi.org/10.1103/PhysRevB.102.125124} {\bibfield  {journal} {\bibinfo
   {journal} {Phys. Rev. B}\ }\textbf {\bibinfo {volume} {102}},\ \bibinfo
  {pages} {125124} (\bibinfo {year} {2020})}\BibitemShut {NoStop}%
\bibitem [{\citenamefont {Stefanini}\ and\ \citenamefont
  {Marino}(2023)}]{stefanini2023orthogonality}%
  \BibitemOpen
  \bibfield  {author} {\bibinfo {author} {\bibfnamefont {M.}~\bibnamefont
  {Stefanini}}\ and\ \bibinfo {author} {\bibfnamefont {J.}~\bibnamefont
  {Marino}},\ }\href@noop {} {\bibinfo {title} {Orthogonality catastrophe
  beyond luttinger liquid from post-selection}} (\bibinfo {year} {2023}),\
  \Eprint {https://arxiv.org/abs/2310.00039} {arXiv:2310.00039
  [cond-mat.stat-mech]} \BibitemShut {NoStop}%
\bibitem [{\citenamefont {Scarlatella}\ \emph {et~al.}(2021)\citenamefont
  {Scarlatella}, \citenamefont {Clerk}, \citenamefont {Fazio},\ and\
  \citenamefont {Schir\'o}}]{scarlatella2021dynamical}%
  \BibitemOpen
  \bibfield  {author} {\bibinfo {author} {\bibfnamefont {O.}~\bibnamefont
  {Scarlatella}}, \bibinfo {author} {\bibfnamefont {A.~A.}\ \bibnamefont
  {Clerk}}, \bibinfo {author} {\bibfnamefont {R.}~\bibnamefont {Fazio}},\ and\
  \bibinfo {author} {\bibfnamefont {M.}~\bibnamefont {Schir\'o}},\ }\bibfield
  {title} {\bibinfo {title} {Dynamical mean-field theory for markovian open
  quantum many-body systems},\ }\href
  {https://doi.org/10.1103/PhysRevX.11.031018} {\bibfield  {journal} {\bibinfo
  {journal} {Phys. Rev. X}\ }\textbf {\bibinfo {volume} {11}},\ \bibinfo
  {pages} {031018} (\bibinfo {year} {2021})}\BibitemShut {NoStop}%
\bibitem [{\citenamefont {Anders}\ and\ \citenamefont
  {Schiller}(2005)}]{anders2005realtime}%
  \BibitemOpen
  \bibfield  {author} {\bibinfo {author} {\bibfnamefont {F.~B.}\ \bibnamefont
  {Anders}}\ and\ \bibinfo {author} {\bibfnamefont {A.}~\bibnamefont
  {Schiller}},\ }\bibfield  {title} {\bibinfo {title} {Real-time dynamics in
  quantum-impurity systems: A time-dependent numerical renormalization-group
  approach},\ }\href {https://doi.org/10.1103/PhysRevLett.95.196801} {\bibfield
   {journal} {\bibinfo  {journal} {Phys. Rev. Lett.}\ }\textbf {\bibinfo
  {volume} {95}},\ \bibinfo {pages} {196801} (\bibinfo {year}
  {2005})}\BibitemShut {NoStop}%
\bibitem [{\citenamefont {Heidrich-Meisner}\ \emph {et~al.}(2009)\citenamefont
  {Heidrich-Meisner}, \citenamefont {Feiguin},\ and\ \citenamefont
  {Dagotto}}]{heidrich2009realtime}%
  \BibitemOpen
  \bibfield  {author} {\bibinfo {author} {\bibfnamefont {F.}~\bibnamefont
  {Heidrich-Meisner}}, \bibinfo {author} {\bibfnamefont {A.~E.}\ \bibnamefont
  {Feiguin}},\ and\ \bibinfo {author} {\bibfnamefont {E.}~\bibnamefont
  {Dagotto}},\ }\bibfield  {title} {\bibinfo {title} {Real-time simulations of
  nonequilibrium transport in the single-impurity anderson model},\ }\href
  {https://doi.org/10.1103/PhysRevB.79.235336} {\bibfield  {journal} {\bibinfo
  {journal} {Phys. Rev. B}\ }\textbf {\bibinfo {volume} {79}},\ \bibinfo
  {pages} {235336} (\bibinfo {year} {2009})}\BibitemShut {NoStop}%
\bibitem [{\citenamefont {Schwarz}\ \emph {et~al.}(2018)\citenamefont
  {Schwarz}, \citenamefont {Weymann}, \citenamefont {von Delft},\ and\
  \citenamefont {Weichselbaum}}]{schwarz2018nonequilibrium}%
  \BibitemOpen
  \bibfield  {author} {\bibinfo {author} {\bibfnamefont {F.}~\bibnamefont
  {Schwarz}}, \bibinfo {author} {\bibfnamefont {I.}~\bibnamefont {Weymann}},
  \bibinfo {author} {\bibfnamefont {J.}~\bibnamefont {von Delft}},\ and\
  \bibinfo {author} {\bibfnamefont {A.}~\bibnamefont {Weichselbaum}},\
  }\bibfield  {title} {\bibinfo {title} {Nonequilibrium steady-state transport
  in quantum impurity models: A thermofield and quantum quench approach using
  matrix product states},\ }\href
  {https://doi.org/10.1103/PhysRevLett.121.137702} {\bibfield  {journal}
  {\bibinfo  {journal} {Phys. Rev. Lett.}\ }\textbf {\bibinfo {volume} {121}},\
  \bibinfo {pages} {137702} (\bibinfo {year} {2018})}\BibitemShut {NoStop}%
\bibitem [{\citenamefont {Kohn}\ and\ \citenamefont
  {Santoro}(2022)}]{Kohn_2022}%
  \BibitemOpen
  \bibfield  {author} {\bibinfo {author} {\bibfnamefont {L.}~\bibnamefont
  {Kohn}}\ and\ \bibinfo {author} {\bibfnamefont {G.~E.}\ \bibnamefont
  {Santoro}},\ }\bibfield  {title} {\bibinfo {title} {Quench dynamics of the
  anderson impurity model at finite temperature using matrix product states:
  entanglement and bath dynamics},\ }\href
  {https://doi.org/10.1088/1742-5468/ac729b} {\bibfield  {journal} {\bibinfo
  {journal} {Journal of Statistical Mechanics: Theory and Experiment}\ }\textbf
  {\bibinfo {volume} {2022}},\ \bibinfo {pages} {063102} (\bibinfo {year}
  {2022})}\BibitemShut {NoStop}%
\bibitem [{\citenamefont {Wauters}\ \emph {et~al.}(2023)\citenamefont
  {Wauters}, \citenamefont {Chung}, \citenamefont {Maffi},\ and\ \citenamefont
  {Burrello}}]{wauters2023simulations}%
  \BibitemOpen
  \bibfield  {author} {\bibinfo {author} {\bibfnamefont {M.~M.}\ \bibnamefont
  {Wauters}}, \bibinfo {author} {\bibfnamefont {C.-M.}\ \bibnamefont {Chung}},
  \bibinfo {author} {\bibfnamefont {L.}~\bibnamefont {Maffi}},\ and\ \bibinfo
  {author} {\bibfnamefont {M.}~\bibnamefont {Burrello}},\ }\href@noop {}
  {\bibinfo {title} {Simulations of the dynamics of quantum impurity problems
  with matrix product states}} (\bibinfo {year} {2023}),\ \Eprint
  {https://arxiv.org/abs/2304.13756} {arXiv:2304.13756 [cond-mat.str-el]}
  \BibitemShut {NoStop}%
\bibitem [{\citenamefont {Dorda}\ \emph {et~al.}(2014)\citenamefont {Dorda},
  \citenamefont {Nuss}, \citenamefont {von~der Linden},\ and\ \citenamefont
  {Arrigoni}}]{dorda2014auxiliary}%
  \BibitemOpen
  \bibfield  {author} {\bibinfo {author} {\bibfnamefont {A.}~\bibnamefont
  {Dorda}}, \bibinfo {author} {\bibfnamefont {M.}~\bibnamefont {Nuss}},
  \bibinfo {author} {\bibfnamefont {W.}~\bibnamefont {von~der Linden}},\ and\
  \bibinfo {author} {\bibfnamefont {E.}~\bibnamefont {Arrigoni}},\ }\bibfield
  {title} {\bibinfo {title} {Auxiliary master equation approach to
  nonequilibrium correlated impurities},\ }\href
  {https://doi.org/10.1103/PhysRevB.89.165105} {\bibfield  {journal} {\bibinfo
  {journal} {Phys. Rev. B}\ }\textbf {\bibinfo {volume} {89}},\ \bibinfo
  {pages} {165105} (\bibinfo {year} {2014})}\BibitemShut {NoStop}%
\bibitem [{\citenamefont {Chen}\ \emph
  {et~al.}(2019{\natexlab{a}})\citenamefont {Chen}, \citenamefont {Cohen},\
  and\ \citenamefont {Galperin}}]{chen2019auxiliary}%
  \BibitemOpen
  \bibfield  {author} {\bibinfo {author} {\bibfnamefont {F.}~\bibnamefont
  {Chen}}, \bibinfo {author} {\bibfnamefont {G.}~\bibnamefont {Cohen}},\ and\
  \bibinfo {author} {\bibfnamefont {M.}~\bibnamefont {Galperin}},\ }\bibfield
  {title} {\bibinfo {title} {Auxiliary master equation for nonequilibrium
  dual-fermion approach},\ }\href
  {https://doi.org/10.1103/PhysRevLett.122.186803} {\bibfield  {journal}
  {\bibinfo  {journal} {Phys. Rev. Lett.}\ }\textbf {\bibinfo {volume} {122}},\
  \bibinfo {pages} {186803} (\bibinfo {year} {2019}{\natexlab{a}})}\BibitemShut
  {NoStop}%
\bibitem [{\citenamefont {Chen}\ \emph
  {et~al.}(2019{\natexlab{b}})\citenamefont {Chen}, \citenamefont {Arrigoni},\
  and\ \citenamefont {Galperin}}]{Chen_2019}%
  \BibitemOpen
  \bibfield  {author} {\bibinfo {author} {\bibfnamefont {F.}~\bibnamefont
  {Chen}}, \bibinfo {author} {\bibfnamefont {E.}~\bibnamefont {Arrigoni}},\
  and\ \bibinfo {author} {\bibfnamefont {M.}~\bibnamefont {Galperin}},\
  }\bibfield  {title} {\bibinfo {title} {Markovian treatment of non-markovian
  dynamics of open fermionic systems},\ }\href
  {https://doi.org/10.1088/1367-2630/ab5ec5} {\bibfield  {journal} {\bibinfo
  {journal} {New Journal of Physics}\ }\textbf {\bibinfo {volume} {21}},\
  \bibinfo {pages} {123035} (\bibinfo {year} {2019}{\natexlab{b}})}\BibitemShut
  {NoStop}%
\bibitem [{\citenamefont {Strathearn}\ \emph {et~al.}(2018)\citenamefont
  {Strathearn}, \citenamefont {Kirton}, \citenamefont {Kilda}, \citenamefont
  {Keeling},\ and\ \citenamefont {Lovett}}]{strathearn2018efficient}%
  \BibitemOpen
  \bibfield  {author} {\bibinfo {author} {\bibfnamefont {A.}~\bibnamefont
  {Strathearn}}, \bibinfo {author} {\bibfnamefont {P.}~\bibnamefont {Kirton}},
  \bibinfo {author} {\bibfnamefont {D.}~\bibnamefont {Kilda}}, \bibinfo
  {author} {\bibfnamefont {J.}~\bibnamefont {Keeling}},\ and\ \bibinfo {author}
  {\bibfnamefont {B.~W.}\ \bibnamefont {Lovett}},\ }\bibfield  {title}
  {\bibinfo {title} {Efficient non-markovian quantum dynamics using
  time-evolving matrix product operators},\ }\href
  {https://doi.org/10.1038/s41467-018-05617-3} {\bibfield  {journal} {\bibinfo
  {journal} {Nature Communications}\ }\textbf {\bibinfo {volume} {9}},\
  \bibinfo {pages} {3322} (\bibinfo {year} {2018})}\BibitemShut {NoStop}%
\bibitem [{\citenamefont {Gribben}\ \emph {et~al.}(2022)\citenamefont
  {Gribben}, \citenamefont {Rouse}, \citenamefont {Iles-Smith}, \citenamefont
  {Strathearn}, \citenamefont {Maguire}, \citenamefont {Kirton}, \citenamefont
  {Nazir}, \citenamefont {Gauger},\ and\ \citenamefont
  {Lovett}}]{gribben2022exact}%
  \BibitemOpen
  \bibfield  {author} {\bibinfo {author} {\bibfnamefont {D.}~\bibnamefont
  {Gribben}}, \bibinfo {author} {\bibfnamefont {D.~M.}\ \bibnamefont {Rouse}},
  \bibinfo {author} {\bibfnamefont {J.}~\bibnamefont {Iles-Smith}}, \bibinfo
  {author} {\bibfnamefont {A.}~\bibnamefont {Strathearn}}, \bibinfo {author}
  {\bibfnamefont {H.}~\bibnamefont {Maguire}}, \bibinfo {author} {\bibfnamefont
  {P.}~\bibnamefont {Kirton}}, \bibinfo {author} {\bibfnamefont
  {A.}~\bibnamefont {Nazir}}, \bibinfo {author} {\bibfnamefont {E.~M.}\
  \bibnamefont {Gauger}},\ and\ \bibinfo {author} {\bibfnamefont {B.~W.}\
  \bibnamefont {Lovett}},\ }\bibfield  {title} {\bibinfo {title} {Exact
  dynamics of nonadditive environments in non-markovian open quantum systems},\
  }\href {https://doi.org/10.1103/PRXQuantum.3.010321} {\bibfield  {journal}
  {\bibinfo  {journal} {PRX Quantum}\ }\textbf {\bibinfo {volume} {3}},\
  \bibinfo {pages} {010321} (\bibinfo {year} {2022})}\BibitemShut {NoStop}%
\bibitem [{\citenamefont {J\o{}rgensen}\ and\ \citenamefont
  {Pollock}(2019)}]{jorgensen2019exploiting}%
  \BibitemOpen
  \bibfield  {author} {\bibinfo {author} {\bibfnamefont {M.~R.}\ \bibnamefont
  {J\o{}rgensen}}\ and\ \bibinfo {author} {\bibfnamefont {F.~A.}\ \bibnamefont
  {Pollock}},\ }\bibfield  {title} {\bibinfo {title} {Exploiting the causal
  tensor network structure of quantum processes to efficiently simulate
  non-markovian path integrals},\ }\href
  {https://doi.org/10.1103/PhysRevLett.123.240602} {\bibfield  {journal}
  {\bibinfo  {journal} {Phys. Rev. Lett.}\ }\textbf {\bibinfo {volume} {123}},\
  \bibinfo {pages} {240602} (\bibinfo {year} {2019})}\BibitemShut {NoStop}%
\bibitem [{\citenamefont {Thoenniss}\ \emph
  {et~al.}(2023{\natexlab{a}})\citenamefont {Thoenniss}, \citenamefont
  {Sonner}, \citenamefont {Lerose},\ and\ \citenamefont
  {Abanin}}]{thoenniss2023efficient}%
  \BibitemOpen
  \bibfield  {author} {\bibinfo {author} {\bibfnamefont {J.}~\bibnamefont
  {Thoenniss}}, \bibinfo {author} {\bibfnamefont {M.}~\bibnamefont {Sonner}},
  \bibinfo {author} {\bibfnamefont {A.}~\bibnamefont {Lerose}},\ and\ \bibinfo
  {author} {\bibfnamefont {D.~A.}\ \bibnamefont {Abanin}},\ }\bibfield  {title}
  {\bibinfo {title} {Efficient method for quantum impurity problems out of
  equilibrium},\ }\href {https://doi.org/10.1103/PhysRevB.107.L201115}
  {\bibfield  {journal} {\bibinfo  {journal} {Phys. Rev. B}\ }\textbf {\bibinfo
  {volume} {107}},\ \bibinfo {pages} {L201115} (\bibinfo {year}
  {2023}{\natexlab{a}})}\BibitemShut {NoStop}%
\bibitem [{\citenamefont {Thoenniss}\ \emph
  {et~al.}(2023{\natexlab{b}})\citenamefont {Thoenniss}, \citenamefont
  {Lerose},\ and\ \citenamefont {Abanin}}]{thoenniss2023nonequilibrium}%
  \BibitemOpen
  \bibfield  {author} {\bibinfo {author} {\bibfnamefont {J.}~\bibnamefont
  {Thoenniss}}, \bibinfo {author} {\bibfnamefont {A.}~\bibnamefont {Lerose}},\
  and\ \bibinfo {author} {\bibfnamefont {D.~A.}\ \bibnamefont {Abanin}},\
  }\bibfield  {title} {\bibinfo {title} {Nonequilibrium quantum impurity
  problems via matrix-product states in the temporal domain},\ }\href
  {https://doi.org/10.1103/PhysRevB.107.195101} {\bibfield  {journal} {\bibinfo
   {journal} {Phys. Rev. B}\ }\textbf {\bibinfo {volume} {107}},\ \bibinfo
  {pages} {195101} (\bibinfo {year} {2023}{\natexlab{b}})}\BibitemShut
  {NoStop}%
\bibitem [{\citenamefont {Ng}\ \emph {et~al.}(2023)\citenamefont {Ng},
  \citenamefont {Park}, \citenamefont {Millis}, \citenamefont {Chan},\ and\
  \citenamefont {Reichman}}]{ng2023realtime}%
  \BibitemOpen
  \bibfield  {author} {\bibinfo {author} {\bibfnamefont {N.}~\bibnamefont
  {Ng}}, \bibinfo {author} {\bibfnamefont {G.}~\bibnamefont {Park}}, \bibinfo
  {author} {\bibfnamefont {A.~J.}\ \bibnamefont {Millis}}, \bibinfo {author}
  {\bibfnamefont {G.~K.-L.}\ \bibnamefont {Chan}},\ and\ \bibinfo {author}
  {\bibfnamefont {D.~R.}\ \bibnamefont {Reichman}},\ }\bibfield  {title}
  {\bibinfo {title} {Real-time evolution of anderson impurity models via tensor
  network influence functionals},\ }\href
  {https://doi.org/10.1103/PhysRevB.107.125103} {\bibfield  {journal} {\bibinfo
   {journal} {Phys. Rev. B}\ }\textbf {\bibinfo {volume} {107}},\ \bibinfo
  {pages} {125103} (\bibinfo {year} {2023})}\BibitemShut {NoStop}%
\bibitem [{\citenamefont {Park}\ \emph {et~al.}(2024)\citenamefont {Park},
  \citenamefont {Ng}, \citenamefont {Reichman},\ and\ \citenamefont
  {Chan}}]{park2024tensor}%
  \BibitemOpen
  \bibfield  {author} {\bibinfo {author} {\bibfnamefont {G.}~\bibnamefont
  {Park}}, \bibinfo {author} {\bibfnamefont {N.}~\bibnamefont {Ng}}, \bibinfo
  {author} {\bibfnamefont {D.~R.}\ \bibnamefont {Reichman}},\ and\ \bibinfo
  {author} {\bibfnamefont {G.~K.-L.}\ \bibnamefont {Chan}},\ }\href@noop {}
  {\bibinfo {title} {Tensor network influence functionals in the
  continuous-time limit: connections to quantum embedding, bath discretization,
  and higher-order time propagation}} (\bibinfo {year} {2024}),\ \Eprint
  {https://arxiv.org/abs/2401.12460} {arXiv:2401.12460 [cond-mat.str-el]}
  \BibitemShut {NoStop}%
\bibitem [{\citenamefont {Schir{\'o}}(2010)}]{schiroPRB2009}%
  \BibitemOpen
  \bibfield  {author} {\bibinfo {author} {\bibfnamefont {M.}~\bibnamefont
  {Schir{\'o}}},\ }\bibfield  {title} {\bibinfo {title} {Real-time dynamics in
  quantum impurity models with diagrammatic {{Monte Carlo}}},\ }\href
  {https://doi.org/10.1103/PhysRevB.81.085126} {\bibfield  {journal} {\bibinfo
  {journal} {Physical Review B - Condensed Matter and Materials Physics}\
  }\textbf {\bibinfo {volume} {81}},\ \bibinfo {pages} {85126} (\bibinfo {year}
  {2010})}\BibitemShut {NoStop}%
\bibitem [{\citenamefont {Schir{\'o}}\ and\ \citenamefont
  {Fabrizio}(2009)}]{schiroFabrizioPRB2009}%
  \BibitemOpen
  \bibfield  {author} {\bibinfo {author} {\bibfnamefont {M.}~\bibnamefont
  {Schir{\'o}}}\ and\ \bibinfo {author} {\bibfnamefont {M.}~\bibnamefont
  {Fabrizio}},\ }\bibfield  {title} {\bibinfo {title} {Real-time diagrammatic
  {{Monte Carlo}} for nonequilibrium quantum transport},\ }\href
  {https://doi.org/10.1103/PhysRevB.79.153302} {\bibfield  {journal} {\bibinfo
  {journal} {Phys. Rev. B}\ }\textbf {\bibinfo {volume} {79}},\ \bibinfo
  {pages} {153302} (\bibinfo {year} {2009})}\BibitemShut {NoStop}%
\bibitem [{\citenamefont {M{\"u}hlbacher}\ and\ \citenamefont
  {Rabani}(2008)}]{muhlbacherRabaniPRL2008}%
  \BibitemOpen
  \bibfield  {author} {\bibinfo {author} {\bibfnamefont {L.}~\bibnamefont
  {M{\"u}hlbacher}}\ and\ \bibinfo {author} {\bibfnamefont {E.}~\bibnamefont
  {Rabani}},\ }\bibfield  {title} {\bibinfo {title} {Real-time path integral
  approach to nonequilibrium many-body quantum systems},\ }\href
  {https://doi.org/10.1103/PhysRevLett.100.176403} {\bibfield  {journal}
  {\bibinfo  {journal} {Physical Review Letters}\ }\textbf {\bibinfo {volume}
  {100}},\ \bibinfo {pages} {176403} (\bibinfo {year} {2008})}\BibitemShut
  {NoStop}%
\bibitem [{\citenamefont {Werner}\ \emph {et~al.}(2009)\citenamefont {Werner},
  \citenamefont {Oka},\ and\ \citenamefont {Millis}}]{Werner_Keldysh_09}%
  \BibitemOpen
  \bibfield  {author} {\bibinfo {author} {\bibfnamefont {P.}~\bibnamefont
  {Werner}}, \bibinfo {author} {\bibfnamefont {T.}~\bibnamefont {Oka}},\ and\
  \bibinfo {author} {\bibfnamefont {A.~J.}\ \bibnamefont {Millis}},\ }\bibfield
   {title} {\bibinfo {title} {Diagrammatic {{Monte Carlo}} simulation of
  nonequilibrium systems},\ }\href {https://doi.org/10.1103/PhysRevB.79.035320}
  {\bibfield  {journal} {\bibinfo  {journal} {Physical Review B - Condensed
  Matter and Materials Physics}\ }\textbf {\bibinfo {volume} {79}},\ \bibinfo
  {pages} {35320} (\bibinfo {year} {2009})}\BibitemShut {NoStop}%
\bibitem [{\citenamefont {Cohen}\ \emph {et~al.}(2015)\citenamefont {Cohen},
  \citenamefont {Gull}, \citenamefont {Reichman},\ and\ \citenamefont
  {Millis}}]{cohen2015taming}%
  \BibitemOpen
  \bibfield  {author} {\bibinfo {author} {\bibfnamefont {G.}~\bibnamefont
  {Cohen}}, \bibinfo {author} {\bibfnamefont {E.}~\bibnamefont {Gull}},
  \bibinfo {author} {\bibfnamefont {D.~R.}\ \bibnamefont {Reichman}},\ and\
  \bibinfo {author} {\bibfnamefont {A.~J.}\ \bibnamefont {Millis}},\ }\bibfield
   {title} {\bibinfo {title} {Taming the dynamical sign problem in real-time
  evolution of quantum many-body problems},\ }\href
  {https://doi.org/10.1103/PhysRevLett.115.266802} {\bibfield  {journal}
  {\bibinfo  {journal} {Phys. Rev. Lett.}\ }\textbf {\bibinfo {volume} {115}},\
  \bibinfo {pages} {266802} (\bibinfo {year} {2015})}\BibitemShut {NoStop}%
\bibitem [{\citenamefont {Bertrand}\ \emph {et~al.}(2019)\citenamefont
  {Bertrand}, \citenamefont {Parcollet}, \citenamefont {Maillard},\ and\
  \citenamefont {Waintal}}]{corentin2019quantum}%
  \BibitemOpen
  \bibfield  {author} {\bibinfo {author} {\bibfnamefont {C.}~\bibnamefont
  {Bertrand}}, \bibinfo {author} {\bibfnamefont {O.}~\bibnamefont {Parcollet}},
  \bibinfo {author} {\bibfnamefont {A.}~\bibnamefont {Maillard}},\ and\
  \bibinfo {author} {\bibfnamefont {X.}~\bibnamefont {Waintal}},\ }\bibfield
  {title} {\bibinfo {title} {Quantum monte carlo algorithm for
  out-of-equilibrium green's functions at long times},\ }\href
  {https://doi.org/10.1103/PhysRevB.100.125129} {\bibfield  {journal} {\bibinfo
   {journal} {Phys. Rev. B}\ }\textbf {\bibinfo {volume} {100}},\ \bibinfo
  {pages} {125129} (\bibinfo {year} {2019})}\BibitemShut {NoStop}%
\bibitem [{\citenamefont {Ma\ifmmode~\check{c}\else \v{c}\fi{}ek}\ \emph
  {et~al.}(2020)\citenamefont {Ma\ifmmode~\check{c}\else \v{c}\fi{}ek},
  \citenamefont {Dumitrescu}, \citenamefont {Bertrand}, \citenamefont {Triggs},
  \citenamefont {Parcollet},\ and\ \citenamefont {Waintal}}]{macek2020quantum}%
  \BibitemOpen
  \bibfield  {author} {\bibinfo {author} {\bibfnamefont {M.}~\bibnamefont
  {Ma\ifmmode~\check{c}\else \v{c}\fi{}ek}}, \bibinfo {author} {\bibfnamefont
  {P.~T.}\ \bibnamefont {Dumitrescu}}, \bibinfo {author} {\bibfnamefont
  {C.}~\bibnamefont {Bertrand}}, \bibinfo {author} {\bibfnamefont
  {B.}~\bibnamefont {Triggs}}, \bibinfo {author} {\bibfnamefont
  {O.}~\bibnamefont {Parcollet}},\ and\ \bibinfo {author} {\bibfnamefont
  {X.}~\bibnamefont {Waintal}},\ }\bibfield  {title} {\bibinfo {title} {Quantum
  quasi-monte carlo technique for many-body perturbative expansions},\ }\href
  {https://doi.org/10.1103/PhysRevLett.125.047702} {\bibfield  {journal}
  {\bibinfo  {journal} {Phys. Rev. Lett.}\ }\textbf {\bibinfo {volume} {125}},\
  \bibinfo {pages} {047702} (\bibinfo {year} {2020})}\BibitemShut {NoStop}%
\bibitem [{\citenamefont {N\'u\~nez Fern\'andez}\ \emph
  {et~al.}(2022)\citenamefont {N\'u\~nez Fern\'andez}, \citenamefont {Jeannin},
  \citenamefont {Dumitrescu}, \citenamefont {Kloss}, \citenamefont {Kaye},
  \citenamefont {Parcollet},\ and\ \citenamefont
  {Waintal}}]{nunez2022learning}%
  \BibitemOpen
  \bibfield  {author} {\bibinfo {author} {\bibfnamefont {Y.}~\bibnamefont
  {N\'u\~nez Fern\'andez}}, \bibinfo {author} {\bibfnamefont {M.}~\bibnamefont
  {Jeannin}}, \bibinfo {author} {\bibfnamefont {P.~T.}\ \bibnamefont
  {Dumitrescu}}, \bibinfo {author} {\bibfnamefont {T.}~\bibnamefont {Kloss}},
  \bibinfo {author} {\bibfnamefont {J.}~\bibnamefont {Kaye}}, \bibinfo {author}
  {\bibfnamefont {O.}~\bibnamefont {Parcollet}},\ and\ \bibinfo {author}
  {\bibfnamefont {X.}~\bibnamefont {Waintal}},\ }\bibfield  {title} {\bibinfo
  {title} {Learning feynman diagrams with tensor trains},\ }\href
  {https://doi.org/10.1103/PhysRevX.12.041018} {\bibfield  {journal} {\bibinfo
  {journal} {Phys. Rev. X}\ }\textbf {\bibinfo {volume} {12}},\ \bibinfo
  {pages} {041018} (\bibinfo {year} {2022})}\BibitemShut {NoStop}%
\bibitem [{\citenamefont {Erpenbeck}\ \emph {et~al.}(2023)\citenamefont
  {Erpenbeck}, \citenamefont {Gull},\ and\ \citenamefont
  {Cohen}}]{erpenbeck2023quantum}%
  \BibitemOpen
  \bibfield  {author} {\bibinfo {author} {\bibfnamefont {A.}~\bibnamefont
  {Erpenbeck}}, \bibinfo {author} {\bibfnamefont {E.}~\bibnamefont {Gull}},\
  and\ \bibinfo {author} {\bibfnamefont {G.}~\bibnamefont {Cohen}},\ }\bibfield
   {title} {\bibinfo {title} {Quantum monte carlo method in the steady state},\
  }\href {https://doi.org/10.1103/PhysRevLett.130.186301} {\bibfield  {journal}
  {\bibinfo  {journal} {Phys. Rev. Lett.}\ }\textbf {\bibinfo {volume} {130}},\
  \bibinfo {pages} {186301} (\bibinfo {year} {2023})}\BibitemShut {NoStop}%
\bibitem [{\citenamefont {Dzhioev}\ and\ \citenamefont
  {Kosov}(2011)}]{dzhioev2011}%
  \BibitemOpen
  \bibfield  {author} {\bibinfo {author} {\bibfnamefont {A.~A.}\ \bibnamefont
  {Dzhioev}}\ and\ \bibinfo {author} {\bibfnamefont {D.~S.}\ \bibnamefont
  {Kosov}},\ }\bibfield  {title} {\bibinfo {title} {{Super-fermion
  representation of quantum kinetic equations for the electron transport
  problem}},\ }\href {https://doi.org/10.1063/1.3548065} {\bibfield  {journal}
  {\bibinfo  {journal} {The Journal of Chemical Physics}\ }\textbf {\bibinfo
  {volume} {134}},\ \bibinfo {pages} {044121} (\bibinfo {year} {2011})},\
  \Eprint
  {https://arxiv.org/abs/https://pubs.aip.org/aip/jcp/article-pdf/doi/10.1063/1.3548065/13788434/044121\_1\_online.pdf}
  {https://pubs.aip.org/aip/jcp/article-pdf/doi/10.1063/1.3548065/13788434/044121\_1\_online.pdf}
  \BibitemShut {NoStop}%
\bibitem [{\citenamefont {Harbola}\ and\ \citenamefont
  {Mukamel}(2008)}]{HARBOLA2008191}%
  \BibitemOpen
  \bibfield  {author} {\bibinfo {author} {\bibfnamefont {U.}~\bibnamefont
  {Harbola}}\ and\ \bibinfo {author} {\bibfnamefont {S.}~\bibnamefont
  {Mukamel}},\ }\bibfield  {title} {\bibinfo {title} {Superoperator
  nonequilibrium green’s function theory of many-body systems; applications
  to charge transfer and transport in open junctions},\ }\href
  {https://doi.org/https://doi.org/10.1016/j.physrep.2008.05.003} {\bibfield
  {journal} {\bibinfo  {journal} {Physics Reports}\ }\textbf {\bibinfo {volume}
  {465}},\ \bibinfo {pages} {191} (\bibinfo {year} {2008})}\BibitemShut
  {NoStop}%
\bibitem [{\citenamefont {Arrigoni}\ and\ \citenamefont
  {Dorda}(2018)}]{Arrigoni2018}%
  \BibitemOpen
  \bibfield  {author} {\bibinfo {author} {\bibfnamefont {E.}~\bibnamefont
  {Arrigoni}}\ and\ \bibinfo {author} {\bibfnamefont {A.}~\bibnamefont
  {Dorda}},\ }\bibinfo {title} {Master equations versus keldysh green's
  functions for correlated quantum systems out of equilibrium},\ in\ \href
  {https://doi.org/10.1007/978-3-319-94956-7_4} {\emph {\bibinfo {booktitle}
  {Out-of-Equilibrium Physics of Correlated Electron Systems}}},\ \bibinfo
  {editor} {edited by\ \bibinfo {editor} {\bibfnamefont {R.}~\bibnamefont
  {Citro}}\ and\ \bibinfo {editor} {\bibfnamefont {F.}~\bibnamefont
  {Mancini}}}\ (\bibinfo  {publisher} {Springer International Publishing},\
  \bibinfo {address} {Cham},\ \bibinfo {year} {2018})\ pp.\ \bibinfo {pages}
  {121--188}\BibitemShut {NoStop}%
\bibitem [{\citenamefont {Werner}\ \emph {et~al.}(2023)\citenamefont {Werner},
  \citenamefont {Lotze},\ and\ \citenamefont
  {Arrigoni}}]{werner2023configuration}%
  \BibitemOpen
  \bibfield  {author} {\bibinfo {author} {\bibfnamefont {D.}~\bibnamefont
  {Werner}}, \bibinfo {author} {\bibfnamefont {J.}~\bibnamefont {Lotze}},\ and\
  \bibinfo {author} {\bibfnamefont {E.}~\bibnamefont {Arrigoni}},\ }\bibfield
  {title} {\bibinfo {title} {Configuration interaction based nonequilibrium
  steady state impurity solver},\ }\href
  {https://doi.org/10.1103/PhysRevB.107.075119} {\bibfield  {journal} {\bibinfo
   {journal} {Phys. Rev. B}\ }\textbf {\bibinfo {volume} {107}},\ \bibinfo
  {pages} {075119} (\bibinfo {year} {2023})}\BibitemShut {NoStop}%
\bibitem [{\citenamefont {Scarlatella}\ and\ \citenamefont
  {Schirò}(2024)}]{scarlatella2023selfconsistent}%
  \BibitemOpen
  \bibfield  {author} {\bibinfo {author} {\bibfnamefont {O.}~\bibnamefont
  {Scarlatella}}\ and\ \bibinfo {author} {\bibfnamefont {M.}~\bibnamefont
  {Schirò}},\ }\bibfield  {title} {\bibinfo {title} {{Self-consistent
  dynamical maps for open quantum systems}},\ }\href
  {https://doi.org/10.21468/SciPostPhys.16.1.026} {\bibfield  {journal}
  {\bibinfo  {journal} {SciPost Phys.}\ }\textbf {\bibinfo {volume} {16}},\
  \bibinfo {pages} {026} (\bibinfo {year} {2024})}\BibitemShut {NoStop}%
\bibitem [{\citenamefont {Gull}\ \emph {et~al.}(2011)\citenamefont {Gull},
  \citenamefont {Millis}, \citenamefont {Lichtenstein}, \citenamefont
  {Rubtsov}, \citenamefont {Troyer},\ and\ \citenamefont
  {Werner}}]{Gull_RMP11}%
  \BibitemOpen
  \bibfield  {author} {\bibinfo {author} {\bibfnamefont {E.}~\bibnamefont
  {Gull}}, \bibinfo {author} {\bibfnamefont {A.~J.}\ \bibnamefont {Millis}},
  \bibinfo {author} {\bibfnamefont {A.~I.}\ \bibnamefont {Lichtenstein}},
  \bibinfo {author} {\bibfnamefont {A.~N.}\ \bibnamefont {Rubtsov}}, \bibinfo
  {author} {\bibfnamefont {M.}~\bibnamefont {Troyer}},\ and\ \bibinfo {author}
  {\bibfnamefont {P.}~\bibnamefont {Werner}},\ }\bibfield  {title} {\bibinfo
  {title} {Continuous-time monte~carlo methods for quantum impurity models},\
  }\href {https://doi.org/10.1103/RevModPhys.83.349} {\bibfield  {journal}
  {\bibinfo  {journal} {Rev. Mod. Phys.}\ }\textbf {\bibinfo {volume} {83}},\
  \bibinfo {pages} {349} (\bibinfo {year} {2011})}\BibitemShut {NoStop}%
\bibitem [{\citenamefont {Prosen}(2008)}]{Prosen_2008}%
  \BibitemOpen
  \bibfield  {author} {\bibinfo {author} {\bibfnamefont {T.}~\bibnamefont
  {Prosen}},\ }\bibfield  {title} {\bibinfo {title} {Third quantization: a
  general method to solve master equations for quadratic open fermi systems},\
  }\href {https://doi.org/10.1088/1367-2630/10/4/043026} {\bibfield  {journal}
  {\bibinfo  {journal} {New Journal of Physics}\ }\textbf {\bibinfo {volume}
  {10}},\ \bibinfo {pages} {043026} (\bibinfo {year} {2008})}\BibitemShut
  {NoStop}%
\bibitem [{\citenamefont {TAKAHASHI}\ and\ \citenamefont
  {UMEZAWA}(1996)}]{takahashi96}%
  \BibitemOpen
  \bibfield  {author} {\bibinfo {author} {\bibfnamefont {Y.}~\bibnamefont
  {TAKAHASHI}}\ and\ \bibinfo {author} {\bibfnamefont {H.}~\bibnamefont
  {UMEZAWA}},\ }\bibfield  {title} {\bibinfo {title} {Thermo field dynamics},\
  }\href {https://doi.org/10.1142/S0217979296000817} {\bibfield  {journal}
  {\bibinfo  {journal} {International Journal of Modern Physics B}\ }\textbf
  {\bibinfo {volume} {10}},\ \bibinfo {pages} {1755} (\bibinfo {year}
  {1996})},\ \Eprint
  {https://arxiv.org/abs/https://doi.org/10.1142/S0217979296000817}
  {https://doi.org/10.1142/S0217979296000817} \BibitemShut {NoStop}%
\bibitem [{\citenamefont {Ojima}(1981)}]{ojima1981}%
  \BibitemOpen
  \bibfield  {author} {\bibinfo {author} {\bibfnamefont {I.}~\bibnamefont
  {Ojima}},\ }\bibfield  {title} {\bibinfo {title} {Gauge fields at finite
  temperatures—“thermo field dynamics” and the kms condition and their
  extension to gauge theories},\ }\href
  {https://doi.org/https://doi.org/10.1016/0003-4916(81)90058-0} {\bibfield
  {journal} {\bibinfo  {journal} {Annals of Physics}\ }\textbf {\bibinfo
  {volume} {137}},\ \bibinfo {pages} {1} (\bibinfo {year} {1981})}\BibitemShut
  {NoStop}%
\bibitem [{\citenamefont {Secl\`i}(2021)}]{secliphd}%
  \BibitemOpen
  \bibfield  {author} {\bibinfo {author} {\bibfnamefont {M.}~\bibnamefont
  {Secl\`i}},\ }\href@noop {} {\emph {\bibinfo {title} {Topology and
  Nonlinearity in Driven-Dissipative Photonic Lattices: Semiclassical and
  Quantum Approaches}}},\ \bibinfo {edition} {1st}\ ed.\ (\bibinfo  {publisher}
  {SISSA},\ \bibinfo {year} {2021})\BibitemShut {NoStop}%
\bibitem [{\citenamefont {McDonald}\ and\ \citenamefont
  {Clerk}(2023)}]{mcdonald2023third}%
  \BibitemOpen
  \bibfield  {author} {\bibinfo {author} {\bibfnamefont {A.}~\bibnamefont
  {McDonald}}\ and\ \bibinfo {author} {\bibfnamefont {A.~A.}\ \bibnamefont
  {Clerk}},\ }\bibfield  {title} {\bibinfo {title} {Third quantization of open
  quantum systems: Dissipative symmetries and connections to phase-space and
  keldysh field-theory formulations},\ }\href
  {https://doi.org/10.1103/PhysRevResearch.5.033107} {\bibfield  {journal}
  {\bibinfo  {journal} {Phys. Rev. Res.}\ }\textbf {\bibinfo {volume} {5}},\
  \bibinfo {pages} {033107} (\bibinfo {year} {2023})}\BibitemShut {NoStop}%
\bibitem [{\citenamefont {{Van Houcke}}\ \emph {et~al.}(2010)\citenamefont
  {{Van Houcke}}, \citenamefont {Kozik}, \citenamefont {Prokof’ev},\ and\
  \citenamefont {Svistunov}}]{VANHOUCKE201095}%
  \BibitemOpen
  \bibfield  {author} {\bibinfo {author} {\bibfnamefont {K.}~\bibnamefont {{Van
  Houcke}}}, \bibinfo {author} {\bibfnamefont {E.}~\bibnamefont {Kozik}},
  \bibinfo {author} {\bibfnamefont {N.}~\bibnamefont {Prokof’ev}},\ and\
  \bibinfo {author} {\bibfnamefont {B.}~\bibnamefont {Svistunov}},\ }\bibfield
  {title} {\bibinfo {title} {Diagrammatic monte carlo},\ }\href
  {https://doi.org/https://doi.org/10.1016/j.phpro.2010.09.034} {\bibfield
  {journal} {\bibinfo  {journal} {Physics Procedia}\ }\textbf {\bibinfo
  {volume} {6}},\ \bibinfo {pages} {95} (\bibinfo {year} {2010})},\ \bibinfo
  {note} {computer Simulations Studies in Condensed Matter Physics
  XXI}\BibitemShut {NoStop}%
\bibitem [{\citenamefont {Krauth}(2007)}]{krauth06}%
  \BibitemOpen
  \bibfield  {author} {\bibinfo {author} {\bibfnamefont {W.}~\bibnamefont
  {Krauth}},\ }\href@noop {} {\emph {\bibinfo {title} {Statistical Mechanics:
  Algorithms and Computations:}}}\ (\bibinfo  {publisher} {Oxford},\ \bibinfo
  {year} {2007})\BibitemShut {NoStop}%
\bibitem [{\citenamefont {Albert}\ and\ \citenamefont
  {Jiang}(2014)}]{albert2014symmetries}%
  \BibitemOpen
  \bibfield  {author} {\bibinfo {author} {\bibfnamefont {V.~V.}\ \bibnamefont
  {Albert}}\ and\ \bibinfo {author} {\bibfnamefont {L.}~\bibnamefont {Jiang}},\
  }\bibfield  {title} {\bibinfo {title} {Symmetries and conserved quantities in
  lindblad master equations},\ }\href
  {https://doi.org/10.1103/PhysRevA.89.022118} {\bibfield  {journal} {\bibinfo
  {journal} {Phys. Rev. A}\ }\textbf {\bibinfo {volume} {89}},\ \bibinfo
  {pages} {022118} (\bibinfo {year} {2014})}\BibitemShut {NoStop}%
\bibitem [{\citenamefont {Misra}\ and\ \citenamefont
  {Sudarshan}(1977)}]{misra1977a}%
  \BibitemOpen
  \bibfield  {author} {\bibinfo {author} {\bibfnamefont {B.}~\bibnamefont
  {Misra}}\ and\ \bibinfo {author} {\bibfnamefont {E.~C.~G.}\ \bibnamefont
  {Sudarshan}},\ }\bibfield  {title} {\bibinfo {title} {The {{Zeno}}'s paradox
  in quantum theory},\ }\href@noop {} {\bibfield  {journal} {\bibinfo
  {journal} {Journal of Mathematical Physics}\ }\textbf {\bibinfo {volume}
  {18}},\ \bibinfo {pages} {756} (\bibinfo {year} {1977})}\BibitemShut
  {NoStop}%
\bibitem [{\citenamefont {Chaudhari}\ \emph {et~al.}(2022)\citenamefont
  {Chaudhari}, \citenamefont {Kelly}, \citenamefont {Valencia-Tortora},\ and\
  \citenamefont {Marino}}]{Chaudhari_2022}%
  \BibitemOpen
  \bibfield  {author} {\bibinfo {author} {\bibfnamefont {A.~P.}\ \bibnamefont
  {Chaudhari}}, \bibinfo {author} {\bibfnamefont {S.~P.}\ \bibnamefont
  {Kelly}}, \bibinfo {author} {\bibfnamefont {R.~J.}\ \bibnamefont
  {Valencia-Tortora}},\ and\ \bibinfo {author} {\bibfnamefont {J.}~\bibnamefont
  {Marino}},\ }\bibfield  {title} {\bibinfo {title} {Zeno crossovers in the
  entanglement speed of spin chains with noisy impurities},\ }\href
  {https://doi.org/10.1088/1742-5468/ac8e5d} {\bibfield  {journal} {\bibinfo
  {journal} {Journal of Statistical Mechanics: Theory and Experiment}\ }\textbf
  {\bibinfo {volume} {2022}},\ \bibinfo {pages} {103101} (\bibinfo {year}
  {2022})}\BibitemShut {NoStop}%
\bibitem [{\citenamefont {Secl\`{\i}}\ \emph {et~al.}(2022)\citenamefont
  {Secl\`{\i}}, \citenamefont {Capone},\ and\ \citenamefont
  {Schir\`o}}]{secli2022steady}%
  \BibitemOpen
  \bibfield  {author} {\bibinfo {author} {\bibfnamefont {M.}~\bibnamefont
  {Secl\`{\i}}}, \bibinfo {author} {\bibfnamefont {M.}~\bibnamefont {Capone}},\
  and\ \bibinfo {author} {\bibfnamefont {M.}~\bibnamefont {Schir\`o}},\
  }\bibfield  {title} {\bibinfo {title} {Steady-state quantum zeno effect of
  driven-dissipative bosons with dynamical mean-field theory},\ }\href
  {https://doi.org/10.1103/PhysRevA.106.013707} {\bibfield  {journal} {\bibinfo
   {journal} {Phys. Rev. A}\ }\textbf {\bibinfo {volume} {106}},\ \bibinfo
  {pages} {013707} (\bibinfo {year} {2022})}\BibitemShut {NoStop}%
\bibitem [{\citenamefont {Kamenev}(2011)}]{kamenev_2011}%
  \BibitemOpen
  \bibfield  {author} {\bibinfo {author} {\bibfnamefont {A.}~\bibnamefont
  {Kamenev}},\ }\href {https://doi.org/10.1017/CBO9781139003667} {\emph
  {\bibinfo {title} {Field Theory of Non-Equilibrium Systems}}}\ (\bibinfo
  {publisher} {Cambridge University Press},\ \bibinfo {year}
  {2011})\BibitemShut {NoStop}%
\end{thebibliography}


%

\end{document}